\documentclass[12pt,english]{article}
\usepackage[T1]{fontenc}
\usepackage[latin9]{inputenc}
\usepackage[a4paper]{geometry}
\usepackage{libertine}
\usepackage{libertinust1math}
\usepackage[T1]{fontenc}
\usepackage{babel}
\usepackage{array}
\usepackage{verbatim}
\usepackage{rotfloat}
\usepackage{mathtools}
\usepackage{url}
\usepackage{multirow}
\usepackage{amsmath}
\usepackage{amssymb}
\usepackage{graphicx}
\usepackage{setspace}
\usepackage[authoryear]{natbib}
\usepackage[unicode=true,
 bookmarks=true,bookmarksnumbered=false,bookmarksopen=false,
 breaklinks=false,pdfborder={0 0 1},backref=false,colorlinks=false]
 {hyperref}
\hypersetup{pdftitle={Use Lyx for typesetting: A template for empirical research projects},
 pdfauthor={First Author and Second Author}}

\usepackage[T1]{fontenc}  % using tilde in url
\usepackage[makeroom]{cancel}  % strike words
\usepackage{tikz}  % circle a letter
\usepackage[final]{showlabels}

\usepackage{threeparttable}  % table with footnote
\usepackage{multirow}
\makeatletter

\providecommand{\tabularnewline}{\\}

\hypersetup{
    colorlinks=true,       % false: boxed links; true: colored links
    linkcolor=blue,          % color of internal links
    citecolor=blue,        % color of links to bibliography
    filecolor=blue,      % color of file links
    urlcolor=blue,           % color of external links
    breaklinks=true
}

\usepackage{booktabs}
\usepackage{longtable}
\usepackage{pdflscape}

\newcommand{\bfI}{\ensuremath{\mathbf{I}}}
\makeatother

\begin{document}
\begin{onehalfspace}

\title{Deep generative modeling for financial time series with application in VaR: a comparative review 
   \thanks{Corresponding author: Steve Guo 
     (\href{mailto:steve.guo@wellsfargo.com}{steve.guo@wellsfargo.com}).  
     All authors are with Wells Fargo.  Email addresses are at the end of the paper.  
    We thank the following for providing helpful comments on earlier drafts, Paul Romanelli and Jeffrey Kung.\newline
    \textcopyright 2024 Wells Fargo Bank, N. A. All rights reserved.  
    The views expressed in this publication are our personal views and do not necessarily reflect the views of Wells Fargo Bank, N.A., its parent company, affiliates and subsidiaries. 
   }
}

\end{onehalfspace}

\author{Lars Ericson, Xuejun Zhu, Xusi Han, Rao Fu,\\ Shuang Li, Steve Guo, Ping Hu}

\date{This Version: January 18, 2024}

\maketitle

% manual add after \maketitle
% ------------------------- no page number of page 1 --------------------------------------
\thispagestyle{empty} 
% -----------------------------------------------------------------------------------------------

\medskip{}
%\begin{FlushLeft}
%\RaggedRight% \begin{flushleft} replaced

\normalfont

\begin{abstract}
\begin{spacing}{0.9}
\noindent

In the financial services industry, forecasting the risk factor
distribution conditional on the history and the current market environment
is the key to market risk modeling in general and value at risk (VaR) model in particular. As one of the most widely adopted VaR models in commercial banks, Historical simulation (HS) uses the empirical distribution of daily returns in a historical window
as the forecast distribution of risk factor returns in the next day.
The objectives for financial time series generation are to generate synthetic data paths with good variety, and similar distribution and dynamics to the original historical data. In this paper, we apply multiple existing deep generative methods (e.g., CGAN, CWGAN, Diffusion, and Signature WGAN) for conditional time series generation, propose and test two new methods for conditional
multi-step time series generation, namely Encoder-Decoder CGAN and Conditional TimeVAE. Furthermore, we introduce a comprehensive framework with a set of KPIs to measure the quality of the generated
time series for financial modeling. The KPIs cover distribution distance, autocorrelation and backtesting.  All models (HS, parametric and neural networks) are tested on both historical USD yield curve data and additional data simulated from GARCH and CIR processes.  The study shows that top performing models are HS, GARCH and CWGAN models.  Future research directions in this area are also discussed.

\vspace{1cm}

\medskip{}

\noindent \textbf{Journal-classification:} C15, C32, C45, C53
\end{spacing}
\end{abstract}
\clearpage
\pagenumbering{arabic} 

\section{Introduction\label{sec:Intro}}
In financial risk management and analytics, it is a matter of great importance
to forecast the future
\footnote{Given that the financial markets are efficient generally, forecast in this context means a distribution forecast rather than a point forecast.}
. For example, forecasting the timing and direction
of stock market moves is important for investment decisions.  Forecasting
the correlation of returns is important for asset allocation decisions
(in the context of the Markowitz mean-variance framework).  And forecasting
the distribution of profit and loss (P\&L) of a portfolio is important
for market and counterparty risk management and regulatory capital
calculation using Value at Risk (VaR) and even Expected Shortfall
measures. VaR is a quantile of the portfolio P\&L distribution in
a future period. P\&L distribution is calculated from the risk factor
(return) distribution for a given portfolio using valuation models. Forecasting the risk
factor distribution conditional on the history and the current market
environment is the key to a good VaR model. One of the most widely
used VaR models at commercial banks adopts historical simulation (HS)
(\cite{perignon2010}), which uses the empirical distribution of daily
returns in a historical window as the forecast distribution of risk
factor returns in the next day. See \cite{bgv1999} and \cite{bg2001} for
descriptions of historical simulation and the improved filtered historical
simulation (FHS). However, the HS method is based on limited historical
scenarios and may not provide an accurate description of the
tails of the distributions for VaR calculation. 

Generative Artificial Intelligence (or Generative AI) refers to deep-learning models that can generate high-quality text, images, and other content based on the data they were trained on
\footnote{\url{https://research.ibm.com/blog/what-is-generative-AI}}.  Generative AI has the potential to bring breakthrough changes to many industries. Recently, an AI chatbot, ChatGPT, a generative large language model (LLM) introduced by OpenAI, is redefining the business of online searching and content creation. In image processing, Generative Adversarial Networks (GANs) have been a success to generate real-like images and enhance image resolutions. After image and text applications were developed, GANs were expanded to financial time series generation \cite{https://doi.org/10.48550/arxiv.1904.11419}, which allows us to enrich the available data for model development and testing. These upsides have lead to a high level of investment in GAN technology in 2022 and 2023
\footnote{\url{https://www.nytimes.com/2023/01/07/technology/generative-ai-chatgpt-investments.html}.}
. These are some of the institutions that have publicized GAN applications, for example, Fujitsu generates applicant-friendly loan denial explanations (\cite{Fujitsu20}). UBS uses synthetic data to enable usage or sharing of information
while protecting real data, to address gaps and weaknesses of real
data, to create data around rare events from crises to fraud that
is scarce by nature, to overcome training data shortages (\cite{Johnson20}). 
Interestingly, there is a legal case involving creation and use of synthetic customer data
\footnote{See ``Fake Accounts And Fake Data: The Good, The Bad And The Preventable'' \url{https://www.forbes.com/sites/forbestechcouncil/2023/04/24/fake-accounts-and-fake-data-the-good-the-bad-and-the-preventable/?sh=680c34886678}}. JP Morgan has been engaging in high profile research and methodology development of deep hedging
\footnote{\url{https://www.risk.net/derivatives/equity-derivatives/7921526/jp-morgan-testing-deep-hedging-of-exotics}}.  Deep hedging is outside the scope of this study.

The objectives for financial time series generation are to generate
synthetic data that look real and with good variability. If the
synthetic data are too different from real data, models that are trained
on synthetic data may not be generalizable to real data. However, if
the synthetic data are too close to real data (e.g. having a very
high correlation), then they may not be useful complements
to real data. Furthermore, for VaR and market risk modeling, the key
focus is to forecast the conditional distribution of P\&L for a specific
future horizon, given the history and recent market conditions. In
this paper, we apply multiple existing deep learning methods
for conditional multi-step time series generation, and they are described in Section 2. We propose and test two new deep learning methods for conditional
multi-step time series generation in Section 2, which are Encoder-Decoder
CGAN (\cite{Fu22}) and Conditional TimeVAE (VAE for short). We test these methods on both simulated data (Sections 3 and 4.4) and real USD yield curves (Section 4.5).

Furthermore, we introduce a comprehensive framework
to measure the quality of the generated time series for financial
modeling. For example,
financial markets have episodes of high and low market volatility.
Once in a high (low) volatility market episode, it is expected that
the market stays in the high (low) volatility episode for certain period
of time. This phenomenon is called volatility clustering. It is important
that the forecast of return distribution be conditional on the volatility
regime as well as other market information. For this study, only models that generate conditional distribution are considered. The key performance indicators (KPIs) for model comparison cover distribution distance, autocorrelation and backtesting. More details are given in Section 4.  Model testing results are presented in Section 4, and we conclude in Section 5.

\section{Methodology review\label{sec:methodology review}}
This study covers three categories of models for forecasting (or simulating)
future distribution of risk factors.
\begin{itemize}
\item The first category is historical simulation, which is widely used
by commercial banks for forecasting short term distributions for VaR purpose. This category includes plain historical simulation
and improved filtered historical simulation (FHS). 
\item The second category is parametric models. This category includes widely
used parametric models such as autoregressive models (AR), GARCH,
Vasicek model and the popular Nelson-Siegel (NS) representation for yield curves (\cite{dieboldli2006}).
AR and GARCH can be used to model yields or returns, while Vasicek model and NS representation is applied to yield level in general.
\item The third category is deep learning models. This category includes
the vanilla conditional GAN (CGAN) model, CGAN with Wasserstein distance (CWGAN),
CGAN with LSTM layers, Diffusion model, Signature CWGAN model and
Time VAE models.  Since only conditional models are covered, there is a C (as in CGAN) for conditional in each deep learning model name.
\end{itemize}
Machine learning, deep learning and neural network are closely related and have been used interchangeably in general.  We use machine learning in a narrow sense: it refers to traditional machine learning models with XGBoost and Random Forecast as examples.  A deep learning model is ``deep'' when its structure is composed of many layers of neural networks.  In this study, machine learning models are not covered, and deep learning and neural networks are used interchangeably.

Due to time constraint, this study does not include transformer deep learning model.

We focus on GANs which reproduce the whole distribution of one or more market observable prices or rates.  Recent work on Tail-GANs focuses on generating the tail distribution alone, either of the observable or of portfolios or trading strategies constructed from the observable (\cite{cont2023tailgan}).\footnote{We appreciate Prof. Rama Cont for bringing his research to our attention.} Tail distribution-only generation is beyond the scope of this study.

The 14 models covered in this study are listed in Table~\ref{tab:modellist}.  In Section~\ref{subsec:parametric models}, Vasicek model is also discussed.  Vasicek model is a continuous time model and has similar properties as an AR(1) model with time series data.  In limited model testing, these two models perform similarly.  Therefore, only the simpler AR(1) model is listed in Table~\ref{tab:modellist} and included in comprehensive model testing.

\begin{table}[htbp]
  \centering
  \caption{List of models}
    %\begin{tabular}{clll}
\begin{tabular}{p{0.05\linewidth}  p{0.20\linewidth}  p{0.2\linewidth} p{0.3\linewidth}}
\toprule
    No.   & Model Category & Model Name & Note \\
\midrule
    1     & Historical simulation (HS) & PHS    & Plain Historical simulation \\
\midrule
    2     & Historical simulation (HS) & FHS   & Filtered historical simulation (EWMA) \\
\midrule
    3     & Parametric (PM) & AR    & Autoregressive model of order 1 of yields \\
\midrule
    4     & Parametric (PM) & AR\_RET & Autoregressive model of order 1 of return of yields \\
\midrule
    5     & Parametric (PM) & GARCH & AR(1)+GARCH-normal model of yields \\
\midrule
    6     & Parametric (PM) & GARCH\_RET & AR(1)+GARCH-normal model of return of yields \\
\midrule
    7     & Parametric (PM) & GARCHt\_RET & AR(1)+GARCH-t (with t distribution) model of return of yields \\
\midrule
    8     & Parametric (PM) & NS\_VS & Nelson-Siegel 3 factor model with Vasicek dynamics \\
\midrule
    9     & Neural network (NN) & CGAN-FC  & Conditional GAN with fully connected layers \\
\midrule
    10    & Neural network (NN) & CGAN-LSTM & Conditional GAN with LSTM layers \\
\midrule
    11    & Neural network (NN) & CWGAN & Conditional GAN with Wasserstein loss \\
\midrule
    12    & Neural network (NN) & DIFFUSION & Diffusion model \\
\midrule
    13    & Neural network (NN) & SIG & Signature CWGAN with CNN layers \\
\midrule
    14    & Neural network (NN) & VAE & Conditional Time VAE model \\
\bottomrule
    \end{tabular}%
  \label{tab:modellist}%
\end{table}%

\begin{figure}
\centering
\includegraphics[scale=0.2]{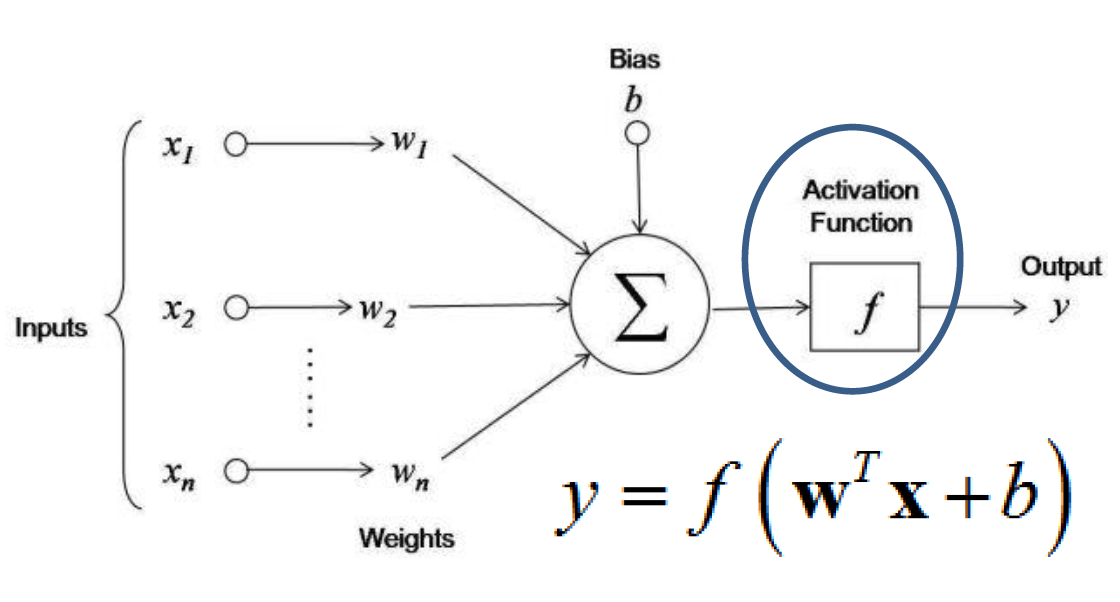}
\caption{A basic neuron representation\label{fig:neu}}
\end{figure}

Common notations in the paper are collected into Table \ref{table_2:notations}.  To introduce terminology on neural networks, a basic neuron representation
\footnote{Source, \url{https://jameskle.com/writes/neural-networks-101}}
 is shown in Figure~\ref{fig:neu}.  The input to the neuron is $x_1, \cdots, x_n$, which are combined with weights $w_i$ and bias $b$ (or slope and intercept in linear regression) to form the raw output $\sum_{i=1}^n x_iw_i + b$.   Since $w_i$ and $b$ can be any real numbers, the raw output may fall outside the expected range for an application (e.g. outside the range [0,255] for pixel values, or [0, 1] for probability estimate).  The raw output is passed to an activation function $f$ to convert the raw output into the expected range.  The following are several widely used activation functions.\\
\textbf{ReLU (rectified linear unit)} or positive part\\
\begin{equation}
f(x)=\max(0,x)
\end{equation}
This function converts the raw output from the real line to the positive real line.\\
\textbf{Sigmoid or logistic function}\\
\begin{equation}
f(x)=\frac{1}{1+e^{-x}}
\end{equation}
This function converts the raw output from the real line to the unit interval $(0, 1)$, appropriate for probability estimates.

Most activation functions are non-linear and inject an element of non-linearity to the neuron.  Interestingly, there is a linear activation function that is appropriate for some applications and is used in our study,
\begin{equation}
f(x)=x
\end{equation}
In our application, the outputs are expected to have mean zero and unit variance and cover the whole real line.  A linear activation function is suitable for such cases.

A neural network or deep learning model consists in multiple (and potentially many) neurons side by side (width of a neural network) or sequentially (depth of a neural network) or both, with each neuron having its own weight, bias and activation function. Activation functions for the hidden layers are almost always non-linear.

The representation in Figure~\ref{fig:neu} is for a basic neuron.  There are enhanced representations for special applications, for example, convoluted neural network (CNN) for image application and long short-term memory (LSTM) for recurrent data including natural language processing (NLP) and time series application.

\begin{table}
\centering{}\label{table_2:notations}\caption{List of notations and model parameters}
%\begin{tabular}{|l|l|}
\begin{tabular}{p{0.15\linewidth}  p{0.85\linewidth}}
\hline 
Notation & Meaning\tabularnewline
\hline 
\hline 
$t$ & Date of the scenario\tabularnewline
\hline 
$d$ & Number of risk factors in multivariate time series\tabularnewline
\hline 
$R_{t}$ & Level of $d$-dimensional risk factor at time $t$ \tabularnewline
\hline 
$R_{t}(\tau)$ & Convenient representation of yield curve.  Level of risk factor (yield) of tenor $\tau$ at time $t$.  The number of different $\tau$ values is $d$.\tabularnewline
\hline 
$R_{t}^{(i)}$ & Level of risk factor $i$ or tenor $\tau_i, i\in\{1,2,...,d\}$ at time $t$ \tabularnewline
\hline 
$\tilde{R}_{t}$ & Synthetic level produced by generative model for $d$-dimensional
risk factor and time $t$ \tabularnewline
\hline 
$x_{t}$ & Return of $d$-dimensional risk factor at time $t$, $x_t=R_t-R_{t-1}$. \tabularnewline
\hline 
$x_{t,i:j}$ & Return $x$ for the period $t+i,\cdots,t+j$\tabularnewline
\hline 
$\tilde{x}_{t}$ & Synthetic return produced by generative model for $d$-dimensional
risk factor and time $t$ \tabularnewline
\hline 
$y$ & Conditions used as input to the generative model\tabularnewline
\hline 
$z$ & Random noise for generative model\tabularnewline
\hline 
$G$ & Generator in GAN\tabularnewline
\hline 
$D$ & Discriminator in GAN\tabularnewline
\hline 
$\sigma_{t}$ & Volatility of return for date t\tabularnewline
\hline 
$\tilde{\sigma}_{t}$ & Volatility forecast of return for date t\tabularnewline
\hline 
$\varepsilon_{t}$ & White noise for date t\tabularnewline
\hline 
$\mathbb{P}_{r}$ & Real data distribution\tabularnewline
\hline 
$\mathbb{P}_{g}$ & Synthetic (generated) data distribution\tabularnewline
\hline 
$E[...]$ & Expectation of a variable/function \tabularnewline
\hline 
\end{tabular}
\end{table}

\subsection{Historical simulation models\label{subsec:historical simulation model}}

Historical simulation (HS) and and the improved filtered historical
simulation (FHS) are widely used by commercial banks for VaR models.
They are simple nonparametric models that use empirical distribution
for distribution forecast.

\subsubsection{Plain Historical Simulation (PHS) model}

In the historical simulation (HS) model, we draw daily returns from
the empirical distribution (or empirical distribution function (EDF)
\footnote{Or empirical cumulative distribution function (ECDF)).}
.
It is simple to implement and naturally takes into account the correlation
(and more general forms of dependence) among variables. It is widely used by
commercial banks for their VaR models, (\citet{perignon2010}).

EDF is an estimate of the distribution that generates the data and
converges to the underlying distribution under certain assumptions
\footnote{See \url{http://faculty.washington.edu/yenchic/17Sp\_403/Lec1\_EDF.pdf}}
.
However, EDF is an estimate of the unconditional distribution of the
data, and does not take into account the serial dependence (over time)
in the data. For example, there are regimes of high volatility and
low volatility in financial markets. Volatility regimes tend to
be persistent in the sense that the market does not switch from high
to low volatility regimes daily.  Once in a regime, the market
tends to stay in that regime for a period of time. If we know that
the market is in a high volatility regime today, then in the next
few days, the market tends to stay in the high volatility regime.
This volatility clustering is captured by the widely used GARCH volatility model, and can be used to improve the distribution
forecast, which is the idea behind the filtered historical simulation (FHS) model. To differentiate from FHS, this simple HS is called plain
HS (PHS).

HS for one day horizon is straightforward, while HS for multiple day
ahead scenario is not obvious. The mechanics of HS is described below.

For a given business date $t_{0}$, a sample of daily returns of size
251 for $t_{0}+1$ from EDF is formed using daily returns of previous
251 days: $x(t_{0}-i),i=0,1,\cdots,250$
\footnote{Due to data processing delays, the time window for selecting the historical scenarios for a bank may not be the most recent 251 days. For example, the time window could be $t_{0}-251,\cdots,t_{0}-1$.}
.  In this paper, an year is assumed to have 251 business dates.  It is common to use 250 and 252 business dates as well.

Similarly, a sample of two-day returns of size 251 for period $(t_{0},t_{0}+2)$
from EDF is formed using previous 251 points of overlapping two-day
returns: $x(t_{0}-i,2),i=0,1,\cdots,250$.  For this illustration, the notation is abused a little bit when the second index is used to indicate the returns horizon ($x(t_{0}-i,2)$ for 2-day return), which is the sum of two daily returns.  The two day return
is calculated as $x(t_{0}-i,2)=x(t_{0}-i)+x(t_{0}-i-1)$.  See Table~\ref{tab:hs2} for an illustration.

% Table generated by Excel2LaTeX from sheet 'HS'
\begin{table}[htbp]
  \centering
  \caption{Illustration of historical simulation for 1-day and 2-day returns}
    \begin{tabular}{clll}
\toprule
%    \multicolumn{1}{r}{\multirow{2}[0]{*}{scenario}} & \multicolumn{3}{c}{Future date} \\
\multirow{3}{*}{Scenario} & \multicolumn{3}{c}{Future date}\\
\cline{2-4}
          & $t_0+1$ & $t_0+1$ & $t_0+2$ \\
\cline{2-4}
          & 1-day & 2-day & 1-day (implied) \\
\midrule
    1     & $x(t_{0})$ & $x(t_{0}-1)+x(t_{0})$ & $x(t_{0}-1)$ \\
    2     & $x(t_{0}-1)$ & $x(t_{0}-2)+x(t_{0}-1)$ & $x(t_{0}-2)$ \\
    3     & $x(t_{0}-2)$ & $x(t_{0}-3)+x(t_{0}-2)$ & $x(t_{0}-3)$ \\
    4     & $x(t_{0}-3)$ & $x(t_{0}-4)+x(t_{0}-3)$ & $x(t_{0}-4)$ \\
    5     & $x(t_{0}-4)$ & $x(t_{0}-5)+x(t_{0}-4)$ & $x(t_{0}-5)$\\
\bottomrule
    \end{tabular}%
  \label{tab:hs2}%
\end{table}%

By this example, a one day return for $t_0+1$ is $x(t_{0}-i,1)=x(t_{0}-i)$, a two day return for period $(t_{0},t_{0}+2)$ is
$x(t_{0}-i,2)=x(t_{0}-i-1)+x(t_{0}-i)$.  By this logic, a one day (``forward'') return for date $t_0+2$ is naturally taken as the difference between the two-day return and the one-day return above, namely, $x(t_{0}-i,2)-x(t_{0}-i)=x(t_{0}-i-1),i=0,1,\cdots,250$. See Table~\ref{tab:hs2} again for an illustration.

Selection of scenarios for dates $t_0+1,t_0+2$ and $t_0+3$ is illustrated in Table \ref{tab:hs}, with each row in the table representing a simulation path.  The benefit of HS is that it is easy to implement, and is not limited by dimensionality of the data.  The drawback is that it is limited to historical (return) scenarios.  It can be seen in Table \ref{tab:hs} that HS scenarios are just historical data in reverse order of time.  As a result, it is expected to capture autocorrelation, dependence across tenors and distribution of historical returns. 
% manual add 
% -------------------------------------------------------------------------
% Table generated by Excel2LaTeX from sheet 'HS'
\begin{table}[htbp]
  \centering
  \caption{Illustration of historical simulation for 1-day returns}
    \begin{tabular}{clll}
\toprule
%    \multicolumn{1}{r}{\multirow{2}[0]{*}{scenario}} & \multicolumn{3}{c}{Future date} \\
\multirow{2}{*}{Scenario} & \multicolumn{3}{c}{Future date}\\
\cline{2-4}
          & $t_{0}+1$ & $t_{0}+2$ & $t_{0}+3$ \\
\midrule
    1     & $x(t_{0})$ & $x(t_{0}-1)$ & $x(t_{0}-2)$ \\
    2     & $x(t_{0}-1)$ & $x(t_{0}-2)$ & $x(t_{0}-3)$ \\
    3     & $x(t_{0}-2)$ & $x(t_{0}-3)$ & $x(t_{0}-4)$ \\
    4     & $x(t_{0}-3)$ & $x(t_{0}-4)$ & $x(t_{0}-5)$ \\
    5     & $x(t_{0}-4)$ & $x(t_{0}-5)$ & $x(t_{0}-6)$\\
\bottomrule
    \end{tabular}%
  \label{tab:hs}%
\end{table}%

% -------------------------------------------------------------------------
%
\subsubsection{Filtered Historical Simulation (FHS) model}
FHS takes into account volatility clustering and generates improved
conditional distribution forecasts with volatility scaling. The idea
is simple (\citet{bg2001}):
\begin{enumerate}
\item For given historical market data $R_t$, calculate returns $x_t$ and estimate
of volatility $\sigma_t$ for each date $t$. Volatility estimate
can be from a generalized autoregressive
conditional heteroskedasticity (GARCH) model or exponentially weighted moving average (EWMA) model.  GARCH model is described later in Section~\ref{subsec:AR+GARCH-model}.  For EWMA, see the well known VaR website
\footnote{Value at Risk Theory and Practice, \url{https://www.value-at-risk.net/exponentially-weighted-moving-average-ewma/}.}.
\item Calculate devolatized (devol) return $\hat{x}_t=x_t/\sigma_t$
\item For the problem of forecasting next day conditional distribution at
a given business date $t_{0}$, make a forecast of volatility for
next day $\tilde{\sigma}_{t_{0}+1}$
\item Select 251 historical devol scenarios in the previous year $\hat{x}_{t_{0}-i},i=0,1,\cdots,250$ 
\item Revolatize the selected scenarios $\tilde{x}^{i}_{t_{0}}=\hat{x}_{t_{0}-i}\tilde{\sigma}_{t_{0}+1},i=0,1,\cdots,250$
\item $\tilde{x}^{i}_{t_{0}}$ is a set of 251 data points that form
the EDF for date $t_{0}+1$.
\end{enumerate}
Note that FHS involves devol (\#2) and revol (\#5) steps, which are
key steps for capturing volatility persistence. In the devol step,
raw returns are standardized for volatility to form a pool of (nearly)
i.i.d. returns for future scenarios. In the revol step, the
selections from the scenario pool is scaled by future volatility forecast,
so that if the market is in a high (low) volatility regime, the i.i.d. scenarios are scaled up (down).

% ---------------- manual add -------------------------------------------
% illustration of HS and FHS
FHS is a simple form of conditional model with the conditional variance
estimated with EWMA or GARCH model. Similar to Table~\ref{tab:hs}, selection of scenarios for FHS
is illustrated in Table~\ref{tab:fhs}.

% Table generated by Excel2LaTeX from sheet 'HS'
\begin{table}[htbp]
  \centering
  \caption{Illustration of filtered historical simulation (FHS) for 1-day returns}
    \begin{tabular}{clll}
\toprule
%    \multicolumn{1}{r}{\multirow{2}[0]{*}{scenario}} & \multicolumn{3}{c}{Future date} \\
\multirow{2}{*}{Scenario} & \multicolumn{3}{c}{Future date}\\
\cline{2-4}
          & $t_{0}+1$ & $t_{0}+2$ & $t_{0}+3$ \\
\midrule
    1     & $\hat{x}(t_{0})\tilde{\sigma}(t_0+1)$ & $\hat{x}(t_{0}-1)\tilde{\sigma}(t_0+2)$ & $\hat{x}(t_{0}-2)\tilde{\sigma}(t_0+3)$ \\
    2     & $\hat{x}(t_{0}-1)\tilde{\sigma}(t_0+1)$ & $\hat{x}(t_{0}-2)\tilde{\sigma}(t_0+2)$ & $\hat{x}(t_{0}-3)\tilde{\sigma}(t_0+3)$ \\
    3     & $\hat{x}(t_{0}-2)\tilde{\sigma}(t_0+1)$ & $\hat{x}(t_{0}-3)\tilde{\sigma}(t_0+2)$ & $\hat{x}(t_{0}-4)\tilde{\sigma}(t_0+3)$ \\
    4     & $\hat{x}(t_{0}-3)\tilde{\sigma}(t_0+1)$ & $\hat{x}(t_{0}-4)\tilde{\sigma}(t_0+2)$ & $\hat{x}(t_{0}-5)\tilde{\sigma}(t_0+3)$ \\
    5     & $\hat{x}(t_{0}-4)\tilde{\sigma}(t_0+1)$ & $\hat{x}(t_{0}-5)\tilde{\sigma}(t_0+2)$ & $\hat{x}(t_{0}-6)\tilde{\sigma}(t_0+3)$\\
\bottomrule
    \end{tabular}%
  \label{tab:fhs}%
\end{table}%

\subsection{Parametric models\label{subsec:parametric models}}

Another widely used simulation framework is based on Monte Carlo
approach. Each model herein can be written in a parametric form with
a theoretical distribution or fixed moving process, governed either
by stochastic dynamics or time series. We explore a number of popular parametric
models. We use these as benchmarks/challenge models against the neural network models.
In the subsequent context, we simply specify the marginal dynamics
of a single risk factor for simplicity, i.e. we use $R_{t}$ and
$x_{t}$ instead of $R_{t}^{(i)}$ and $x_{t}^{(i)}$. The correlation
of different risk factors can be established through the standard
Gaussian copula approach for returns or random noise.

\subsubsection{Vasicek model\label{subsec:Vasicek-model}}

The Vasicek model is a mean-reverting stochastic process (\cite{vasicek1977}). It has a
long history of applications for interest rate movements and fixed-income
risk factor modeling in mathematical finance. 

The Vasicek model is usually used to model stochastic process with mean reverting behavior, for example, the level of interest rates.

The Vasicek model specifies the following stochastic differential
equation (SDE) of $R_{t}$.
\begin{equation}
dR_{t}=\kappa(\theta-R_{t})dt+\sigma dW_{t}\label{eq:Vasicek diffential}
\end{equation}
where $\kappa>0$ is the mean-reverting speed, $\theta$ is the long-term
mean level of $R_{t}$, and $\sigma$ is the volatility of the Brownian
motion $dW_{t}$. With the initial condition $R_{0}$ given and without
loss of generality, for an even partition
over $(0,T]:0=t_{0}<t_{1}=\delta<t_{2}=2\delta<\cdots<t_{n}=n\delta=T$,
the SDE (\ref{eq:Vasicek diffential}) can be solved iteratively as
the following.
\begin{equation}
R_{i+1}=R_{i}e^{-\kappa\delta}+\theta(1-e^{-\kappa\delta})+\sigma\sqrt{\frac{1-e^{-2\kappa\delta}}{2\kappa}}z_{t+1}\label{eq:Vasicek integral}
\end{equation}
where $R_{i}=R_{t_{i}}$ and $z_{t+1}$ is a random sample from standard
normal distribution.

Note that in simulation process, the parameters $\kappa,\theta,\sigma$
are first calibrated from historical data, for instance based on maximum-likelihood
estimate (see \citet{VasicekCalibration}), and then plugged into
equation (\ref{eq:Vasicek integral}) for future path generation with
pre-defined $\delta$, which could be 1-day, 2-day etc. In the multi-dimensional
case, i.e. there are a number of risk factors $R^{(1)},R^{(2)},\cdots,R^{(d)}$,
the correlation among these factors needs to be factored in the vector
version of equation (\ref{eq:Vasicek integral}) by correlating $d$
iid samples drawn from $N(0,1)$ to form the noise vector.

%\subsubsection{AR+GARCH model\label{subsec:AR+GARCH-model}}
\subsubsection{Autoregressive (AR) model\label{subsec:AR}}

In statistics, econometrics and signal processing, an autoregressive
(AR) model is a representation of a type of random process; as such,
it is used to describe certain time-varying processes in nature, economics,
behavior, etc. The autoregressive model specifies that the output
variable depends linearly on its own lagged values and on a stochastic
term (an imperfectly predictable term); thus the model is in the form
of a stochastic difference equation.

The autoregressive model is usually used to model stationary time series, for example, financial returns.

The notation $AR(p)$ indicates an autoregressive model of order $p$.
The $AR(p)$ model is defined as:
\begin{equation}
x_{t}=\phi_0+\sum_{i=1}^{p}\phi_{i}x_{t-i}+\varepsilon_{t}\label{eq:AR}
\end{equation}
where $\phi_{1,}\cdots,\phi_{p}$ are the parameters of the model,
and $\varepsilon_{t}\sim N(0,\sigma^2)$ is white noise. Here again, the $\{\phi\}_{i=1}^{p}$are
calibrated from historical time series data and used for future simulation. When $p=1$, the model is simplified to the popular AR(1) model,
\begin{equation}
x_{t}=\phi_{0}+\phi_{1}x_{t-1}+\varepsilon_{t}\label{eq:AR1}
\end{equation}

To assess the relationship between the above $AR(1)$ model and Vasicek
model, assume an AR(1) for $R_t$ and rewrite (\ref{eq:AR1}) for $R_t$ into the following.
\begin{equation}
R_{t}-R_{t-1}=\underbrace{(1-\phi_{1})}_{\kappa dt}\left (\underbrace{\frac{\phi_{0}}{1-\phi_{1}}}_{\theta}-R_{t-1} \right )+\underbrace{\sigma_{\varepsilon_{t}}}_{\sigma\sqrt{dt}}Z_t\label{eq:AR1_rw}
\end{equation}
Compared with Vasicek model (\ref{eq:Vasicek diffential}), it is easy to see (\ref{eq:AR1_rw}) is special a case of
(\ref{eq:Vasicek diffential}) with $\kappa dt=1-\phi_{1},\theta=\frac{\phi_{0}}{1-\phi_{1}},$
and $\sigma \sqrt{dt}=\sigma_{\varepsilon_{t}}$. But note
that this equivalence is under the assumption that $dt$ can be set
to 1 day. However $dt$ in differential form (\ref{eq:Vasicek diffential})
should be infinitesimally small in the sense of calculus. Therefore
the following distinctions should be made between Vasicek and $AR(1)$:
\begin{itemize}
\item The Vasicek model is a general continuous-time model while $AR(1)$
is a typical time-series model in discrete time. 
\item In view of (\ref{eq:AR1_rw}), when $\phi_{1}=1$, it is still an $AR(1)$
model but it no longer satisfies the assumption of a Vasicek model.  In fact, with $\phi_{1}=1$, it is a unit root process (\cite{df1979}), which is a random walk in discrete time and nonstationary.  For modeling, a unit root process is usually converted to a stationary process first, for example, by calculating returns.
\item Following further with the point above, when $\phi_{1}>1$, the Vasicek
model would break as $\kappa<0$. Instead of mean-reverting, it becomes
mean-avoiding (if the random variable is above the mean, it increases, moving further away from the mean) so that the solution (\ref{eq:Vasicek integral})
is divergent (or explosive).
\end{itemize}
In model testing later, Vasicek and AR(1) models have similar performance.  For such comparison, Vasicek and AR(1) models are used to model the same variable, which is $R_t$.  The simpler AR(1) model is selected for further comparison with other models.
\subsubsection{GARCH model\label{subsec:AR+GARCH-model}}
For AR(1) model \eqref{eq:AR1}, it is assumed that the error variance $\sigma^2$ is constant.  The autoregressive conditional heteroskedasticity (ARCH) model is
a statistical model for time series data that describes the variance
of the current error term (or innovation) as a function of the actual
sizes of the previous time periods' error terms. The ARCH model is
appropriate when the error variance in a time series follows an autoregressive
(AR) model; if an autoregressive moving average (ARMA) model is assumed
for the error variance, the model is a generalized autoregressive
conditional heteroskedasticity (GARCH) model. See \cite{arch} and \cite{garch}. In general, the $GARCH(p,q)$
has the following specification:
\begin{gather}
\varepsilon_{t}\sim N(0,\sigma_{t}^{2})\\
\sigma_{t}^{2}=\omega+\sum_{i=1}^{q}\alpha_{i}\varepsilon_{t-i}^{2}+\sum_{i=1}^{p}\beta_{i}\sigma_{t-i}^{2}\label{eq:GARCH}
\end{gather}
where $\omega,\alpha_{i},\beta_{i}$ are constants to be calibrated.
In applications, GARCH(1,1) model with $p=1$ and $q=1$ often works quite well. In this case, combining equations (\ref{eq:AR})
and (\ref{eq:GARCH}) yields the $AR(1)+GARCH(1,1)$ model below.
\begin{equation}
\begin{cases}
x_{t}=\phi_{0}+\phi_{1}x_{t-1}+\varepsilon_{t};\\
\varepsilon_{t}=\sigma_{t}Z_{t},Z_{t}\sim N(0,1)\\
\sigma_{t}^{2}=\omega+\alpha\varepsilon_{t-1}^{2}+\beta\sigma_{t-1}^{2}
\end{cases}\label{eq:AR+GARCH}
\end{equation}
Note that $AR(1)+GARCH(1,1)$ is a good benchmark to Vasicek model
in that it enables a version of stochastic volatility.  The autoregressive nature of conditional variance in GARCH captures volatility clustering in financial time series.
\subsubsection{Nelson-Siegel yield curve representation}
\begin{figure}
\centering
\includegraphics[scale=0.5]{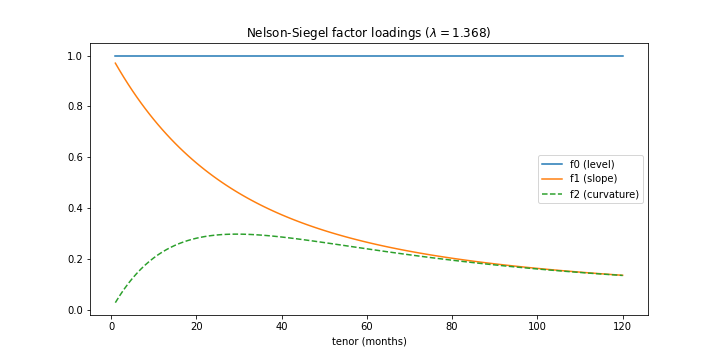}
\caption{Nelson-Siegel factor loadings\label{fig:ns3f}}
\end{figure}

The Nelson-Siegel (NS) representation of a curve is used for yield curve modeling in financial markets. Instead of being named a model, it is indeed a
representation of yield curves $R_{t}(\tau)$ in term structure ($\tau$) dimension rather
than the evolution process in time dimension as in \ref{subsec:Vasicek-model} and \ref{subsec:AR+GARCH-model}.
A curve $R_{t}(\tau)$ is approximated with parameters $\beta_{0},\beta_{1},\beta_{2},\lambda$ as the following (\cite{ns1987} as reformulated in \cite{dieboldli2006}),
\begin{equation}
R_{t}(\tau;\lambda)\approx\beta_{0}+\beta_{1}\left(\frac{1-e^{-\frac{\tau}{\lambda}}}{\tau/\lambda}\right)+\beta_{2}\left(\frac{1-e^{-\frac{\tau}{\lambda}}}{\tau/\lambda}-e^{-\frac{\tau}{\lambda}}\right)\label{eq:NS}
\end{equation}
Here the four parameters $\beta_{0},\beta_{1},\beta_{2},\lambda$ are
calibrated for any given curve with term structure $R_{t}(\tau_i),i=1,2,\cdots,d$ on date $t$.
The rationale of approximation (\ref{eq:NS}) relies on the fact that it captures
the key attributes of a curve such as level, slope and curvature.
In addition, with $\beta_{0},\beta_{1},\beta_{2},\lambda$ known, it
is straight-forward to obtain $R_{t}(\tau)$ for any value $\tau$.  To highlight the dependence of $\beta_{0},\beta_{1},\beta_{2}$ on time $t$, \eqref{eq:NS} is rewritten as
\begin{equation}
R_{t}(\tau;\lambda)\approx\beta_{0t}f_0(\tau)+\beta_{1t}\underbrace{\left(\frac{1-e^{-\frac{\tau}{\lambda}}}{\tau/\lambda}\right)}_{f_1(\tau)}+\beta_{2t}\underbrace{\left(\frac{1-e^{-\frac{\tau}{\lambda}}}{\tau/\lambda}-e^{-\frac{\tau}{\lambda}}\right)}_{f_2(\tau)}\label{eq:NSt}
\end{equation}
where $f_0(\tau)\equiv 1$.  In this representation, $\beta_{0t},\beta_{1t},\beta_{2t}$ are the three factors, and $f_0(\tau), f_1(\tau)$ and $f_2(\tau)$ are the factor loadings on tenor $\tau$.  The three factor loadings are plotted in  Figure~\ref{fig:ns3f}.  $f_0(\tau)$ is the loading for a level factor, with $f_1(\tau)$ the loading of a slope factor and $f_2(\tau)$ the loading for a curvature factor.

As noted in \cite{dieboldli2006}, $\lambda$ affects the decay rate of the slope and the peak of the curvature $f_2(\tau)$.  $\lambda$ is calibrated so that the curvature loading is maximized at 30 months.  A benefit of fixing $\lambda$ to a value (not time variant) is that the other parameters  $\beta_{0t},\beta_{1t},\beta_{2t}$ can be estimated with OLS.  This is a common practice followed in several studies
\footnote{Our notation is slightly different from the notation in \cite{dieboldli2006}.  $\lambda=1/\lambda^*$ (in the reference paper).  In the paper $\lambda^*=0.069$.  However,  the reference paper uses tenor in months, converting to using annual tenor, $\lambda^{*M}\tau^M=[\lambda^{*M}*12]\tau^Y = [0.069\times 12]\tau^Y =0.7308 \tau^Y$. In our study, $\lambda=1/0.7308=1.368$.}
.

A direct usage of Nelson-Siegel representation is similar to PCA (principal
component analysis) in the sense that the original curve $R_{t}$ has $d$
tenors and the approximated $R_{t}$ only has three parameters ($\lambda$ is often fixed and not calibrated). When $d>3$,
this is a reduction in dimension in the simulation of $R_{t}$. With historical yield curve data, the parameters $\beta_{0t},\beta_{1t},\beta_{2t}$ can be calibrated for every date $t$.  Dynamics of these parameters can be modeled using Vasicek model (\ref{eq:Vasicek integral}) or $AR+GARCH$
model (\ref{eq:AR+GARCH}).  Once the parameters are simulated for a future period, the yield curve $R_{t}(\tau)$ can be simulated through Nelson-Siegel
representation.  In this study, Nelson-Siegel representation is applied to the yield curve in levels.

Model testing results show that NS representation with Vasicek dynamics for the factors $\beta_{0t},\beta_{1t},\beta_{2t}$ performs better than that with AR+GARCH dynamics. Therefore, for model testing in Section~\ref{sec:performance comparison}, only NS with Vasicek dynamics is presented in model comparison.
\subsection{Deep generative models \label{subsec:deep learning models}}

Given samples from a data distribution, our goal is to generate synthetic
data as close to the real data distribution as possible. The key methodology
lies in the way the distance between two probability
distributions is quantified. Based on the approach to measuring the data distribution,
deep generative models can be widely categorized as likelihood-based
and likelihood-free models. Among the NN models considered, variational autoencoder (VAE) and Diffusion models are examples
of likelihood-based models, while GAN is a unique likelihood-free
model. 

GAN is inspired by game theory: the two models, a generator and a discriminator,
compete with each other via an adversarial training process. We use
conditional GAN (CGAN) to learn conditional distributions in our study.
The generator learns to produce samples that resemble real data,
while the discriminator is simultaneously trained to distinguish between
real and synthetic samples. As the two models are trained together
in a zero-sum game, improvements to one model come at the expense
of the other model. It is thus challenging to train a GAN model and
it may suffer from issues of mode collapse and vanishing gradients.
To address the issues with GAN training, WGAN was proposed using Wasserstein
distance as an alternative loss function. There are also variations
in GAN with different architectures in the generator and discriminator.
In LSTM CGAN (LSTM for short), the generator adopts an encoder-decoder structure to capture
the volatility dynamics and more complex autocorrelation structures
in time series. Signature Conditional Wasserstein GAN (SIGCWGAN) uses
an autoregressive network for the generator and uses the mean squared error (MSE) distance between signatures
of real and synthetic data as the discriminator. 

Besides likelihood-free models, we also evaluate likelihood-based
models including VAE and diffusion models. The idea of VAE is to learn
a low-dimensional latent representation of the training data called
latent variables (variables which can not be directly observed but are
rather inferred through a mathematical model), which we assume to
have generated our actual training data. Diffusion model is inspired
by non-equilibrium thermodynamics. Diffusion model maps data to noise
through the successive addition of Gaussian noise in the forward process,
and iteratively removes noise in the reverse process. New samples
are generated by simply passing randomly sampled noise through the
learned denoising process.

A simple way to characterize likelihood-free and likelihood-based models is that likelihood-free models do not take into account the statistical property in the data and just treat the data similar to images, while likelihood-based models utilizes the statistical properties in the data.

We use historical data as continuous conditions to all deep generative
models evaluated in this paper. Some of the continuous conditions
significantly enhance the model performance on financial time series.
Some previous papers for generative models directly use levels of
financial time series (such as stock prices and trading volume) for training and generation \citet{Desai2021,Srinivasan2022}.
In our study, we calculate arithmetic returns or log returns
from levels (as discussed in section \ref{subsec:testing data}),
and model returns in all deep generative models. The usage of returns
is more appropriate for financial time series than levels, as returns
are more stationary than levels and more widely used in financial applications. We provide a
brief summary of deep generative models covered in our study
below. 

The three groups of deep learning models (GAN, VAE and Diffusion) are described below.  For a preview of the size of the neural networks fitted to USD Libor curve data, the number of model parameters are presented in Table~\ref{tab:sizeUSDYC1}.\\

\begin{table}
\centering
\caption{Number of model parameters for USD Libor dataset}
\label{tab:sizeUSDYC1}
\begin{tabular}{lrrrr}
\toprule
    Model & GEN/DEC &   DIS/ENC &     TOTAL & Code Library \\
\midrule
  CGAN-FC & 137,434 &   112,385 &   249,819 &   Tensorflow \\
    CWGAN & 797,658 &   158,465 &   956,123 &        Torch \\
DIFFUSION &  39,279 &         0 &    39,279 &      GluonTS \\
CGAN-LSTM &   9,929 &     8,262 &    18,191 &   Tensorflow \\
      SIG &  11,612 & 1,781,730 & 1,793,342 &        Torch \\
      VAE &  86,316 &   156,420 &   242,736 &   Tensorflow \\
\bottomrule
\end{tabular}
\end{table}

To facilitate discussion of model architecture, the data structure for deep learning model is described first.
\subsubsection{Data structure\label{subsec:datastructure}}
 Imaging application is an important use case for the development of deep learning models and helps to shape the deep learning models for later applications.  For an application, the image size is usually fixed.  For example, images in the well known MNIST
\footnote{See \url{http://yann.lecun.com/exdb/mnist/}.}
 image dataset are of fixed size $28 \times 28$.  The setup of neural network for financial time series follows image application.
Since deep learning models usually expect a dataframe or sequence of a fixed size,
time series data in 2-dimensional arrays are sliced into time windows
of fixed size. For a typical model in the paper, the 9-tenor USD yield curve
\footnote{For any business date, there are nine interest rates for different maturities.  An yield curve is defined as yield (interest rate) as a function of time to maturity or tenor.}
is sliced (or windowed/sequenced) into windows of 20 business dates.  Each sequence is a 2-dimensional array (of dimension $20 \times 9$), which is similar to an image in deep learning
models for imaging applications. This is called sliding or rolling windows (of length 20) in statistical
modeling. Concatenating all sequences to an array forms a 3-dimensional
array, which are used in model fitting.  
 The first 10 days in the sequence are used as condition and the second 10 days in the window are used as target for prediction (or ``label'' for supervised learning). 
The 3-dimensional array is similar to a collection of images.  Each 2-dimensional array ($20 \times 9$)  is called a dataframe, a window, a slice or a sequence.  In general, a sequence is of dimension $(p+q)\times d$, with $p$ the condition length, $q$ the target sequence length, and $d$ the number of time series (features).  For USD yield curve application in this study, $p=10, q=10, d=9$.
\subsubsection{GAN Models\label{subsec:conditional GAN}}
\paragraph{CGAN FC}

A CGAN method is developed and tested by \citet{https://doi.org/10.48550/arxiv.1904.11419}
for multidimensional financial time series generation, where the conditions
can be both categorical and continuous random variables. Similar to GAN, CGAN is also a minmax game \eqref{eq:ganloss} on a cost function between generator (G) and discriminator (D), where
both G and D are neutral network models. The inputs of G are $z$
and $y$, where $z$ is a random variable from uniform or Gaussian
distribution, and $y$ is the condition variable. $\mathbb{P}_{r}$
is the real data distribution and $\mathbb{P}_{g}$ is the model distribution
implicitly defined by $\tilde{x}=G(z,y)$

\begin{equation}
\underset{G}{min}\underset{D}{max}\underset{x\thicksim\mathbb{P}_{r}}{\mathbb{E}}[D(x,y)]+\underset{\tilde{x}\thicksim\mathbb{P}_{g}}{\mathbb{E}}[log(1-D(\tilde{x},y))]\label{eq:ganloss}
\end{equation}

For the training of the discriminator D, the objective is to identify
the correct label (True/False) of the data, making $D(\tilde{x},y)$
\footnote{Note that $y$ appears twice in the formula. $\tilde{x}=G(z,y)$ shows the conditional generator and $D(G,y)$ shows the conditional discmininator. }
 close to zero for
synthetic (fake) data, and $D(x,y)$ close to 1 for real data, thus maximizing the
loss function. For the training of the generator G, the objective
is to fool the discriminator, and make $D(\tilde{x},y)$ ($\tilde{x}=G(x,y)$) close to 1 for
synthetic (fake) data, thus minimizing the loss function. This explains the minmax
notation in the loss function \eqref{eq:ganloss}.

Here the conditions $y$ and noises $z$ are non-linearly mapped to the outputs
by multiple fully connected layers with ReLU activation functions
for both generator and discriminator in CGAN. A clipping method is
used in discriminator to stabilize the training process against spikes
in the weights of the neural network layers. Both simulation studies
and empirical backtesting show promising model performance for various
autocorrelation structure and volatility dynamics \citet{https://doi.org/10.48550/arxiv.1904.11419}.

CGAN FC refers to CGAN with fully connected layers.

\paragraph{CGAN LSTM}
In the previous sub-sections, we introduced the usage of CGAN in time
series generation, which is conditional on given historical data.
These historical data are added into the generator of CGAN with the
noise variables by fully connected layers with ReLU activation functions
in general. In the previous studies \citet{https://doi.org/10.48550/arxiv.1904.11419},
we found that the proposed CGAN structure is relatively more sufficient
for AR types of time series with simple autocorrelation structures,
while it may be challenging to learn the volatility dynamics as GARCH
type time series. In order to capture the volatility dynamics and
more complex autocorrelation structures, we propose an encoder-decoder
structure for the generator of CGAN. The encoder is trained to convert
conditional raw data into required information, and the decoder
is trained to decode the trained conditions and use it to generate
the conditional outcomes. We propose to use separate LSTM layers
(\citet{Yu_2021}) for both encoder and decoder in the generator.
The LSTM layers capture both long- and short-term dependences and
generate sequence outputs with shared weights along time, which is
a natural choice for time series generation. We only use single-direction
LSTM in the generator, as the time series is generated along the forward
time direction. There are two LSTMs used in G, where the first
LSTM takes the conditions as inputs and generate the long- and short-term
state variables as the initial state variables of the second LSTM.
Suppose the conditions are from $x_{1}...,x_{T},$ then for $t\in[1,T]$,
and a typical LSTM cell has three gates: input, forget, and output, which
decide whether or not to allow new input in, forget old information,
and affect output at current time-step ( \citet{Yu_2021}). 
\begin{gather}
i_{1,t}=\sigma(w_{1,i}[a_{1,t-1},x_{t}]+b_{1,i})\\
f_{1,t}=\sigma(w_{1,f}[a_{1,t-1},x_{t}]+b_{1,f})\\
o_{1,t}=\sigma(w_{1,o}[a_{1,t-1},x_{t}]+b_{1,o})\\
\tilde{c}_{1,t}=tanh(w_{1,c}[a_{1,t-1},x_{t}]+b_{1,c})\\
c_{1,t}=i_{1,t}\centerdot\tilde{c}_{1,t}+f_{1,t}\centerdot c_{1,t-1}\\
a_{1,t}=o_{1,t}\centerdot tanh(c_{1,t})
\end{gather}
where $i_{1,t}$, $f_{1,t}$ and $o_{1,t}$ denote the input, forget
and output gates of the first LSTM respectively. $a_{1,t}$ is the
short-term state variable, while $c_{1,t}$ is the long-term state
variable passing along the time in the first LSTM layer training process.
$w_{1,.}$'s and $b_{1,.}$'s are the weights and biases of the first
LSTM. The first LSTM will process the conditions $x_{1}...,x_{t},$
and output the final long- and short-term state variable $c_{1,T}$
and $a_{1,T}$.  $a_{1,T}$ is also the ultimate output of the LSTM layer.
\footnote{The gates serve as latent variables that define the internal structure of LSTM.}
.

The second LSTM takes $c_{1,T}$ and $a_{1,T}$ as the initial state
variables, and takes noises , $z_{1}...,z_{S},$ as inputs to generate
real-like outputs of the time series. 

Under the LSTM-CGAN structure, the discriminator leverages LSTM layers
in order to compete with the generator with LSTM layers. For the time
series $y_{1}...,y_{S+T}$ from either real or generated data, the
discriminator LSTM has a similar structure to the LSTM layers
discussed above. More details of the structure are given in Appendix.
The training process follows the same minmax game and uses the same
loss function as CGAN. 
\paragraph{CWGAN}
The Wasserstein distance is a distance function defined
between two probability distributions $P$ and $Q$ on a given metric
space $M$. If each distribution is viewed as a unit amount of earth
(soil) piled on $M$, the metric is the minimum \textquotedbl{}cost\textquotedbl{}
of turning one pile into the other, which is assumed to be the amount
of earth that needs to be moved times the mean distance it has to
be moved. Because of this analogy, the metric is known in computer
science as the \textit{earth mover's distance}. 

For two empirical distributions $P$ with samples $p_{1},\ldots,p_{n}$
and $Q$ with samples $q_{1},\ldots,q_{n}$, the Earth-Mover (EM)
or Wasserstein-1 distance is
\begin{equation}
W(P,Q)=\inf_{\gamma\in\prod{(P,Q)}}E_{(p,q)\sim\gamma}[\lVert p-q\rVert]
\end{equation}
where $\prod{(P,Q)}$ denotes the set of all joint distributions $\gamma(p,q)$
whose marginals are $P$ and $Q$. For a given $p,q$ the joint distribution
$\gamma(p,q)$ tells how much \textquotedbl{}mass\textquotedbl{} must
be transported from $p$ to $q$ to transform $P$ into $Q$. The
EM distance is then the cost of the optimal transport plan (\citet{https://doi.org/10.48550/arxiv.1701.07875}).
Intuitively, given two distributions, one distribution $P$ can be
seen as a mass of earth properly spread in space, the other distribution
$Q$ as a collection of holes in that same space. Then, the Wasserstein
distance measures the least amount of work needed to fill the holes
with earth. $\lVert p-q\rVert$ measures the distance between $p$
and $q$. The infimum is over all permutations of elements in the
two distributions. Wasserstein distance is used in both Wasserstein
GAN (WGAN) and Signature Conditional Wasserstein GAN (SIGCWGAN). 

Generative Adversarial Networks (GANs) are challenging to train, and
oftentimes it causes the model convergence issue such as mode collapse
and vanishing gradients \citet{https://doi.org/10.48550/arxiv.1701.07875}.
To address the issues in GAN training, WGAN was proposed using Wasserstein
distance as a new cost function. Compared to GAN, WGAN improves the
stability of learning, gets rid of the mode collapse issue, and provides
meaningful learning curves useful for debugging and hyperparameter
searches \citet{https://doi.org/10.48550/arxiv.1701.07875}. The loss
function of WGAN is constructed using the Kantorovich-Rubinstein duality
as:
\begin{equation}
\underset{G}{min}\underset{D\in\boldsymbol{\mathit{\mathcal{\mathbf{\mathcal{D}}}}}}{max}\underset{x\thicksim\mathbb{P}_{r}}{\mathbb{E}}[D(x)]-\underset{\tilde{x}\thicksim\mathbb{P}_{g}}{\mathbb{E}}[D(\tilde{x}))]
\end{equation}
where $\mathcal{D}$ is the set of 1-Lipschitz functions
\footnote{Explain the meaning of Lipschitz functions.}
,$\mathbb{P}_{r}$
is the real data distribution, and $\mathbb{P}_{g}$ is the model distribution
implicitly defined by $\tilde{x}=G(z)$, $z\sim N(0,1)$. 

There are two approaches to enforcing the Lipschitz constraint in WGAN:
weight clipping and gradient penalty (GP). In our analysis, we evaluate
the conditional version of both WGAN and WGANGP, which are CWGAN and
CWGANGP, respectively. CWGAN achieves better backtesting performance
than CWGANGP. Therefore, only CWGAN is covered in model comparison in this study.  CWGAN takes additional conditions $y$ as input to both
the generator and critic. For CWGAN, the discriminator is called the critic because the discriminator is no longer a binary classifier (True/False) as in CGAN.
The loss function for CWGAN is:
\begin{equation}
\underset{G}{min}\underset{D\in\boldsymbol{\mathit{\mathcal{\mathbf{\mathcal{D}}}}}}{max}\underset{x\thicksim\mathbb{P}_{r}}{\mathbb{E}}[D(x,y)]-\underset{\tilde{x}\thicksim\mathbb{P}_{g}}{\mathbb{E}}[D(\tilde{x},y))]\label{eq:CWGANloss}
\end{equation}
The critic calculates the distance between the real and synthetic data as in \eqref{eq:CWGANloss}.
We set the clipping threshold to 0.01, so the weights of the critic
lie within a compact space $[-0.01,0.01].$

\paragraph{Conditional Signature Wasserstein GAN}

Instead of using neural networks to do classification in the discriminator,
Signature Conditional Wasserstein GAN (SIGCWGAN, or SIG for short) calculates the distance
between the signature of real and synthetic paths as the discriminator.
The signature $S$ of a time series sequence $X$ is a nonlinear function $S(X)$ sufficient to capture certain features
of the time series (think of principal component analysis as an example).
Details of signature and the loss function Conditional Signature Wasserstein-1
metric (C-Sig-W1) are discussed in the appendix.

A model is fitted to predict future signature from the past signature,
$S_{t, 1:q}=L(S_{t, -p:0})$ with predicted values $\hat{S}_{t, 1:q}=\hat{L}(S_{t, -p:0})$. The generator network generates random
values of $X$ from past conditions $X_{t, -p:0}$ and random noise
$Z, \tilde{X}=G(Z,X_{t, -p:0})$.  The generated sample has signature $S(\tilde{X})$. The objective of the generator
is to match the expected future signature of synthetic data $E[S(\tilde{X})]$ with the
fitted future signature of the actual data $\hat{S}_{t, 1:q}$,
thus minimizing $\left(E[S(\tilde{X})]-\hat{S}_{t, 1:q}\right)^{2}$.

The signature function $S$ is given. The first step in training SIGCWGAN
is to fit the signature forecast function $L$. The second step is
to optimize the generator $G$ given loss function calculated from
signatures of real and synthetic data. Both $S$ and $L$ are given
in the fitting of the generator model $G$. If the fitting of $L$
in the first step is not good, then the optimization in the second
step will be optimizing to the wrong target $\hat{S}$. In the original paper,
$L$ is suggested to be a linear function. To improve model fit in
the first step and for the overall model, we also implement feed forward
(fully connected) networks and CNN as the signature forecast function
$L$.

Note that this is a typical forecast problem for $\hat{S}_{t, 1:q}$,
only the mean (of the signature of the synthetic and actual distribution) is used,
and the distribution itself is not used directly.

Prior to taking the signature $S(X)$ of $X$, $X$ is \textit{augmented}
in the sense of \citet{morrill2020generalised,chevyrev2016primer},
with the following transformations: scaling, cumulative sum, and adding lags and leads.  These augmentations are intended to normalize and enrich the time
series with additional information. This is similar to adding quadratic and interaction terms in regression.  We provide one example of the
data transformation process in the appendix. 

In the case of 9-tenor USD yield curve data, there are 2,824
training sequences which are 20-day windows of returns (rate changes).  The first 10 days in the window are used as condition and the second 10 days in the window are used as target for prediction.  
The condition matrix is of size $10\times9$.  Applying
the augmentations above, we get a matrix of size $3\times36.$
The augmented matrix of size $3\times36$ for a single sequence is
slightly larger than the input matrix of size $10\times9$.  There is a depth parameter for augmentation.  At depth 1, the signature is of size 36.  At depth 2, the signature is of size 1332. At depth 3, the signature is of size 47,988. The size of the
signature is highly nonlinear in the depth. The original source for
this model, \citet{Sabate21}, uses depth 2 for most time series. We
feel that depth 3 might be hard to fit or lead to overfitting. So our
model uses depth 2.

To generate a synthetic time series, the generator $G$ uses an \textit{autoregressive
feed-forward neural network} (ARFNN) which is built from layers of \textit{residual
blocks} and a forward function that implements
the autoregressive behavior. Details of ARFNN configuration are discussed in the appendix.
\subsubsection{Likelihood-base deep generative models\label{subsec:Encoder-decoder structure}}
\paragraph{Variational Autoencoder (VAE)}
Variational Autoencoder is one of the latent based models \citet{Kingma_2019}.
A variational autoencoder can be defined as being an autoencoder whose
training is regularised to avoid overfitting and to ensure that the latent
space has good properties that enable generative process. Other than
traditional autoencoder that provides a point estimate that might
cause severe overfitting among input data, the encoder in variational
autoencoder constructs a distribution of the embedding in the latent
space. The goal of variational autoencoder is a sample that accurately
represents the distribution of original data, given input data $X$
that consist of $N$ i.i.d. samples of some continuous or discrete variable
$x$. A two step process is applied. The first step generates value
$z$ from prior distribution $p_{\theta}(z)$, and the second step
is to generate value $x$ from conditional distribution or likelihood $p_{\theta}(x|z)$.
Assuming that both the prior $p_{\theta}(z)$ and likelihood $p_{\theta}(x|z)$
are differentiable, the VAE is composed of three components:
input data, output data and latent space. As such, the encoder is
applied to model probabilistic posterior distribution $p_{\theta}(x|z)$,
which is approximated by $q_{\phi}(z|x)$ to ease computation
and ensure tractability, while the decoder is to model the conditional
likelihood $p_{\theta}(x|z)$. 

Since latent space is modeled by a distribution, a specific distribution
needs to be assumed. In the case of Gaussian Variational Autoencoder,
both probabilistic posterior distribution $q_{\phi}(z|x)$ and prior
distribution $p_{\theta}(z)$ are assumed to be Gaussian. 

The loss function of VAE is called Evidence Lower Bound loss function
(ELBO) and given in the following formula.
\begin{equation}
L_{\theta,\phi}=-E_{q_{\phi(z|x)}}[\log p_{\theta}(x|z)]+D_{KL}(q_{\phi}(z|x),p_{\theta}(z))
\end{equation}
where the first term $-E_{q_{\phi(z|x)}}[\log p_{\theta}(x|z)]$ is
the negative log-likelihood of data given $z$ sampled from $q_{\phi}(z|x)$.
The second term $D_{KL}(q_{\phi}(z|x),p_{\theta}(z))$ is the KL-Divergence
loss between the encoded latent space distribution and the prior. 

In our case, we adapt the TimeVAE model (VAE for short) \citet{Desai2021}
implementation
\footnote{published on GitHub as \url{https://github.com/abudesai/timeVAE}}
.
This model is an implementation of VAE on time series data, with deep learning
layers such as dense and convolutional layers, as well as custom layers
to model time-series specific components such as level, trend, and seasonal patterns. 

In our study, we further improve VAE
with couple of steps listed below.

\emph{Network enhancement: LSTM }

Other than the existing network structure in VAE \citet{Desai2021}
(such as dense, convolutional, trend block, seasonality block, etc),
we add LSTM layers which is a good candidate to capture longer time
dependency of data, as enhancement to the model.

\emph{Continuous conditional VAE}

To enable longer horizon generation of synthetic data, a continuous
conditional VAE framework is proposed and implemented. The basic idea
behind the condition is to take advantage of the most currently
available information and concatenate with the input data (either real
time series data in encoder or noise input as in the decoder) to make
the generated data align better with real data. With the condition
$y$, the loss function of conditional VAE ( or CVAE) becomes
\begin{equation}
L_{\theta,\phi}=-E_{q_{\phi(z|x)}}[logp_{\theta}(x|z,y)]+D_{KL}(q_{\phi}(z|x,y),p_{\theta}(z|y))
\end{equation}
%
%
% ------------------------------------------- Diffusion -----------------------------------------------
%
%
\paragraph{Diffusion}
Diffusion originates from thermodynamics to describe the moving of particles from regions of higher concentration to lower concentration. In quantitative finance, diffusion appears most commonly in the form of Brownian motion with linearly increasing variance.  As a deep learning model, diffusion has constant variance through a variance scheduler $\beta_t$ as in \eqref{eq:bmdiff}.

Diffusion models have emerged as a powerful class of generative models
recently. Well known deployed applications include DALL.E 2, Stable Diffusion and Midjourney.  For example, the model can generate realistic images from simple text input such as ``vibrant California poppies''
\footnote{\url{https://www.unite.ai/diffusion-models-in-ai-everything-you-need-to-know}}.  

These models generate high quality samples and often outperform
generative adversarial networks (GANs) in the challenging task of
image synthesis \citet{NEURIPS2020_4c5bcfec,NEURIPS2021_49ad23d1}.
Diffusion models are a class of latent variable models inspired by
considerations from non-equilibrium thermodynamics \citet{NEURIPS2020_4c5bcfec}
\footnote{There are three categories of diffusion models, (1) denoising diffusion probabilistic models, (2) noise conditioned score based generative models, (3) stochastic differential equations.  \url{https://www.unite.ai/diffusion-models-in-ai-everything-you-need-to-know/}.  See \url{https://sander.ai/2022/01/31/diffusion.html?ref=assemblyai.com} for informal technical discussion.}
.
There are two processes in a diffusion model: a forward process that
maps data to noise, and a reverse process that performs iterative
denoising. New samples are subsequently generated by first sampling
from a simple prior distribution (e.g., standard Gaussian), followed
by ancestral sampling through the reverse process. 

In the forward process, an image (the data) is converted to noise in $N$ successive steps, $x_0\rightarrow x_N$.  Each step is done through Gaussian normal distributions, also known as forward diffusion kernel (FDK),
\begin{flalign}
x_t&=\sqrt{1-\beta_t}x_{t-1}+\sqrt{\beta_t}\epsilon_t,\epsilon_t\sim N(0,1)\label{eq:fdk}\\
x_t|x_{t-1} &\sim N(\sqrt{1-\beta_t}x_{t-1},\beta_t\bfI)\quad \mbox{conditional}\\
x_t &\sim N(0,\bfI)\quad \mbox{unconditional}
\end{flalign}
For perspective, \eqref{eq:fdk} can be converted into the usual SDE form (with $dt=1$):
\begin{equation}
dx_t=-\underbrace{(1-\sqrt{1-\beta_t})}_{\kappa_t}x_t dt + \underbrace{\sqrt{\beta_t}}_{\sigma_t} dW_t\label{eq:bmdiff}
\end{equation}
This is a controlled mean reversion process with mean zero, time varying $\kappa$ and $\sigma$ to keep the variance of the process constant. 

The forward process $q(x^{1:N}|x^{0})$ is fixed to a Markov chain
\footnote{In economic and finance applications, Markov Chain with finite state space is often used in Markov switching model.  In the current context, the Markov Chain has infinite uncountable number of states due to the use of normal distribution in each step.}
 that describes the transition from step $n-1$ to step $n: q(x_t|x_{n-1})$.
It gradually adds Gaussian noise (``corrupting'' the data in the noising process) to the data following a variance
schedule $\beta_{1}$, ..., $\beta_{N}$. In our experiment, we follow
the practice in TimeGrad \citet{https://doi.org/10.48550/arxiv.2101.12072}:
we set the number of diffusion steps $N=100$, and use a linear variance
schedule starting from $\beta_{1}=1\times10^{-4}$ till $\beta_{N}=0.1$. 

The distribution in the reverse process is denoted as $p_{\theta}(x^{0:N})$. It is defined as a Markov chain with learned Gaussian transitions
starting at $p(x^{N})\sim\mathcal{N}(\mathbf{0},\text{\ensuremath{\mathbf{I}}})$:
\begin{gather}
p_{\theta}(x^{0:N})\coloneqq p(x^{N})\prod_{n=N}^{1}p_{\theta}(x^{n-1}|x^{n})\\
p_{\theta}(x^{n-1}|x^{n})\coloneqq\mathcal{N}(\mu_{\theta}(x^{n},n),\Sigma_{\theta}(x^{n},n)\text{\ensuremath{\mathbf{I}}})\label{eq:3}
\end{gather}
In this setup, the reverse diffusion kernel (RDK) is also Gaussian.  With this reverse Markov chain, we can generate a data sample $x^{0}$
by first sampling a noise vector $x^{N}\sim p(x^{N})$, then iteratively
sample from the learnable transition kernel $x^{n-1}\sim p_{\theta}(x^{n-1}|x^{n})$
until $n=1$. The mean and variance of the Gaussian kernel $\mu_{\theta}$ and $\Sigma_{\theta}$ are parameterzied by $\theta$ and defined by a neural network to be specified later.
This is a critical drawback of the diffusion model:
it requires iterations through $N$ time steps to produce a high quality
sample. It is much slower than GANs, which only needs one pass through
a network. As noted in \citet{https://doi.org/10.48550/arxiv.2010.02502},
diffusion model takes 20 hours to sample 50k images of size $32\times32$,
while GAN takes less than a minute. It is consistent with what we
observed in our analysis.  
%Later research shows that a reparameterization of the model can speed up model testing
%\footnote{\url{https://learnopencv.com/denoising-diffusion-probabilistic-models/}}
%.

The diffusion model we test in our evaluation is TimeGrad, an autoregressive
model for multivariate probabilistic time series forecasting. Temporal
dynamics are modeled by the autoregressive recurrent neural network
(RNN) architecture from \citet{https://doi.org/10.48550/arxiv.1308.0850,NIPS2014_a14ac55a}.
Maximization of the log likelihood of a complex joint distirbution has much commonality with VAE.  Indeed, for Diffusion model, the log likelihood is also formulated with evidence variational lower bound (ELBO).  Details of TimeGrad architectures are discussed in appendix. In training
process, we randomly sample context and adjoining prediction sized
windows from the training data. The network minimizes the difference
between real noise $\epsilon$ and predicted noise $\epsilon_{\theta}$
for time step $t$ and step index $n$:
\begin{equation}
\mathbb{E}_{x_{t}^{0},\epsilon,n}[||\epsilon-\epsilon_{\theta}(\sqrt{\bar{\alpha}_{n}}x_{t}^{0}+\sqrt{1-\bar{\alpha}_{n}}\epsilon,\mathbf{\mathrm{h}}_{t-1},n)||^{2}]\label{eq:11}
\end{equation}
where $\alpha_n=1-\beta_2, \bar{\alpha}_n=\Pi_{s=1}^n \alpha_s$. 
$\epsilon_{\theta}$ is the noise predicted from RNN, which takes
three inputs: $x^{n}=\sqrt{\bar{\alpha}_{n}}x_{t}^{0}+\sqrt{1-\bar{\alpha}_{n}}\epsilon$,
the hidden state $\mathbf{\mathrm{h}}_{t-1}$, and step index $n$.
In inference step, we sample random noise $x_{T+1}^{N}\sim\mathcal{N}(\mathbf{0},\text{\ensuremath{\mathbf{I}}})$
and iterate through $N$ time steps in reverse process to generate
a sample $x_{T+1}^{0}$, which is passed autoregressively to the RNN
(together with possible covariates $\mathbf{\mathrm{c}}_{T+1}$) to obtain
the next hidden state $\mathbf{\mathrm{h}}_{T+1}$. This process is
repeated until the desired forecast horizon has been reached.

\subsubsection{Network architecture structure\label{subsec:network architecture structure}}

\begin{figure}
\includegraphics[scale=0.2]{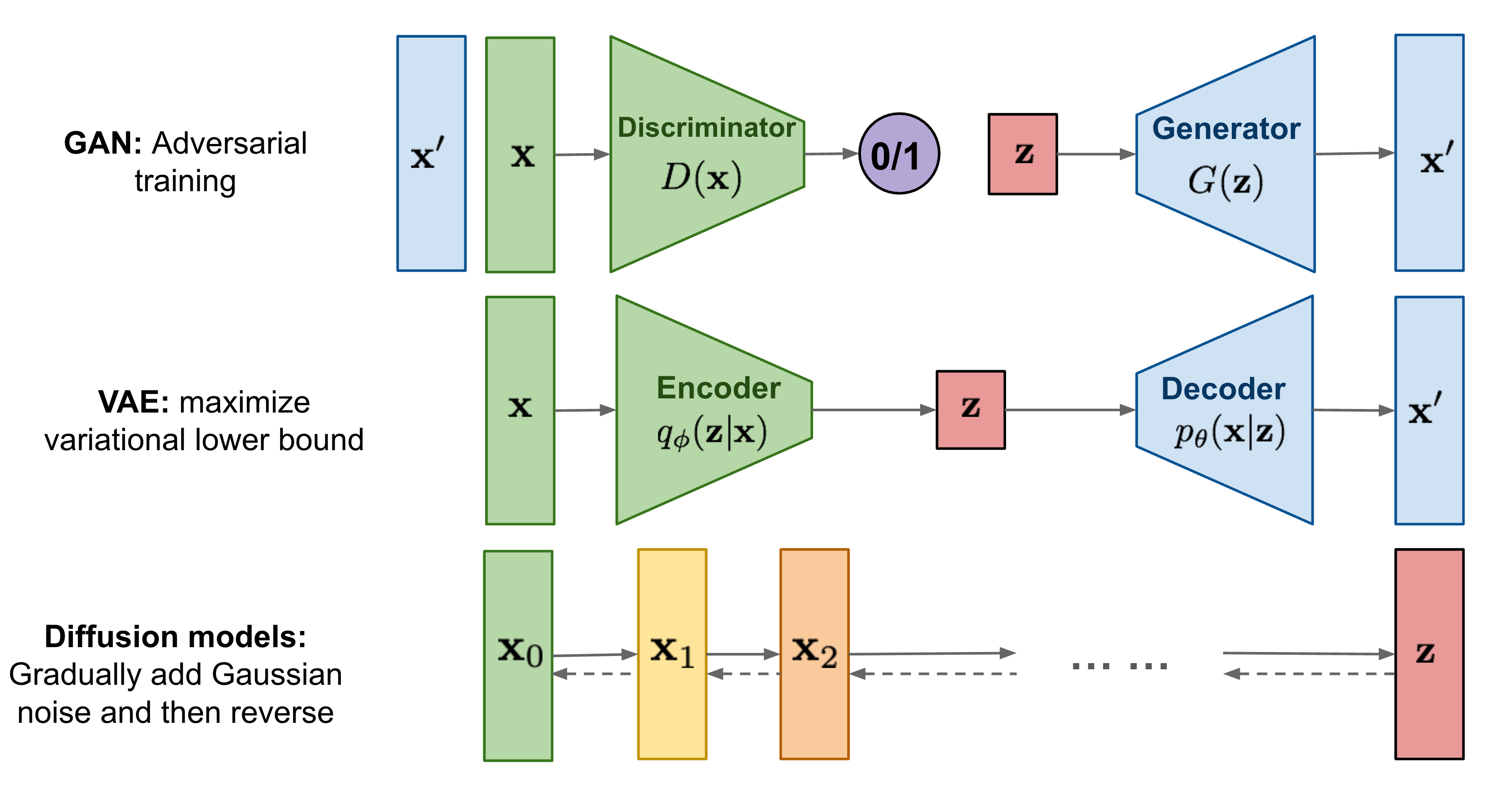}
\caption{Architecture comparison among three groups of generative models\label{fig:Architecture-comparison-among}}
\end{figure}

The high-level architecture comparison of the major three groups of deep generative
models (GAN, VAE and Diffusion) can be found in Figure \ref{fig:Architecture-comparison-among}.
There are quite some review work done on GAN models used in
time series data, as in \citet{https://doi.org/10.48550/arxiv.2107.11098},\citet{10.1145/3559540}.
Basically all that generative models do is to build up and train a mapping
between input data $x$ and random noise $z$ through different architectures
and building blocks, and a generator takes random numbers to generate
synthetic samples
\footnote{A linear regression model is a trivial case.  Let $y=bx+\epsilon, \epsilon\sim iid N(0,\sigma^2)$.  The mapping from data to noise is $y-bx$, and the generator takes noise $\epsilon$ in simulation, $bx+\epsilon$. }
.

Given sample data, our goal is to approximate a data distribution
as closely as possible. In other words, we have to evaluate and optimize
the distance between the data distribution and model distribution.
Different approaches can be applied to this end, based on the way
to measure distribution distance. One way is likelihood-based models, which seek
to learn a model that assigns a high likelihood to the observed data
samples. Most of the models in this class apply KL divergence, which
is statistically efficient, but requires the ability to evaluate
or optimize likelihood  tractably. Another approach is likelihood-free inference
using adversarial training, which conduct a minmax optimization. Therefore
the deep generative models can be widely categorized into two classes:
likelihood-based models and likelihood-free models. The loss function
of each model reflects the architecture. Table~\ref{tab:loss_function_list}
lists the loss functions and explanation of the architecture. The loss functions of different deep generative models  measure the distance between real and synthetic data with different approaches. 

\begin{table}[ht]
\centering
\caption{Loss function of NN models}
\label{tab:loss_function_list}
\begin{tabular}{p{0.15\linewidth} | p{0.4\linewidth}| p{0.45\linewidth}}
\hline 
Model & Loss function & Architecture explained\tabularnewline
\hline 
\hline 
CGAN (FC, LSTM) & 
$\begin{aligned}[t]
&\underset{G}{min}\underset{D}{max}\underset{x\thicksim\mathbb{P}_{r}}{\mathbb{E}}[D(x,y)]\\
&+\underset{\widetilde{x}\thicksim\mathbb{P}_{g}}{\mathbb{E}}[log(1-D(\widetilde{x},y))]
\end{aligned}$
& Conditional GAN optimizes two conflicting objective functions at the
same time: generator and discriminator. \tabularnewline
\hline 
CWGAN & 
$\begin{aligned}[t]
&\underset{G}{min}\underset{D\in\boldsymbol{\mathit{\mathcal{\mathbf{\mathcal{D}}}}}}{max}\underset{x\thicksim\mathbb{P}_{r}}{\mathbb{E}}[D(x,y)]\\
&-\underset{\widetilde{x}\thicksim\mathbb{P}_{g}}{\mathbb{E}}[D(\widetilde{x},y)]
\end{aligned}$
& Conditional GAN with Wasserstein loss improves the convergence of
GAN. Wasserstein distance has the properties that it is continuous
and differentiable and continues to provide a linear gradient.\tabularnewline
\hline 
SIG & 
$\begin{aligned}[t]
&\mathrm{CSigW_{1}}(\mu,\nu;p,q)=\\
&\Big|E_{\mathbb{P}_{r}}[S(X_{t+1:t+q})|x_{t-p+1:t}=x]\\
&-E_{\mathbb{P}_{g}}[S(\hat{X}_{t+1:t+q})|x_{t-p+1:t}=x]\Big|
\end{aligned}$ & Signature CWGAN with CNN layers mininizes the summation of the $\ell_{2}$

norm of the error between the conditional expected signature of future
real path and future synthetic path generated by the generator with a given condition. \tabularnewline
\hline 
VAE & 
$\begin{aligned}[t]
&L_{\theta,\phi}=-E_{q_{\phi(z|x)}}[\log p_{\theta}(x|z,y)]\\
&+D_{KL}(q_{\phi}(z|x,y),p_{\theta}(z|y))
\end{aligned}$
 & Conditional Time VAE model minimizes the reconstruction loss (in real
space) and KL-divergence (in latent space) at the same time.\tabularnewline
\hline 
Diffusion & 
$\begin{aligned}[t]
&-\log p_{\theta}(\text{\ensuremath{\mathbf{x}}}^{0}|\text{\ensuremath{\mathbf{x}}}^{1})\\
&+D_{KL}(q(\text{\ensuremath{\mathbf{x}}}^{N}|\text{\ensuremath{\mathbf{x}}}^{0}),p(\text{\ensuremath{\mathbf{x}}}^{N}))\\
&+\sum_{n=2}^{N}D_{KL}(q(\text{\ensuremath{\mathbf{x}}}^{n-1}|\text{\ensuremath{\mathbf{x}}}^{n},\text{\ensuremath{\mathbf{x}}}^{0}),\\
&p_{\theta}(\text{\ensuremath{\mathbf{x}}}^{n-1}|\text{\ensuremath{\mathbf{x}}}^{n}))
\end{aligned}$
 & Diffusion model optimimizes KL-divergence between Gaussian distributions. Encoder is predetermined, and the goal is to learn a decoder that is the
inverse of this process. \tabularnewline
\hline 
\end{tabular}
\end{table}

\subsection{Single-step vs. multi-step prediction \label{subsec:single-step prediction}}
Traditional time series model such as ARMA predicts the observation
at the next time step $x_{t+1}$ given the previous available observations
$x_{t-k},\ldots,x_{t}.$ This is called single step prediction. This is the case with most traditional statistical models, such as AR(1) model \eqref{eq:AR1}.  Different
from traditional time series prediction, deep learning generative
models normally work in a sequence-to-sequence approach (\citet{https://doi.org/10.48550/arxiv.1805.03714,https://doi.org/10.48550/arxiv.1308.0850,NIPS2014_a14ac55a}).  Image application is an important use case for the development of deep learning models and helps to shape the deep learning models for later applications.  In image generation, the model generates a whole image in one pass, rather than generating the image by part.  For time series generation with deep learning model, similar to image generation, it is customary to generate multiple time series covering multiple days in one pass.

Given the shape of training data, specifically the target sequence length
of data, deep learning models can generate up to the sequential length
of prediction in one sample. As a special case, if the target sequential
length $q$ is 1, it is equivalent to single step prediction. Otherwise,
if target sequence length is $q>1$, multistep forecast is generated in one pass, and the first
observation in the generated sample can be used as single-step prediction. 

It should be noted that the focus of our study is generating distribution forecast of risk factors rather than any point forecast.  The forecast is represented
by drawing Monte Carlo samples from the underlying probabilistic models,
as defined either from real or latent space. With that, the generated
samples (or paths) need to be large enough to reasonably describe the distribution
of generated sample, and the corresponding quantiles
or confidence intervals  at any given step. 

For the generation of any single path, in a multi-step prediction, the model needs to learn to predict a
sequence of future values. Thus, unlike a single-step model, where only
a single future point is predicted, a multi-step model predicts a
sequence of the values into the future. Given that generative model normally
has more than 1 as target sequence length to learn the long term
dynamics of data, the generative model normally generates multi-step
prediction with any given sample.

For illustration, assume the condition length $p=10$ and target sequence length $q=10$, that is, the model uses the previous 10 days to generate forecast for the next 10 days in one pass.

\subsubsection{Direct forecast\label{subsec:direct models}}
Consider the case that up to 5 days forecast (forecast horizon=5) is needed.  Forecasts up to 10 days are produced by the model in one pass.  The first 5 days' forecast satisfy the forecast requirement, and the forecast for days 6 to 10 can be ignored.
\subsubsection{Iterative forecast\label{subsec:interative models}}
Consider the case that up to 20 days forecast is required.  Using 10 lagged values as conditions $y_0=x_{t,-9:0}=(x_{t-9},x_{t-8},\cdots,x_t)$, forecast for the first 10 days covering the period ($t+1,\cdots,t+10$), $\tilde{x}_{t,1:10}=G(z,y_0)$ ($z$ is random noise), is generated in one pass of the model.  Then using $y_1=\tilde{x}_{t,1:10}$ as condition, $\tilde{x}_{t,11:20}=G(z,y_1)$ ($z$ is random noise) is generated in another pass of the model.  This process can be iterated to generate forecast for longer horizon.

Signature (SIG) model has the built-in autoregressive feature, and can generate long horizon forecast without ad hoc iteration outside the model.

\section{Data\label{subsec:testing data}}

Both historical data and simulated data are used for model training and testing.

Since historical data are generated in the real world subject to many factors, its exact property or data generation process (DGP) is unknown.  To enhance model testing, we simulate multiple long time series from two well known models: GARCH and CIR (\cite{cir1985}) model.  Both are widely used models for risk factor modeling in finance.  The simulated data have known autocorrelation and conditional variance properties.  We can use the simulated data to check whether the models can capture these properties in the data.

In fact, our model testing results show that model performance is indeed different when the data have very fat tails (when simulated data follow GARCH-t(3) distribution).  See Section~\ref{sec:simresults} for details.

\subsection{USD yield curve\label{subsec:usd yield curve}}
 Yield curve modeling is an active area of multivariate time series
 modeling for several reasons, (1) yield curve is the driver for many financial products, (2) there is strong dependence among interest rates of different tenors, i.e., term structure, (3) there is interesting time series dependence for yield curves.  USD yield curve (USDYC) is
 the main data for model development testing. Three USDYC datasets with nine tenors are used.  For ease of reference, these three datasets are referred to as USDYC1 (Libor curve 2008-2022), USDYC2 (Par yield 2008-2023) and USDYC3 (Par yield 2000-2023).  See more details below.

 \subsubsection{3 month USD LIBOR curve}
 This dataset is comprised of 9 typical tenors: 3 month, 6 month, and 1,2,3,5,10,20
 and 30 years, of the United States Dollar 3 month LIBOR zero-coupon
 yield curve, for the business dates ranging from Jan. 2, 2008 to Feb. 5, 2022.

 \subsubsection{Treasury Par Yield Curve}
 This is sourced from the Federal Reserve Bank of St Souis (FRED)\footnote{\url{https://fred.stlouisfed.org/}. Also available from Federal Reserve Board's H15 series or Treasury website, \url{https://home.treasury.gov/policy-issues/financing-the-government/interest-rate-statistics}}.  The dataset covers the period 1990-current date. To fully use the
 available nine tenor historical data, our model training and testing
 are done using both 2000-2023 (Jan. 3, 2000 - Feb. 16, 2023) and 2008-2023 (Jan. 2, 2008 - Feb. 16, 2023) time periods.

\begin{figure}
 \centering
 \includegraphics[scale=0.5]{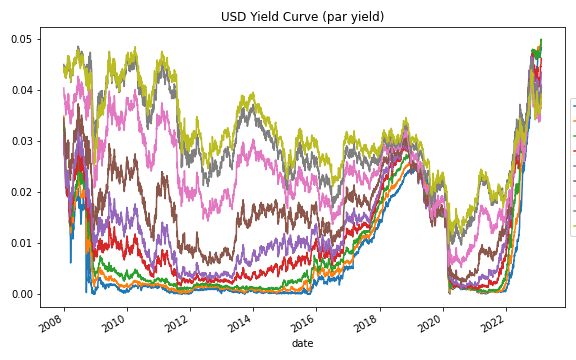}
 \caption{USD Par Yield Curve\label{fig:usdyc}}
 \end{figure}

 For ease of reference, these three datasets are referred to as USDYC1 (2008-2022 Libor curve), USDYC2 (Par yield 2008-2023) and USDYC3 (Par yield 2000-2023).  USDYC3 is plotted in Figure~\ref{fig:usdyc}.
 \subsection{Simulated data for model training and testing\label{subsec:simulated data}}
 \subsubsection{AR(1)+GARCH for rate changes}
 An AR(1) for conditional mean and GARCH(1,1) for conditional variance is used to simulate
 data for model testing. This is designed to assess whether the candidate
 models can capture correlation across tenors, autocorrelation and volatility clustering in the
 returns. In order to limit the model running time for testing,  only bivariate time series are simulated. The simulation model is represented as follows,
 \begin{gather}
 R_{t+1}=R_{t}+x_{t+1}\\
 x_{t+1}=\phi_1 x_{t}+\epsilon_{t+1}\\
 \epsilon_{t+1}=\sigma_{t+1}z_{t+1},z_{t+1}\sim N(0,1), t(3)\\
 \sigma_{t+1}^2=\omega+\alpha \epsilon_{t}^{2}+\beta \sigma_{t}^2
 \end{gather}
 Bivariate time series is generated from the AR(1)+GARCH(1,1) model
 with separate parameters for each time series, and correlation between
 innovations $z_{t}$ of the two time series. The simulated bivariate
 time series covers 30 years of daily data (or length of $250 \times 30$). For ease
 of reference, these two time series are called ``3m'' and ``1y''
 for 3-month and 1-year interest rates. The parameters used in the
 simulation are presented in Table \ref{tbl: param_sim}.  In the table, $\phi_1$ is the autocorrelation in the AR(1) model, $\omega,\alpha,\beta$ are parameters of the GARCH(1,1) model, $\rho$ is the correlation
 between the innovations of the two time series, Variance is the unconditional
 variance of the GARCH model ($\omega/(1-\alpha-\beta)$), and Vol
 is the unconditional volatility ($\sqrt{\omega/(1-\alpha-\beta)}$).
 The parameters are at the scale of interest rates in percentage.
 For example, 0.029 volatility is 2.9\% volatility for daily changes
 in annualized interest rate.

 \begin{table}
 \begin{centering}
 \begin{tabular}{|c|c|c|}
 \hline 
 Param & 3M & 1Y\tabularnewline
 \hline 
 \hline 
 $\phi_1$ & 0.5 & -0.5\tabularnewline
 \hline 
 $\omega$ & 0.000009 & 0.000012\tabularnewline
 \hline 
 $\alpha$ & 0.1742 & 0.0724\tabularnewline
 \hline 
 $\beta$ & 0.8158 & 0.9176\tabularnewline
 \hline 
 $\alpha+\beta$ & 0.9900 & 0.9900\tabularnewline
 \hline 
 $\rho$ & \multicolumn{2}{c|}{0.70}\tabularnewline
 \hline 
 Variance & 0.00086 & 0.00115\tabularnewline
 \hline 
 Vol & 0.029 & 0.034\tabularnewline
 \hline 
 \end{tabular}
 \par\end{centering}
 \caption{Parameters for data simulation from GARCH model}
 \label{tbl: param_sim}
 \end{table}
 \subsubsection{CIR for Interest Rates}
\begin{figure}
 \centering
 \includegraphics[scale=0.5]{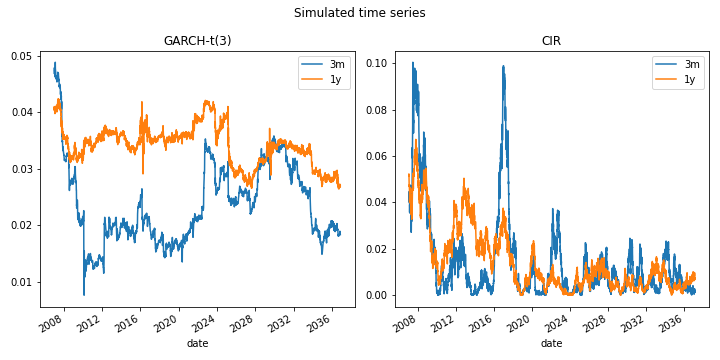}
 \caption{Simulated time series\label{fig:simplot}}
 \end{figure}

 The following typical parameters are used to simulate interest rates
 data from CIR model,
 \begin{gather}
 dR_{t}=\kappa(\theta-R_{t})dt+\sigma \sqrt{R_t}dW_{t}\\
 \kappa_1=0.45,\theta_1=0.02,\sigma_1=0.15\\
 \kappa_2=0.20,\theta_1=0.03,\sigma_1=0.10\\
 dW_{1t}dW_{2t}=\rho dt, \rho=0.60
 \end{gather}
 These are typical parameters calibrated for yield curve.  Again bivariate time series of 30 year daily data are simulated, and the same set of parameters are applied for both time series and uncorrelated with each other
 \footnote{In the simulation algorithm for ``exact'' distribution with normal distribution and $\chi^2$ distribution as components, it is not clear how to introduce correlation across time series in an intuitive way. To introduce correlation between the two simulated series, simulation uses Euler discretimzation.}
 .
 Selected simulated data are plotted in Figure~\ref{fig:simplot}.

\subsubsection{Data preparation pipeline and modeling steps\label{subsec:Training-and-testing} }
The key data preparation and modeling steps are listed below:
\begin{enumerate}
\item Calculating returns.
In statistical modeling, it is customary to transform the times series
data to make its distribution stationary. Transformation for interest
rate time series usually involves calculating daily differences or
absolute returns
\footnote{There is a long standing debate on whether interest rates follow a normal or lognormal model.  A normal model would imply that absolute return is appropriate, while a lognormal model would imply that log-return is appropriate.  In this paper, we follow the intuitive normal model and use absolute returns.}
.
\begin{equation}
x_{t}=R_{t}-R_{t-1}
\end{equation}
Transformation for stock prices usually involves calculating relative
returns or log returns. 
\begin{equation}
x_{t}=log(P_{t})-log(P_{t-1})
\end{equation}
To differentiate the original time series and its returns, the original
time series is usually referred to as ``levels''.

For the 9-tenor USD Libor curve (USDYC1) dataset, from 12/31/2007 to 2/25/2022, there are 3550 business dates, and the returns (rate change) dataset has 3549 business dates (1/2/2008-2/25/2022).  The returns array $x$ is $3549\times 9$.
\item Standardization or normalization.
To improve model fitting, the data are usually standardized (transformed
to mean zero and standard deviation 1.0) or normalized (usually to a
range of {[}0, 1{]} or {[}-1, 1{]}). For image data, all pixel values are in the range $[0, 255]$, normalization to a standard range is preferred.  For random financial data such as yield curve, standardization to zero mean and unit variance is used to scale the returns.
The scaled data are used in model fitting. After synthetic data are generated,
they are ``unscaled'' to the original form for comparison with the
raw returns data.

Note that activation function used in neural networks needs to be consistent with the range of values expected.  For example, if the output range from the network is expected to be $(-\infty,+\infty)$, then an activation function that restricts the output range to be in the range $[0,1]$ is inappropriate.  For this reason, linear activation for the final layer in the generator is used in this paper.

For USDYC1, the standardized returns array $x$ is $3549\times 9$.
\item Windowing, time slicing or sequencing.
Imaging application is an important use case for the development of deep learning models and helps to shape the deep learning models for later applications.  For an application, the image size is usually fixed.  For example, images in the well known MNIST
\footnote{See \url{http://yann.lecun.com/exdb/mnist/}.}
 image dataset are of fixed size $28 \times 28$.  The setup of neural network for financial time series follows image application.
Since deep learning models usually expect a dataframe or sequence of a fixed size,
time series data in 2-dimensional arrays are sliced into time windows
of fixed size. For a typical model in the paper, the 9-tenor USD yield curve
\footnote{For any business date, there are nine interest rates for different maturities.  An yield curve is defined as yield (interest rate) as a function of time to maturity or tenor.}
is sliced (or windowed/sequenced) into windows of 20 business dates.  Each sequence is a 2-dimensional array (of dimension $20 \times 9$), which is similar to an image in deep learning
models for imaging applications. This is called sliding or rolling windows (of length 20) in statistical
modeling. Concatenating all sequences to an array forms a 3-dimensional
array, which are used in model fitting.  
 The first 10 days in the sequence are used as condition and the second 10 days in the window are used as target for prediction (or ``label'' for supervised learning). 
The 3-dimensional array is similar to a collection of images.  Each 2-dimensional array ($20 \times 9$)  is called a dataframe, a window, a slice or a sequence.  In general, a sequence is of dimension $(p+q)\times d$, with $p$ the condition length, $q$ the target sequence length, and $d$ the number of time series (features).  For USD yield curve application in this study, $p=10, q=10, d=9$.

For USDYC1, the sequenced 3d array is $3530\times 20\times 9$.
\item Train-test (or train-validation) split.
The windowed data are split into train-test datasets randomly
(in the window dimension or the first dimension of the 3d array), with 80\% in train
and 20\% in test.

For USDYC1, there are 2824 rows in the train data and 608 rows in the test data.  In another word, there are 608 test dates in the test data.
\item Model fitting (training).
The model is trained (fitted) on the train dataset.
\item Model validation:
\begin{enumerate}
\item Synthetic data generation.
For each sequence in test dataset, synthetic data are generated.
For example, a test sequence includes data for 20 business dates
from 12/13/2021 to 1/11/2022
\footnote{The $20\times 9$ 20-day test data can be further divided into a $10\times 9$ array for the first 10 days as test condition, and $10\times 9$ array for the second 10 days as target data for validation.}
. The 10-day data from 12/13/2021 to 12/27/2021
are used as condition to generate synthetic data for the 10-day period 12/28/2021 to 1/11/2022.
The synthetic data can be compared with the test data for the same
period (12/28/2021 to 1/11/2022).  This test sequence is used to generate forecast for 10 days after business date 12/27/2021.  For simplicity, we can say that for this test sequence, the test date is 12/27/2021.  This is analogous VaR calculation: suppose the current date is $t_0$ (test date), we need to forecast the distribution for the next 10 dates $t_0+1,\cdots,t_0+10$.

When a large number of copies (or paths) of synthetic data are generated for each day in a period, the synthetic data form an empirical distribution for each date in the period.  Test data can be compared with the synthetic data distribution on corresponding dates to assess the quality of the synthetic data distribution.  This is done as part of backtesting KPI.
\item KPI calculation.
KPIs are calculated by comparing test and synthetic data in statistical distribution and autocorrelation.  The calculation of these KPIs uses the 20\% data reserved for testing or validation.  See Section~\ref{subsec:evaluation metrics} for details.
\item Backtest.
Backtesting is an important part of model validation for VaR models and is usually implemented using data for consecutive dates.  Furthermore, the 20\% data in the test dataset (that covers non-consecutive dates) may not be enough for backtesting.  For these two reasons (non-consecutive dates and not enough data), we perform backtesting using data on every eligible dates.  This gives us more data points to assess the quality of the models.  For this purpose, synthetic data need to be generated for every date, which is why the backest step can be time consuing for some models.  See Table~\ref{tab:timeUSDYC1 (Libor curve)} for the running time of the backest step of each model.  Because backtest uses data on every eligible dates, unlike other KPIs, backtest KPI is not strictly out of sample.
\item Combination of KPIs into a composite score.
Distribution, autocorrelation and backtest KPIs are combined into a composite score.  See Section~\ref{subsec:evaluation metrics}.

Note that all KPIs are calculated using returns.
\end{enumerate}
\end{enumerate}

\section{Performance comparison \label{sec:performance comparison}}
In this section, we discuss details of model testing methodology and results.  The computing platform for model testing is described in Section~\ref{sec:platform}. Hyperparameter tuning is discussed in Section~\ref{subsec:Hyperparameter-tuning}.  As a result, a set of hyperparameter settings and list of model specifications forms the basis for model testing.  KPIs for model comparison are discussed in Section~\ref{subsec:evaluation metrics}.  Model testing results using simulated data are presented in Section~\ref{sec:simresults}, and model results using historical data are presented in Section~\ref{sec:hisresults}.

\subsection{Computing platform\label{sec:platform}}
The computing platform for model testing is as follows: OS: Red Hat Enterprise Linux Server 7.9 (Maipo) fedora, Architecture: x86\_64, Model name: Intel(R) Xeon(R) Gold 6138 CPU @ 2.00GHz, CPU(s): 80, GPU: NVidia V100-PCIE-32GB, Driver Version: 510.47.03, CUDA Driver Version: 11.6, RAM: 527GB.

We add the following notes on the coding for the study:
\begin{itemize}
\item For several neural network models (Diffusion, SIG, VAE), the code from external github is used as the starting point, we then made improvements as needed. See Section~\ref{sec:githubs} for external githubs used.
\item For CGANFC, CGANLSTM and CWGAN, we developed the code by ourselves.
\item For HS and all parametric models, we developed the code by ourselves.
\end{itemize}
For all models, the code is factored so that model training and validation follow the same steps: data scaling, training, generation, backtesting and KPI calculation.

For typical parameter settings, the model running time
\footnote{The code does not track the running time for KPI calculation currently.}
 in minutes for USDYC1 dataset is presented in Table~\ref{tab:timeUSDYC1 (Libor curve)}.  It can be observed that the top three time consuming models are CWGAN, Diffusion and LSTMCGAN.  CWGAN and LSTM are slow in training and Diffusion model is slow in backtesting.  To save time in backtesting step, the number of business dates in backtesting for Diffusion model are reduced from 3028 to 1216.  As a result, the running time for backtesting step for Diffusion model is reduced from 61 minutes to 21 minutes for USDYC1 dataset.  The two sets of backtest dates generate similar backtesting KPI.  As a result, for extensive model testing, a reduced number of backtest dates is used for Diffusion model.

Relative to production grade models with millions or billions of parameters, the NN models in this study are very limited in scale (most of the NN networks have 3$\sim$5 layers only).  This is also a reflection of the short time series for historical data used in the study.

\begin{table}
\centering
\caption{Model running time (in minutes) for dataset USDYC1 (Libor curve)}
\label{tab:timeUSDYC1 (Libor curve)}
\begin{tabular}{lllrrrrr}
\toprule
 No. & CAT &      MODEL &  TRAINING &  GENERATION &  BACKTEST &  KPI &  TOTAL \\
\midrule
   1 &  HS &        FHS &         0 &           0 &         0 &    7 &      7 \\
   2 &  HS &        PHS &         0 &           0 &         0 &    8 &      8 \\
   3 &  PM &         AR &         1 &           0 &         1 &    7 &      9 \\
   4 &  PM &     AR-RET &         1 &           0 &         1 &    7 &      9 \\
   5 &  PM &      GARCH &         7 &           0 &         1 &    7 &     15 \\
   6 &  PM &  GARCH-RET &         7 &           0 &         1 &    7 &     15 \\
   7 &  PM & GARCHt-RET &        11 &           0 &         1 &    7 &     19 \\
   8 &  PM &      NS-VS &         0 &           0 &         1 &    7 &      8 \\
   9 &  NN &    CGAN-FC &        50 &           0 &         0 &    7 &     57 \\
  10 &  NN &  CGAN-LSTM &       109 &           0 &         1 &    7 &    117 \\
  11 &  NN &      CWGAN &       149 &           0 &         0 &    7 &    156 \\
  12 &  NN &  DIFFUSION &        28 &          10 &        21 &    7 &     66 \\
  13 &  NN &        SIG &        79 &           0 &         0 &    8 &     87 \\
  14 &  NN &        VAE &         0 &           0 &         1 &    7 &      8 \\
\bottomrule
\end{tabular}
\end{table}

\subsection{Hyperparameter tuning\label{subsec:Hyperparameter-tuning}}

Model performance depends on hyperparameters that define network structure and training mechanics. Deep learning models have various hyperparameters, including model structural parameters, configuration parameters, as well as data preprocessing/postprocessing
parameters. To ensure consistency across all models that are evaluated,
the same configuration/hyperparameters are applied as much as possible.
The following model configuration and hyperparameters are in the setup
files that can be adjusted for each model run. 

\begin{table}
\begin{tabular}{p{0.3\linewidth}  p{0.7\linewidth}}
\hline 
 \textbf{Hyperparameter} & \textbf{        Note}\tabularnewline
\hline 
\hline 
\multicolumn{2}{l}{\textbf{Data processing}}\tabularnewline
\hline
dataset & Training dataset name\tabularnewline
\hline 
Input data return type & Return type of the input data (absolute or relative)\tabularnewline
\hline 
Traing-test split percent & 80\% train and 20\% test split in general\tabularnewline
\hline 
data split seed & Seed to split the data\tabularnewline
\hline 
\multicolumn{2}{l}{\textbf{Trainer}}\tabularnewline
\hline 
batch size & Batch size of training data\tabularnewline
\hline 
epochs & Training total steps\tabularnewline
\hline 
condition length & Condition length\tabularnewline
\hline 
sequence length & Target sequence length for forecasting\tabularnewline
\hline 
scaler type & data transformation type (MinMaxScaler or StandardScaler)\tabularnewline
\hline 
activation function & Activation function \tabularnewline
\hline 
noise\_dim & noise dimension\tabularnewline
\hline 
\multicolumn{2}{l}{\textbf{Generator}}\tabularnewline
\hline 
generator.seed & Generator seed \tabularnewline
\hline 
n\_synthetic & Number of copies of synthetic data to generate\tabularnewline
\hline 
long path steps & The long path steps for generator\tabularnewline
\hline 
short\_path\_steps & short sequence length of the generated data (for calculating KPI), normally same as training sequence
length\tabularnewline
\hline 
long\_path\_steps & long sequence length of the generated data (for calculating ACF)\tabularnewline
\hline 
\multicolumn{2}{l}{\textbf{Common parameters}}\tabularnewline
\hline 
optimizer & By each model\tabularnewline
\hline 
learning rate & controls how much to change the model in response to the estimated
error each time the model weights are updated\tabularnewline
\hline 
\multicolumn{2}{l}{\textbf{Network parameters}}\tabularnewline
\hline 
network type & By each model\tabularnewline
\hline 
layers & number of layers\tabularnewline
\hline 
clip value & The upper bound of clipping the weights in GAN\tabularnewline
\hline 
hidden\_dim & hidden dimension of the network (often in latent space)\tabularnewline
\hline 
\end{tabular}
\caption{Model configuration and hyperparameter list \label{tab:Model-configuration-and-1}}
\end{table}

Our major efforts over hyperparameter tuning is on the trade-off between
model stability and performance. Overall, we have tested different
sequential length of sliced data to better align to the model use,
and finally chose to set condition length to 10, and target sequential length to 10.

The following hyperparameters and variations are tested for applicable models:
\begin{itemize}
\item GAN: clipping value. Specifically on LSTM CGAN, clip\_value 0.05,
0.075, 0.1, 0.75 are tested and 0.075 was chosen.
\item VAE: weights on different loss function components are tested. 
\item Signature: different prediction model in signature space (predicting future signature from past signature).  Linear, NN with dense layers, and NN with CNN layers are tested.  NN with CNN layers are adopted.  See the oneline appendix for details.
\item Nelson-Siegel representation for yield curve: both AR+GARCH and Vasicek model are considered for the dyamics of the three factors.  NS with Vasicek dynamics performs better and is adopted.  See the oneline appendix for details.
\item AR and Vasicek models
\footnote{Strictly speaking, this is not hyperparameter selection, but this shows how the model selection decision is made.}
: AR and Vasicek models are very close.  Model testing shows that they have similar performance.  The simpler AR model is used in model comparison.  See the oneline appendix for details..
\end{itemize}
To be consistent across all models, we carefully designed network
structure to enable apple-to-apple performance comparison. Despite
the significant differences among all network structures embedded in
the different model architectures, one of the common hyperparameters
is the dimension of the random noise $z$ as input ($noise\_dim$). For
a typical linear three factor model of the yield curve (e.g. Nelson-Siegel
model), we replicate \eqref{eq:NSt} below,
\begin{equation}
R_t(\tau)=b_{0t}+b_{1t}f_1(\tau)+b_{2t}f_2(\tau)\nonumber
\end{equation}
where $R_{\tau}(t)$ is yield for tenor $\tau$ at time $t$, $f_1(\tau)$ is loading of slope factor on tenor $\tau$, and $f_2(\tau)$ is loading of curvature factor on tenor $\tau$. $b_{it},i=0,1,2$ are random factors. With
this three factor model for a nine tenor yield curve, three random numbers $(b_{0t},b_{1t}, b_{2t})$
are used to generate nine yields. 

For some models (CGAN, CWGAN, Diffusion, TimeVAE), $noise\_dim$ is
used to generate random noise for each condition/business date, which is in turn
used to generate a $10\times9$ sequence for a typical model of generating
daily 9 tenor yields for the next 10 days ($step=10$), thus 90 yields. 
\begin{itemize}
\item For USD yield curve, this parameter is set to $noise\_dim=30$, equivalent
to a three factor model. In such cases, 30 random numbers are used
to generate 90 yields, or equivalently, three random numbers are used
to generate nine yields.
\item For simulated two tenor data, $noise\_dim=20$ is used, which is equivalent
to a two factor (or full factor) model. A value lower than 20 would
imply a model that may be too simple and restrictive.
\end{itemize}
For other models (LSTM and Signature), $noise\_dim$ is used to generate
random noise for each step (of 10 steps) in generation, thus $10\times noise\_dim$
random numbers are used to generate a $10\times9$ sequence for a
typical model of generating daily 9 tenor yields for the next 10 days,
thus 90 yields.
\begin{itemize}
\item For USD yield curve, this parameter is set to $noise\_dim=3$, equivalent
to a three factor model.
\item For simulated two tenor data, $noise\_dim=2$ is used, which is equivalent
to a two factor (or full) model.
\end{itemize}
This study does not include comprehensive testing of hyperparameters.

For parametric models, AR models are estimated with 5-year window ($252\times 5$ dates), GARCH models are estimated with 3-year window ($252\times 3$ dates).

\subsection{KPIs for model comparison\label{subsec:evaluation metrics}}

We quantify what it means for a set of generated synthetic samples
to have similar statistical properties (``look like'') to the real
data. Real data here refers to windowed historical data in the testing
dataset, which is 20\% of the available real data in a typical application. Details of the data preparation pipeline are discussed in Section \ref{subsec:Training-and-testing}. We generate both short-path (10 days) and long-path
(502 days) synthetic samples from each generative model. Short-path
synthetic samples use the single-step prediction method. Long-path synthetic samples use the iterative method
 to generate multi-step prediction of 502 days. See Section~\ref{subsec:single-step prediction} for details. Only
the ACF plot (Figures~\ref{fig:Example-of-empirical} and \ref{fig:Example-ACF-ID}) uses long-path synthetic samples. All other plots and
metrics use short-path synthetic samples.

Table~\ref{tab:List-of-KPIs} provides a list of qualitative and quantitative
measures we use to compare the performance of different models.

\begin{table}
\caption{List of measures for model comparison\label{tab:List-of-KPIs}}
\centering{}%
\begin{tabular}{ll}
\toprule
KPI Category & Measure \tabularnewline
\hline 
\multirow{6}{*}{Distribution} & Distribution and ACF plots \tabularnewline
\cmidrule{2-2}
 & Distribution distance (DY) \tabularnewline
\cmidrule{2-2}
 & Earth moving distance (EMD) \tabularnewline
\cmidrule{2-2}
 & *Kolmogorov-Smirnov test (KS) of sample moments \tabularnewline
\cmidrule{2-2}
 & Series distance \tabularnewline
\cline{2-2} 
 &  *Kolmogorov-Smirnov test (KS) of returns \tabularnewline
\hline 
\multirow{3}{*}{Correlation} & Inter-tenor correlation matrix \tabularnewline
\cline{2-2} 
 & ACF score \tabularnewline
\cline{2-2} 
 & *Fisher test of equality of correlation \tabularnewline
\hline 
\multirow{3}{*}{Embedding} & t-SNE \tabularnewline
\cline{2-2} 
 & UMAP \tabularnewline
\cline{2-2} 
 & PCA \tabularnewline
%\hline 
\midrule
\multirow{6}{*}{Backtesting} & u-value histogram \tabularnewline
\cline{2-2} 
 & u-value histogram ranges \tabularnewline
\cline{2-2} 
 & u-value histogram difference from 1.0 \tabularnewline
\cline{2-2} 
 & u-value breach rate (diff from theoretical) from 1.0 \tabularnewline
\cline{2-2} 
 & Envelope plot \tabularnewline
\cline{2-2} 
 &  *Kolmogorov-Smirnov test (KS) of u-values \tabularnewline
\hline 
\multirow{3}{*}{Combination of KPIs}  & KS of moments + KS of returns $\rightarrow$ Distribution (DIST) score \tabularnewline
\cline{2-2}
& Breach rates + KS of u-value $\rightarrow$ Backtest (BT) score \tabularnewline
\cline{2-2} 
 & Distribution + ACF (Fisher test) + BT $\rightarrow$ Composite score \tabularnewline
\bottomrule
\end{tabular}
\begin{tablenotes}\item[*] (Only a subset (*) of KPIs are combined to form Composite score.)\end{tablenotes}
\end{table}

We examine the fidelity of synthetic data to the real data using qualitative and quantitative measures.  
Qualitative measures include empirical distribution visual comparison, t-SNE, PCA and UMAP.
Quantitative measures include distribution distance (Earth Moving Distance, DY and KS distance metrics, Series distance), ACF and backtesting.  Details are described below.

Model rankings vary depending on the KPIs used and the way they are combined to form the composite score.

\subsubsection{Qualitative measure}

\begin{figure}
\centering{}\includegraphics[width=1\columnwidth]{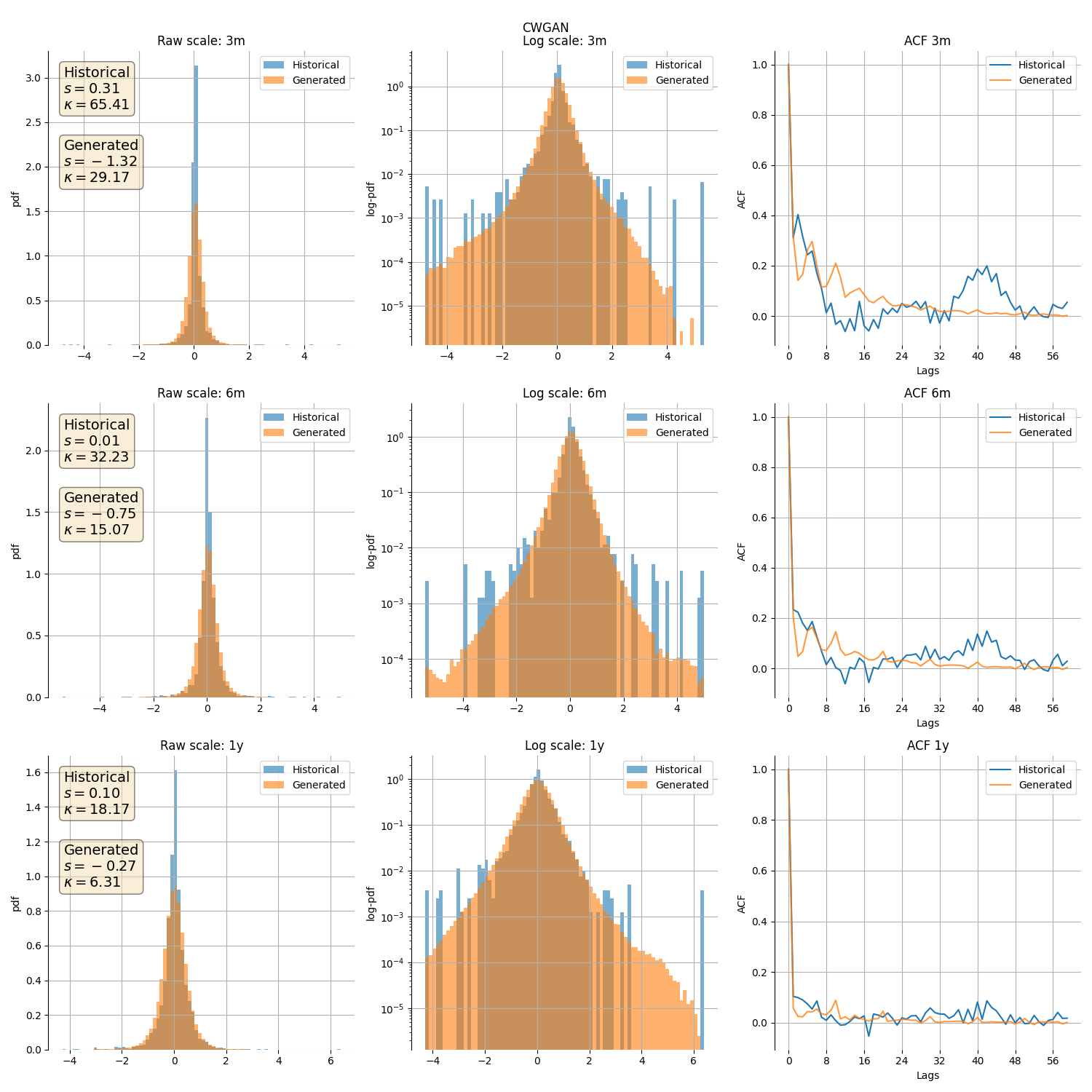}\caption{\label{fig:Example-of-empirical}Example of empirical distribution visualization}
\end{figure}

We visualize the distribution of real versus generated distributions
with and without log scale. Figure \ref{fig:Example-of-empirical}
shows an example visualization of histograms and ACF plots. Real data
in the plots is test samples of windowed historical data. For synthetic
data, the histogram uses short-path synthetic samples, while ACF plot
uses long-path synthetic samples to cover longer lags.

We use three dimensionality reduction techniques to map real data
and synthetic samples onto a two-dimensional space and compare their
distributions. These are visualizations, hence, qualitative measures.

\textbf{Empirical distribution and ACF plot}\\
Examples of empirical distribution of test and synthetic data and ACF are presented in Figure~\ref{fig:Example-of-empirical}. It shows visual comparison of distributions and ACF.

\begin{figure}
\centering{}\includegraphics[width=1\columnwidth]{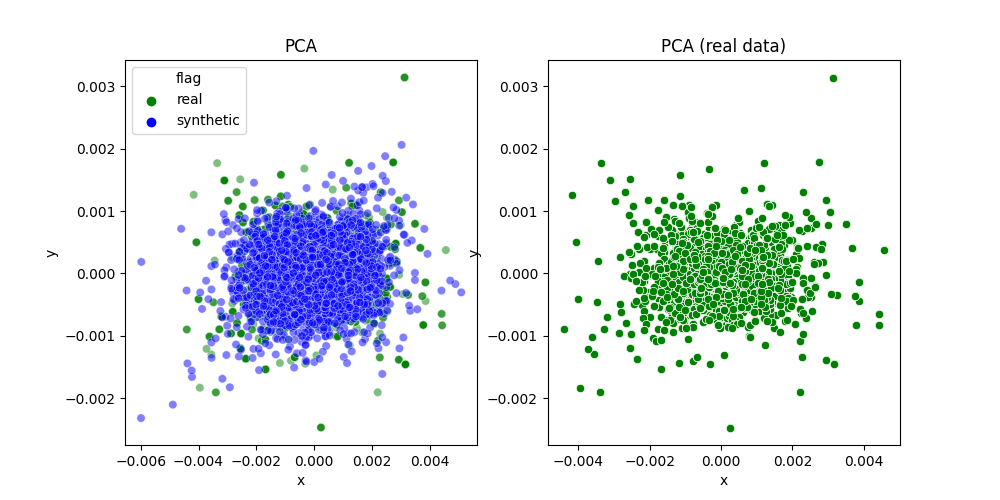}\caption{\label{fig:Example-PCA-visualization}Example of PCA visualization}
\end{figure}

\textbf{PCA}\\
Principal component analysis (PCA) is a statistical technique that performs linear dimensionality reduction for a dataset.   This is accomplished by linearly transforming the data into a new coordinate system where (most of) the variation in the data can be described with fewer dimensions than the initial data. Many studies use the first two or three principal components, and plot the data in two dimensions and to visually identify clusters of closely related data points.\\
We use PCA to project real and synthetic data
onto two-dimensional space for visualization. Figure \ref{fig:Example-PCA-visualization}
shows an example of PCA results. Real data and synthetic data also covers
similar regions in the projection space
\footnote{Two plots are shown because real and synthetic data often show a high degree of overlap and it is hard to tell whether real data are covered by (a subset) or overlapped with synthetic data.}
.

\textbf{t-SNE}\\
t-SNE (t-distributed stochastic neighbor embedding) is introduced
in 2008 in \cite{JMLR:v9:vandermaaten08a}.
It is a non-linear method that maps high-dimensional data into a low-dimensional
space, while preserving the pairwise similarities between data points.
The mathematics of t-SNE involves two main steps. First, t-SNE constructs
a probability distribution over data points in high-dimensional space
using a Gaussian kernel. Second, t-SNE defines a similar probability
distribution over the points in the low-dimensional space, and it
minimizes the Kullback\textendash Leibler divergence (KL divergence)
between the two distributions. Compared with PCA, t-SNE can be characterized as a nonlinear dimensionality reduction technique
\footnote{For a comparion of PCA and t-SNE, see \url{https://www.geeksforgeeks.org/difference-between-pca-vs-t-sne/}}.

\begin{figure}
\begin{centering}
\includegraphics[width=1\columnwidth]{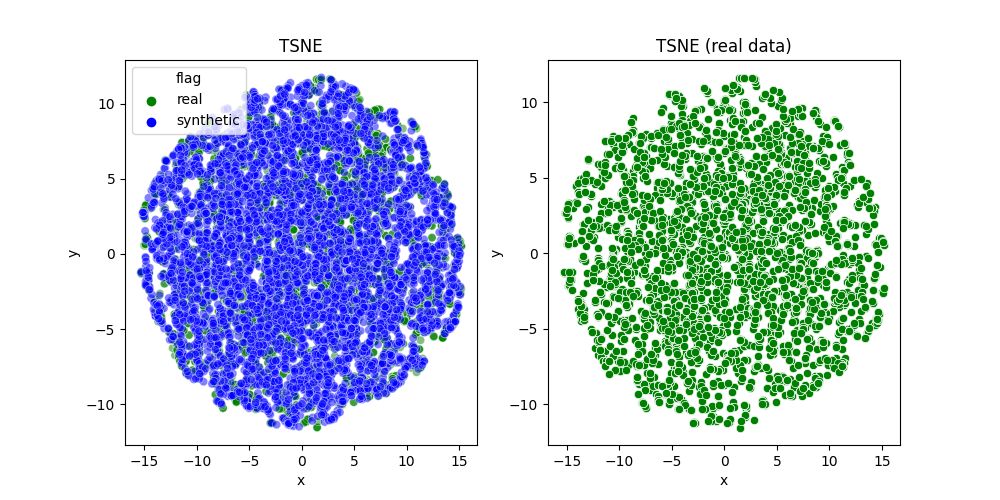}
\par\end{centering}
\caption{\label{fig:Example-TSNE-visualization}Example t-SNE visualization}
\end{figure}

Figure \ref{fig:Example-TSNE-visualization} shows an example t-SNE
visualization. Figure on the left shows the data points for real data
and synthetic data together, while the figure on the right shows only the
points from the real data. We can see the synthetic data covers similar
regions as the real data.

\textbf{UMAP}\\
UMAP (Uniform Manifold Approximation and Projection) was proposed
 in 2018 in \cite{mcinnes2020umap}.
It constructs a topological representation of the high-dimensional
data and then projects it onto a lower-dimensional space. This is
done by constructing a weighted graph of the data, where vertices
represent the data points and the edges represent the relationships
between the points. UMAP is effective in preserving both global and
local structure in high-dimensional data
\footnote{For a detailed description of UMAP, see \url{https://umap-learn.readthedocs.io/en/latest/}}.

t-SNE and UMAP are related and different as follows
\footnote{For more details, see \url{https://blog.bioturing.com/2022/01/14/umap-vs-t-sne-single-cell-rna-seq-data-visualization/}.}
: t-SNE uses a Gaussian probability function to calculate how likely a cell will pick another cell as its neighbor, and repeats this step for all cells. In the low dimension space, cells are rearranged according to these distances, creating the t-SNE plot.  UMAP, in a more clever way, creates a fuzzy graph that accurately reflects the topology (a.k.a shape) of the true high dimensional graph, calculates the weight for edges of this graph, then builds the low dimensional graph mimicking the fuzzy graph.  In another word: while t-SNE moves the graph point-to-point from high to low dimensional space, UMAP makes a fuzzy, but topologically similar graph and compresses it into a lower dimension.

\begin{figure}
\begin{centering}
\includegraphics[width=1\columnwidth]{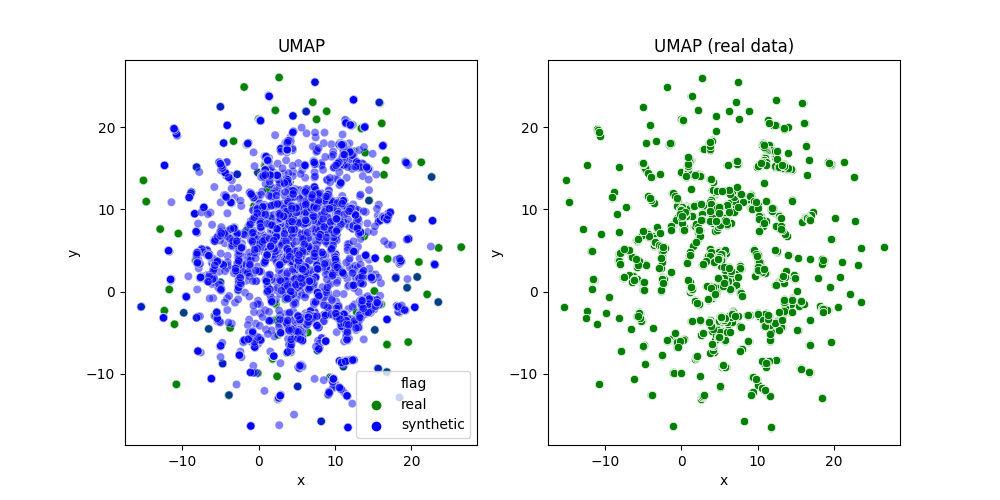}
\par\end{centering}
\caption{\label{fig:Example-UMAP-visualization}Example UMAP visualization}
\end{figure}

We apply UMAP to map the real and synthetic data onto two-dimensional space.
Figure \ref{fig:Example-UMAP-visualization} shows an example of UMAP
visualization. synthetic data cover similar regions as the real data in
2D space.

\subsubsection{Distribution comparison\label{subsec:Real-versus-synthetic}}

\begin{figure}
\centering{}
\includegraphics[width=1\columnwidth]{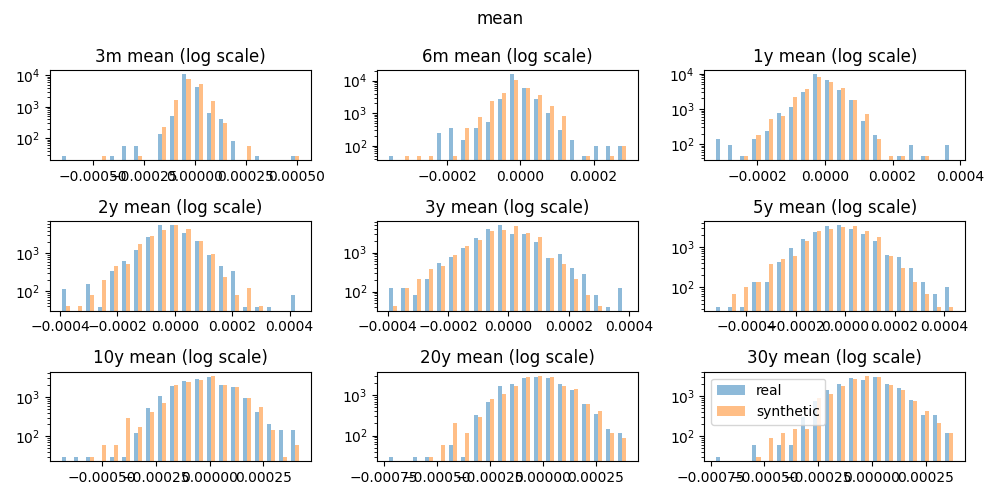}
\caption{Example of histogram of sample mean}
\label{fig:meanhist}
\end{figure}

For each test date $t$, one copy of the windowed return $\tilde{x}_t$ is generated, and this is a $q \times d$ array (typically $q=10$, $d=9$ for yield curve). Sample moments (mean, standard deviation, skewness and kurtosis) for each $\tilde{x}_t$ (in the $q$ dimension) is calculated.  This is calculated for each test date for real data $x_t$ and synthetic data $\tilde{x}_t$, producing a set of sample moments.  Take the sample mean for example.  Suppose there are $S$ test dates in total, this procedure creates an array of size $S\times d$ sample means for the real data, and an array of size $S\times d$ sample means for synthetic data.  The distribution of the sample mean across all $S$ test dates are presented in Figure~\ref{fig:meanhist}.

Sample means of the synthetic data and sample means of real data are tested for whether they are from the same distribution.  Other sample statistics (standard deviation, skewness and kurtosis) can also be used in the testing.  However, since the typical sequence length is $q=10$ days,  skewness and kurtosis calculated from 10 data points may not be reliable.  Therefore, only sample mean and sample standard deviation are used in the testing.

The sample moments are used to calculate distribution distance in Section~\ref{sec:distd}.  The returns are used to calculate series distance in Section~\ref{sec:seriesd}.

\paragraph{Distribution distance\label{sec:distd}}
We measure distribution distance using EMD, DY and KS metrics. Details are provided below.

\textit{Earth Mover or Wasserstein-1 distance (EMD)}
Let $\mathbb{P}_{r}$ denote the historical distribution $X\sim \mathbb{P}_{r}$ and $\mathbb{P}_{g}$
the generated distribution $\tilde{X}\sim\mathbb{P}_{g}$. Let
$\prod({\mathbb{P}_{r}},{\mathbb{P}_{g}})$ denote the set of all
joint probability distributions with marginals ${\mathbb{P}}_{r}$
and ${\mathbb{P}}_{g}$. \textit{\emph{The earth mover distance}}
(\cite{villani2008optimal}) describes how much probability mass has
to be moved to transform ${\mathbb{P}}_{r}$ into ${\mathbb{P}}_{g}$
. Details of EMD are discussed in \cite{villani2008optimal}. 

EMD is defined by: 
\begin{equation}
\mathrm{{EMD}}({\mathbb{P}}_{r},{\mathbb{P}}_{g})=\inf_{\pi\in\prod({\mathbb{P}}_{r},{\mathbb{P}}_{g})}E_{(X,\tilde{X})\sim\pi}(\|X-\tilde{X}\|)
\end{equation}
This is the infimum across all joint distributions.  It is shown \cite{ramdas2015wasserstein} that EMD can be calculated as the following ordinary integral (without the infimum)
\footnote{See \url{https://docs.scipy.org/doc/scipy-1.7.1/reference/generated/scipy.stats.wasserstein\_distance.html}}
:
\begin{equation}
EMD = \int_{-\infty}^\infty |F_X(s)-F_{\tilde{X}}(s)|ds
\end{equation}
where $F_X$ and $F_{\tilde{X}}$ are the CDF of random variables $X$ and $\tilde{X}$ respectively.\\
\textit{Distributional metric: DY metric}\\
\textit{\emph{DY metric}} is proposed in (\cite{DragulescuEtAl02}).
The DY metric is defined by:
\begin{equation}
DY=\sum_{x}|\log{P_{r}(A_{x})}-\log{P_{g}(A_{x})}|
\end{equation}
where $P_{r}$ and $P_{g}$ denote the empirical
probability density function of the historical and generated path. Further, $(A_{x})$ denotes a partitioning of the
real number line such that for all $x$ we (approximately)
have $\log{P_{r}(A_{x})}=\frac{5}{T}$ for $T$ the number of
historical returns. \\
\textit{KS distance}\\
The Kolmogorov-Smirnov (KS) distance measures the similarity between
the empirical CDF of real data and synthetic samples. It is the max
absolute difference of the empirical pdf.\\[2ex]
Sample result for distribution distance is presented in Table~\ref{tab:distdUSDYC1}.  In the table, $KS\_pval$ is the p value (significance level) of the two-sample KS test.  $1-KS\_pval$ is 1 minus $KS\_pval$, which is used in model ranking since a smaller value of $1-KS\_pval$ (a larger value of $KS\_pval$) indicates a better model.
\begin{table}
\centering
\caption{Example distribution distance for dataset USDYC1 (Libor Curve)}
\label{tab:distdUSDYC1}
\begin{tabular}{cclllll}
\toprule
Statistic & Tenor &   EMD &    DY &    KS &  $KS_{\rm pval}$ &  $1-KS_{\rm pval}$ \\
\midrule
     mean &    3m & 0.013 & 1.289 & 0.252 &    0.000 &      1.000 \\
     mean &    6m & 0.012 & 1.095 & 0.191 &    0.000 &      1.000 \\
     mean &    1y & 0.009 & 0.964 & 0.118 &    0.000 &      1.000 \\
     mean &    2y & 0.008 & 0.896 & 0.035 &    0.862 &      0.138 \\
     mean &    3y & 0.009 & 0.846 & 0.043 &    0.635 &      0.365 \\
\bottomrule
\end{tabular}
\end{table}

\paragraph{\textit{\emph{Series distance\label{sec:seriesd}}}}
In model comparison, for simplicity, the two-sample KS test for distribution is used for series distance.  It is calculated using the returns rather than the sample moment of returns.

\subsubsection{\textit{\emph{Inter-tenor correlations\label{subsec:Inter-tenor-correlations-and}}}}

\begin{figure}
\begin{centering}
\includegraphics[width=0.8\columnwidth]{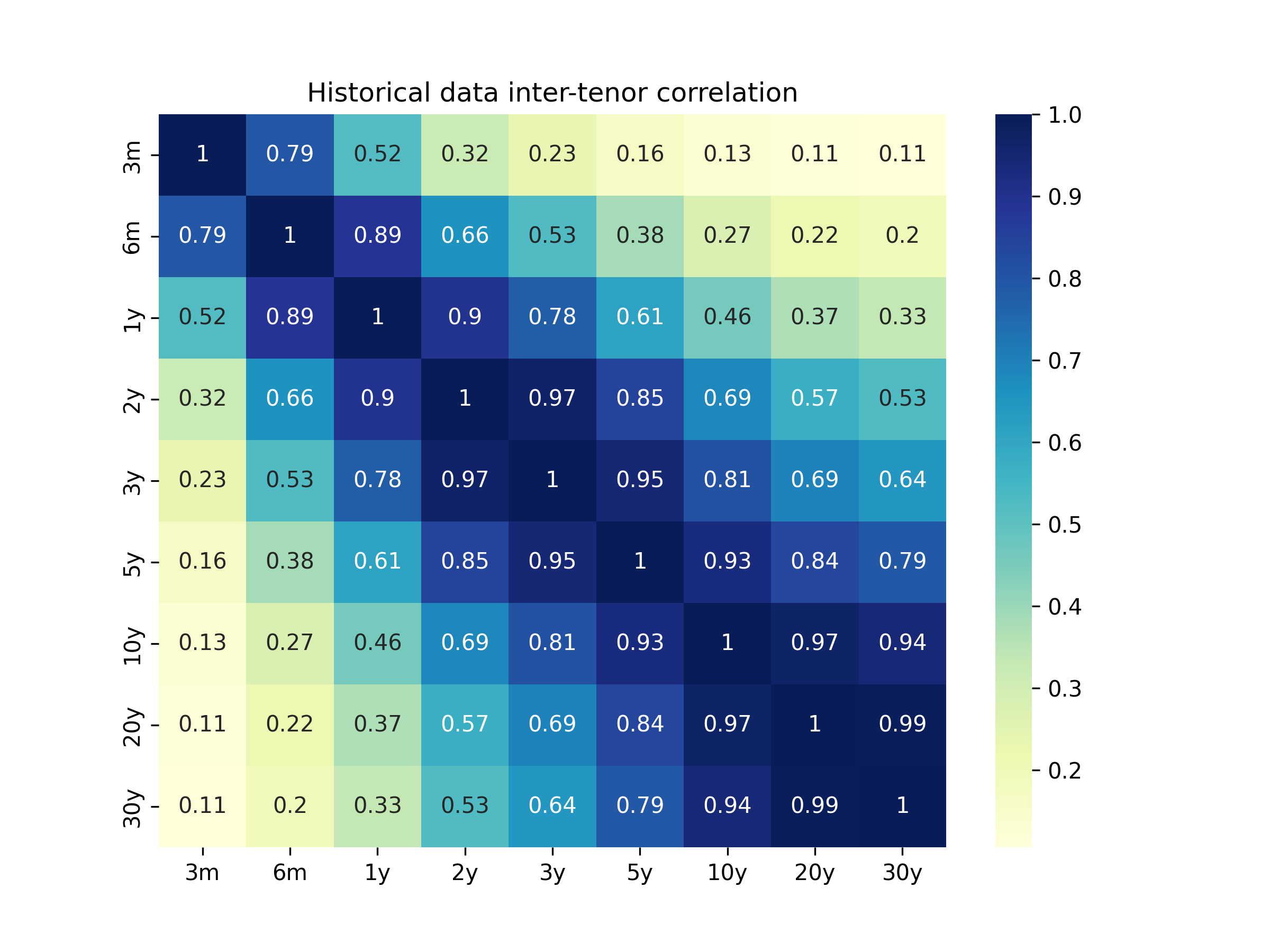}
\par\end{centering}
\caption{\label{fig:Example-historical-data}Example of historical data returns
correlation matrix}
\end{figure}

\begin{figure}
\begin{centering}
\includegraphics[scale=0.5]{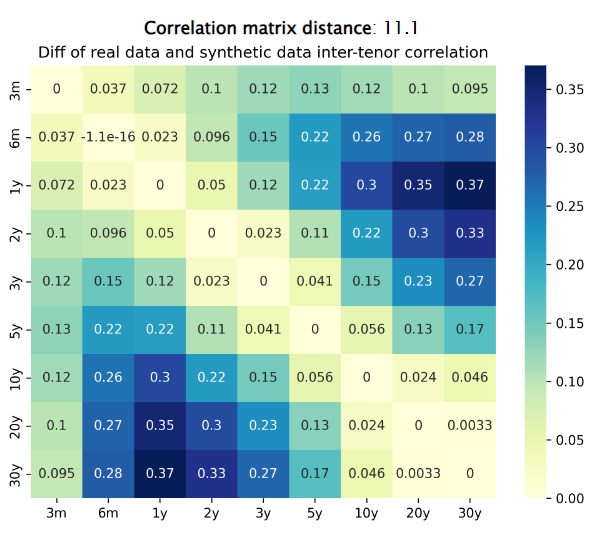}
\par\end{centering}
\caption{\label{fig:correlation_distance}Example of correlation matrix distance visualization}
\end{figure}

We estimate the cross tenor covariance and correlation of real and synthetic returns.  Figure \ref{fig:Example-historical-data} shows an example of cross tenor correlation matrix for USDYC1 dataset. Figure~\ref{fig:correlation_distance} shows the difference between correlation matrices of real and synthetic data.

\subsubsection{Autocorrelation\label{subsec:Real-versus-synthetic-1}}

It is straightforward to calculate autocorrelation (ACF) for a time series $x_t$. 
Let the ACF with lag $l$ be denoted as,
\[
{\mathcal{C}}(l;x)=\mathrm{Corr}(x_{t+l},x_{t})
\]
However we need to go through some tricky steps because the real and synthetic data are in sequence form (3-dimensional array) rather than in time series form (2-dimensional array).  For discussion, let $X$ be a 3-dimensional array, with the first dimension for test date or sequence number, the second dimension the sequence length $q=10$, and the third dimension $d=9$ for tenor.  ACF should be calculated along the second dimension $q=10$ (for dates $t+1,\cdots,t+q$).  However, since the length of the time series (time dimension) is only 10, there isn't enough data to calculate ACF for longer lags.  For illustration, we select one tenor and fix the resulting 2D array $X=(x_{ij}), i=1,2,\cdots,N,j=1,2,\cdots,q$ array.  Prior to model training, the whole dataset is split randomly into train and test datasets.  As a result, the test date that the first dimension of $X$ represents, $t_i$, is not even consecutive, highlighting the fact that ACF is calculated along the second dimension of $X$.  To increase the sample size for ACF calculation, we form all pairs of a given lag from all sequences.  For fixed $i$, lag 1 pairs can be formed from $(X(i,j),X(i,j+1)), j\le q-1$, only 9 pairs for $q=10$.  This is hardly enough to calculate the first order ACF.  The trick is to let $i$ vary and we collect all lag 1 pais from all sequences.  Hypothetically, if there are 600 sequences of length 10, this procedure forms a sample of lag 1 pairs of size $600\times 9 = 5400$, much more data than just 9 observations for first order ACF calculation in one sequence.

\begin{figure}
\centering{}\includegraphics[width=1\columnwidth]{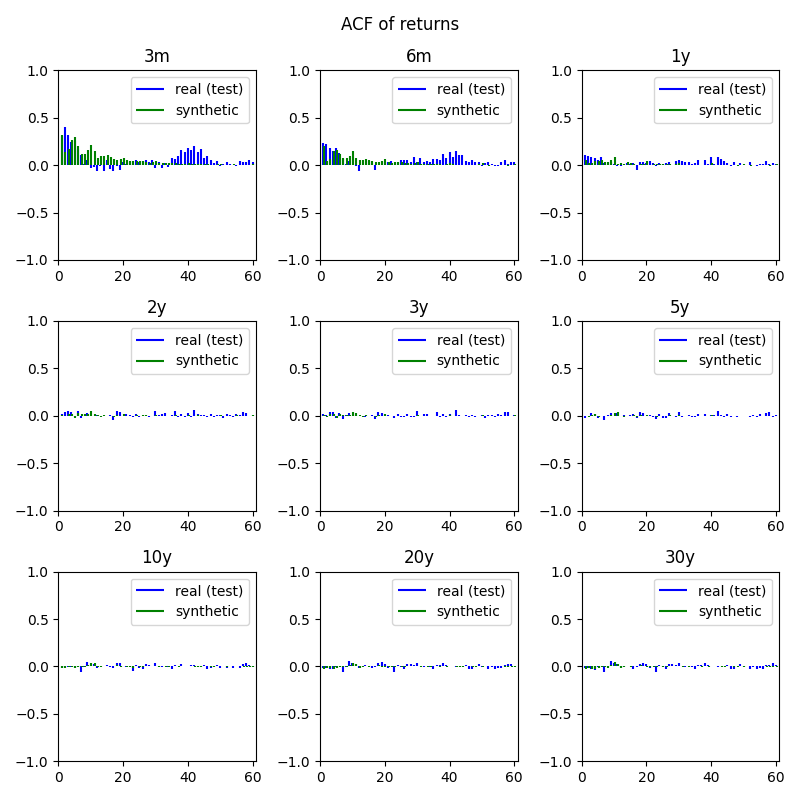}\caption{\label{fig:Example-ACF-ID}Example visualization of real
versus synthetic autocorrelation for returns}
\end{figure}

\begin{figure}
\centering{}\includegraphics[width=1\columnwidth]{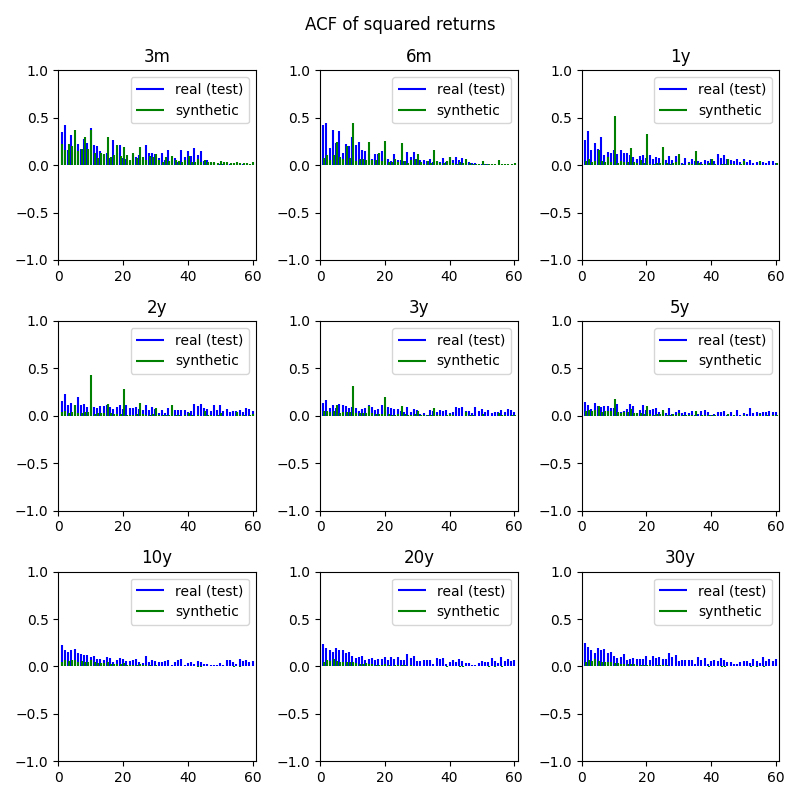}\caption{\label{fig:Example-ACF-SQ}Example visualization of real
versus synthetic autocorrelation for returns squared}
\end{figure}

For any function $f$, ACF can also be calculated for $f(x)$.  For financial time series, $f(x)=|x|$ and $f(x)=x^2$ is particularly relevant since the ACF for these $f(x)$ captures volatility (or variance) clustering.
Figure~\ref{fig:Example-ACF-ID} shows an example visualization of real versus synthetic autocorrelation
for $f(x)=x$, and Figure~\ref{fig:Example-ACF-SQ} for $f(x)=x^2$
\footnote{In addition to the regular ACF, leverage ACF can also be calculated, $\mbox{leverage ACF}=Corr(x_{t+l}^2,x_t)$, which measures the correlation of variance with lagged returns (the leverage effect). For financial time series, volatility clustering is a much stronger feature than the leverage effect.}
.  It can be observed that the ACF for squared returns (Figure~\ref{fig:Example-ACF-SQ}) is much stronger than the ACF for returns (Figure~\ref{fig:Example-ACF-ID}), which is typical of financial time series.

The \textbf{ACF score} $\mathrm{ACF}(f)$ is a summary measure of the ACF difference for several lags.  Let $C(x)$ be a vector of ACF at several lags (since $q=10$, ACF at only lags 1 and 2 are calculated),
\[
C(x)=(C(1,x),C(2,x))
\]
The ACF norm score is just the $L_2$ norm of the difference of the ACF vector of real and synthetic data.
\[
\mathrm{ACF}(f)=\Big\| C(f(x))-C(f(\tilde{x}))\Big\|_{2}
\]

where the function f is applied element-wise to the series. The ACF norm
score is computed for the functions $f(x)=x$, $f(x)=|x|$ and $f(x)=x^{2}$.

To facilitate the calculation of a composite score later, the comparison of ACF (for any pair of correlation coefficients $\rho_1$ and $\rho_2$) is converted into probability space using the Fisher Z-transformation
\footnote{See \url{https://github.com/psinger/CorrelationStats/blob/master/corrstats.py}.}
.  This ACF score is defined as,
\begin{equation}
\mathrm{ACF(\rho_1,\rho_2)}=1-\Phi\left ( \frac{|\rho_1-\rho_2|}{\sqrt{\frac{1}{n_1-3}+\frac{1}{n_2-3}}}\right )\label{eq:fisherz}
\end{equation}
where $\Phi$ is the CDF of a standard normal distribution, and $n_i$ are the number of observations used to calculate $\rho_i$.

For ACF vectors, $C(x)$ (for real data) and $C(\tilde{x})$ (for synthetic data), a vector $\mathrm{ACF(\rho_1,\rho_2)}$ is calculated using the Fisher Z-transformation.  $\mathrm{ACF(\rho_1,\rho_2)}, \rho_1\in C(x),\rho_2\in C(\tilde{x})$ is used in the calculation of the composite score.

\subsubsection{Training loss\label{subsec:Training-loss}}

\begin{figure}
\centering{}\includegraphics[width=1\columnwidth]{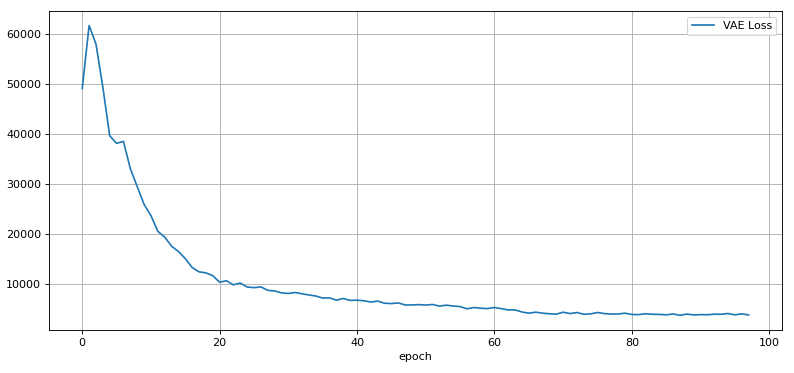}\caption{\label{fig:Example-loss-function}Example loss function visualization}
\end{figure}

We visualize the loss function over training epochs to evaluate convergence
of the model. Figure \ref{fig:Example-loss-function} gives an example
of the loss function for VAE model.

\subsubsection{Backtesting \label{subsec:backtest}}

Backtesting is a common requirement to assess the quality of VaR models
\footnote{For the online version, see \url{https://www.bis.org/bcbs/publ/d457.htm}.}
.

In probability theory, the probability integral transform\footnote{\url{https://en.wikipedia.org/wiki/Probability_integral_transform}}
(PIT) states that, for any random variable $X$ with a continuous
distribution and cumulative density function (CDF) $F(x)$, the random
variable $Y=F(X)$ follows a uniform distribution. Let $X_{i}$ be
a series of random variables with CDF $F_{i}(x)$, and $x_{i}\sim F_{i},i=1,\cdots,N$
be a random sample, then, $y_{i}=F(x_{i}),i=1,\cdots,N$ is an independent
sample from the uniform distribution. Note that each $x_{i}$ is sampled
from its own true distribution. A statistical test for $\{y_{i}\}\sim U[0,1]$
is equivalent to the joint test $x_{i}\sim F_{i}(x),i=1,\cdots,N$.
This idea has been used in the context of backesting market risk VaR
models. See \cite{crnkovic1996,berkowitz2001}.

In our application, for each business date $t$, there is a condition
($x_{t-p},x_{t-p+1},\cdots,x_{t-1}$) used to generate synthetic data
for a range of dates in the future $\hat{x}_{t},\hat{x}_{t+1},\cdots,\hat{x}_{t+q-1}$.
We focus on $x_{t}$ and $\hat{x}_{t}$ (one step ahead generation).
When $N$ independent copies of $\hat{x}_{t}$ are generated, which
is denoted as $\hat{x}_{t}^{j},j=1,\cdots,N$, it forms a distribution
forecast $\hat{F}_{t}$ for $x_{t}$. Let $y_{t}=\hat{F_{t}}(x_{t})$,
which is the probability value calculated for the actual data point $x_{t}$
relative to the distribution forecast $\hat{F}_{t}$. Collecting $y_{t}$
for multiple business dates, then the random sample $y_{j},j=t,t+1,\cdots,t+N-1$
follows uniform distribution under the assumptions that (1) $x_{j}$
is independent for $j=t,t+1,\cdots,t+N-1$, and (2) $\hat{F_{j}}$ is
the true distribution for $x_{j}$,$j=t,t+1,\cdots,t+N-1$.  $y_{t}$ is called u-value instead of p-value for two reasons, (a) $y_{t}$ follows a uniform distribution under the null hypothesis of correct model, (b) this distinguishes the p-value from a statistical test. 

The uniformity of $y_{j}$ can be assessed using different statistical
tests, and the most well known test is the one sample Kolmogorov\textendash Smirnov
test (KS test). As with all statistical tests, the result of the KS
test includes the value of the test statistic and the p-value of the
test statistic in the null distribution. However, the independence
assumption for the KS test may not hold, and the p-value of the KS
test is only approximate.

We also use two intuitive metrics to measure deviation from uniformity.
Create a histogram of u-values $y_{j},j=t,t+1,\cdots,t+N-1$ with
10 equal bins on the unit interval {[}0,1{]}. Under uniformity, all
bars in the histogram are of the same height.  This has three implications,
(1) the standard deviation among the height of the 10 bars is zero;
(2) the range of the bars (the difference between the tallest
bar and the shortest bar) is zero; (3) the deviation of the height of bars from 1.0 is zero. Therefore, the standard deviation, range of the u-value histogram bars, and deviation of the bars from 1.0 are the three intuitive measures we can use as indications of deviation from uniformity: the smaller,
the closer to uniformity by each measure, and the better quality
of the model producing the distribution forecast $\hat{F_{j}}$.

For market risk VaR model, a breach of the VaR model occurs when the realized loss of a trade or portfolio is greater than the VaR forecast at a confidence level for the day.  For example, if the VaR forecast for the day is \$100 million, but the realized loss is \$105 million, then the VaR is breached.  A correct 99\% VaR is expected to be breached with 1\% probability.  In a sample, a breach proportion (or rate) that is different from 1\% is evidence of violation of the VaR model.  This holds similarly for other breaching probabilities (e.g. 2.5\%, 5\%, 10\% in the left tail and 90\%, 95\% and 97.5\% in the right tail).  For a small probability value, e.g. $p=0.5\%$, breach rate in the left tail is counted, while for a large probability level, e.g. $p=95\%$, breach rate in the right tail is counted.
For use of backtesting with breach numbers in regulatory capital calculation for market risk, see \cite{FRTB2019}
\footnote{For the online version, see \url{https://www.bis.org/bcbs/publ/d457.htm}.}
.

A larger number of metrics are generated in backtesting, including (absolute) deviations from theoretical breach rates at 2.5\%, 5\%, 10\%, 90\%, 95\%, 97.5\%, (absolute) deviation of height of u-value histogram bar from 1.0, KS distance of u-value from uniform distribution.  An example of various backtest scores is presented in Table~\ref{tab:btUSDYC1 (Libor curve)}.  KSpval in the table is the p-value from KS test of u-values following a uniform distribution. "DIFF" in the table is the average absolute difference between the height of u-value histogram from 1.0.

To save space, the breach rate for only 5\% (BR05) and 95\% (BR95) are included.  In the actual calculation, 2.5\%, 5\%, 10\% in the left tail and 90\%, 95\% and 97.5\% in the right tail are used.  For each probability level $p, BRp$ indicates the absolute difference between the actual breach rate and expected breach rate.

In the table, KSpval=0 for all six rows shown,indicating that uniform distribution of u-values for these sample periods and 3m tenor are rejected.  DIFF column shows that the histogram is very different from uniform.

\begin{table}
\centering
\caption{Sample backtesting scores for model CWGAN dataset USDYC1 (Libor curve)}
\label{tab:btUSDYC1 (Libor curve)}
\begin{tabular}{cccccllllllll}
\toprule
 J &  DAYS &      START &        END & TENOR &  KSpval &  DIFF &  BR05 &  BR95 \\
\midrule
 1 &     1 & 2010-01-15 & 2012-01-17 &    3m &   0.000 & 0.428 & 0.022 & 0.042 \\
 2 &     1 & 2012-01-18 & 2014-01-10 &    3m &   0.000 & 0.909 & 0.004 & 0.006 \\
 3 &     1 & 2014-01-13 & 2016-01-08 &    3m &   0.000 & 0.829 & 0.002 & 0.016 \\
 1 &    10 & 2010-01-15 & 2012-01-17 &    3m &   0.000 & 0.405 & 0.028 & 0.038 \\
 2 &    10 & 2012-01-18 & 2014-01-10 &    3m &   0.000 & 0.934 & 0.000 & 0.000 \\
 3 &    10 & 2014-01-13 & 2016-01-08 &    3m &   0.000 & 0.925 & 0.000 & 0.028 \\
\bottomrule
\end{tabular}
\end{table}

\subsubsection{Combination of KPIs \label{subsec:Combination-of-KPIs}}

We evaluate the performance of each generative model using multiple
KPIs discussed in previous sections. Each KPI is calculated across
multiple tenors and scores. To reduce the number of
KPIs and make it easier for comparison, we combine KPIs across tenors and across score categories (or subscores).

Ideally, the KPIs can be combined through a statistically sound multivariate distribution.  Given the diverse variety of KPIs considered, it is not intuitive how to do it.  In this study, only KPIs in probability space (e.g. p-values from KS tests, p-values from Fisher's Z-transformation and VaR breach rates) are combined through mean and median across tenors and KPI categories.  It is somewhat ad hoc and arbitrary.  We will continue to explore better ways to combine the KPIs.

\textbf{Backtest}\\
u-value analysis and backtest is performed across multiple tenors and sub-periods
for 1-day and 10-day returns. Eight backtest scores are calculated. See \ref{subsec:backtest} for details.
The summary backtest score is calculated with the following steps,
\begin{enumerate}
\item Calculate eight backtest scores for each tenor and each 2-year sub-period (as well as whole period): KSpval, DIFF, BR025, BR05, BR10, BR90, BR95, BR975 (BRp is the absolute difference between the actual and theoretical u-value breach rate at probability level $p=0.025$,0.05,0.10,0.90,0.95,0.975).  The sub-period analysis is intended to address the case in which a model performs badly in sub-periods (e.g. over- or under-estimate risk), but the problems in sub-periods when combined in the whole period and lead to good performance.
\item Calculate the median of each score across sub-periods.
\item Calculate the average between 1-day and 10-day scores.
\item Calculate the median across tenors.
\item Sum up breach rates at all six probabilities, $BR=BR025+BR05+BR10+BR90+BR95+BR975$.
\item Calculate the sum, $BT = [BR+(1-KSpval)]/2$.  $1-KSpval$ is used in the sum, since a smaller value indicates a better model, which is consistent with the breach rate $BR$.
\end{enumerate}
DIFF is excluded in the calculation of the summary backtesting score $BT$, since DIFF itself is not a probability.

\textbf{Distribution Distance Scores}\\
Distribution distance (EMD, DY, KS, 1-KSpval) of 1-day returns using sample moments (mean and standard deviation) is calculated.  $KSpval$ is the pval from KS test.  $1-KSpval$ is used since a small value of this metric indicates a better model.  The summary score (DistributionDistance) is using only $1-KSpval$, and is the average of scores across sample statistics (mean and standard deviation) and tenors.

\textbf{Series Distance}\\
Series distance is calculated as $1-KSpval$, where $KSpval$ is the p-value from two sample KS test.  Summary series distance is calculated as the average across tenors and between 1-day and 10-day scores.

\textbf{ACF Score}\\
To compare the autocorrelations in real and synthetic data, we calculate p-value (``ACF score'') from Fisher's Z-transformation \eqref{eq:fisherz}, for $f(x)=x$ and $f(x)=x^{2}$ separately.
\footnote{ACF for $f(x)=|x|$ and leverage score are calculated but not used as a KPI.}
. ACF score is further aggregated
first by averaging ACF scores for $x$ and $x^{2}$, and then by averaging across all tenors.  In model comparison, $1-\mbox{ACF}$ is reported and used, since a smaller value of $1-\mbox{ACF}$ indicates a better model, consistent with other KPIs.

\textbf{Composite score}\\
Multiple scores from different categories are combined further as follows.  (1) The average of
Distribution Distance (1-KS-pval from KS test for sample moments) and Series Distance ((1-KS-pval) from KS test for returns) is called Distribution (DIST) score, (2) The sum of Distribution (DIST) score, ACF score, and backtesting score (BT) is called Composite score (COMP).  For all KPIs including Composite score, a smaller value indicates a better model.  For an example of the combined score and model ranking, see Table~\ref{tab:combSIMGARCH2N30Y}.

Model rankings vary depending on the KPIs used and the way they are combined to form Composite score.

\subsection{Results using simulated data\label{sec:simresults}}

Daily data for 30 years are simulated following GARCH model and CIR square root model.  Returns (rate changes) are simulated from GARCH model, and then accumulated over time to get level of rates.  Rate levels are simulated from CIR model and returns are calculated as the daily change.  These simulation models are called data generating process (DGP).  See Section~\ref{subsec:simulated data}. 

Each path from the simulation coves 30 years of daily data.  All applicable models are run for each path.  Since Nelson-Siegel is a term structure model for multiple tenors, it is not included in the model test for simulated dataset with bivariate time series.  For each simulation model, five paths are generated and used in model testing. The 13 models are applied to the simulated levels or returns as appropriate.   The model testing results are presented in this section.

There are several sets of tables for each DGP: (1) KPI and model ranking for one simulation path, Table~\ref{tab:combSIMGARCH2N30Yshort}. (2) Model ranking across all five simulation paths, Table~\ref{tab:combrankGARCH-normalshort}. The ranking is done for each path individually.  (3) The average KPI subscore (distribution, autocorrelation and backtest) across five datasets, Table~\ref{tab:combsubscoreGARCH-normalshort}.  This shows, on average across five paths, how each model performs in each KPI category (or subscore). (4) The ranking by average subscore,  Table~\ref{tab:combranksubscoreGARCH-normalshort}.

More detailed results are available in an extended appendix (available upon request).

\subsubsection{Results using simulated GARCH-normal Dataset}
Each time series follows AR(1) model for its conditional mean and follows GARCH(1,1) model for its conditional variance.  The conditional distribution follows normal distribution.  See Section~\ref{subsec:simulated data} for details.

The models are tested using five simulated paths (five datasets) from AR(1)+GARCH with normal distribution for returns.  Results for one path
\footnote{In the table caption, SIM\_GARCH2N30Y means Path 2 with 30 year data simulated from GARCH-normal model.}
 are presented in Table~\ref{tab:combSIMGARCH2N30Y}.  This is a large table covering 13 models.  Anticipating results later in the paper, we find that several models always perform badly for most datasets.  Therefore, to save space, we present results for only a representative subset of seven models.  These seven models include: (1) Plain historical simulation (PHS), which is simpler and performs better than FHS, (2) GARCH with conditional t distribution (GARCHt-RET), which performs better than GARCH with conditional normal distribution, (3) AR-RET, which is an autoregressive model without conditional heteroscedasticity, and (4) four NN models: CGAN-FC, CWGAN, SIG and VAE.  The results for the complete list of models are available in an extended appendix (available on request).  The shorter version of Table~\ref{tab:combSIMGARCH2N30Y} is presented in Table~\ref{tab:combSIMGARCH2N30Yshort}\\

\begin{table}
\centering
\caption{Model comparison for dataset SIM\_GARCH2N30Y}
\label{tab:combSIMGARCH2N30Y}
\begin{tabular}{cclrrrr}
\toprule
 Rank & Cat &      Model &  DIST &   ACF &    BT &  Composite \\
\midrule
    1 &  HS &        PHS & 0.129 & 0.194 & 0.375 &      0.699 \\
    2 &  HS &        FHS & 0.487 & 0.761 & 0.303 &      1.551 \\
    3 &  NN &    CGAN-FC & 0.609 & 0.817 & 0.542 &      1.968 \\
    4 &  NN &      CWGAN & 0.804 & 0.650 & 0.516 &      1.970 \\
    5 &  PM & GARCHt-RET & 0.947 & 0.670 & 0.496 &      2.113 \\
    6 &  NN &        VAE & 0.746 & 0.909 & 0.464 &      2.120 \\
    7 &  PM &  GARCH-RET & 0.904 & 0.689 & 0.535 &      2.128 \\
    8 &  PM &     AR-RET & 0.998 & 0.637 & 0.554 &      2.188 \\
    9 &  NN &  DIFFUSION & 0.975 & 0.865 & 0.615 &      2.455 \\
   10 &  NN &        SIG & 0.920 & 1.000 & 0.564 &      2.484 \\
   11 &  PM &         AR & 0.999 & 1.000 & 0.600 &      2.599 \\
   12 &  NN &  CGAN-LSTM & 0.991 & 0.900 & 0.806 &      2.697 \\
   13 &  PM &      GARCH & 1.000 & 1.000 & 0.913 &      2.913 \\
\bottomrule
\end{tabular}
\begin{tablenotes}\item[*] KPI/score: the smaller the value, the better the model.  KPIs are calclulated for one dataset. DIST: distribution score, ACF: autocorrelation score, BT: backtesting score.\end{tablenotes}\end{table}

\begin{table}
\centering
\caption{Model comparison for dataset SIM\_GARCH2N30Y}
\label{tab:combSIMGARCH2N30Yshort}
\begin{tabular}{cclrrrr}
\toprule
 Rank & Cat &      Model &  DIST &   ACF &    BT &  Composite \\
\midrule
    1 &  HS &        PHS & 0.129 & 0.194 & 0.375 &      0.699 \\
    2 &  NN &    CGAN-FC & 0.609 & 0.817 & 0.542 &      1.968 \\
    3 &  NN &      CWGAN & 0.804 & 0.650 & 0.516 &      1.970 \\
    4 &  PM & GARCHt-RET & 0.947 & 0.670 & 0.496 &      2.113 \\
    5 &  NN &        VAE & 0.746 & 0.909 & 0.464 &      2.120 \\
    6 &  PM &     AR-RET & 0.998 & 0.637 & 0.554 &      2.188 \\
    7 &  NN &        SIG & 0.920 & 1.000 & 0.564 &      2.484 \\
\bottomrule
\end{tabular}
\begin{tablenotes}\item[*] KPI/score: the smaller the value, the better the model.  KPIs are calclulated for one dataset. DIST: distribution score, ACF: autocorrelation score, BT: backtesting score.\end{tablenotes}\end{table}

KPIs used in the table are described in Section~\ref{subsec:evaluation metrics}.  KPIs measure the distance (difference) between the synthetic data and real data.  The smaller the KPI value, the better.  Columns in the tables are:
\begin{enumerate}
\item Rank: model rank by composite score (the smaller, the better).
\item Cat: model category with historical simulation (HS), parametric model (PM) and neural network (NN)
\item Model: short model name
\item DIST: distribution score
\item ACF: autocorrelation score
\item BT: backtest score.
\item Composite: composite score (DIST+ACF+BT)
\end{enumerate}
The summary comparison across all five paths in Table~\ref{tab:combrankGARCH-normalshort}.  Tables for other paths are available in the extended appendix.  Model ranking is based on Composite score.  Table~\ref{tab:combrankGARCH-normalshort} shows the ranking of models for each of the five simulated datasets (paths) and the (arithmetic) average (AVG) ranking.

\begin{table}
\centering
\caption{Model ranking for dataset GARCH-normal}
\label{tab:combrankGARCH-normalshort}
\begin{tabular}{cclcccccc}
\toprule
 Rank & Cat &      Model &  DS1 &  DS2 &  DS3 &  DS4 &  DS5 &  AVG \\
\midrule
    1 &  HS &        PHS &    1 &    1 &    1 &    1 &    1 &  1.0 \\
    2 &  PM & GARCHt-RET &    4 &    2 &    2 &    3 &    3 &  2.8 \\
    3 &  NN &    CGAN-FC &    2 &    6 &    6 &    2 &    4 &  4.0 \\
    4 &  PM &     AR-RET &    6 &    4 &    3 &    5 &    2 &  4.0 \\
    5 &  NN &      CWGAN &    3 &    5 &    5 &    4 &    5 &  4.4 \\
    6 &  NN &        VAE &    5 &    3 &    4 &    6 &    6 &  4.8 \\
    7 &  NN &        SIG &    7 &    7 &    7 &    7 &    7 &  7.0 \\
\bottomrule
\end{tabular}
\end{table}

The results across the five simulated datasets have some variation and also consistency: Historical simulation (HS) methods are always ranked as the top model.  HS is a commonly used VaR model among commercial banks.  GARCHt-RET is ranked as the 2nd best model overall, but there is some variation in rank across datasets (e.g. \#4 for dataset 1, and \#2 for datasets 2 and 3).

Since the data are simulated using GARCH model, GARCH on returns (or GARCHt-RET) is the correctly specified model
\footnote{Strictly speaking GARCH-normal (GARCH-RET) is the correctly specified model.  GARCH-t distribution (GARCHt-RET) is a general case of GARCH-normal and can be loosely regarded as correctly specified.}
, which may which may explain its good performance. Even though GARCHt-RET is the correctly specified model, it does not beat PHS performance because of the noise from the (GARCH) parameter estimation.

AR on returns (AR-NET) shares the AR part with the GARCH model, but it does not have conditional heteroscedasticity.  It is ranked the 4th model in performance.

CGAN-FC is ranked the 3rd model in performance, better than the AR-RET model, and the other three NN models.

In the table above, the ranking is based on Composite score for each individual dataset.  In Table~\ref{tab:combsubscoreGARCH-normalshort}, we present the average subscore in each performance (or KPI) category as well as the composite score across five datasets. The model ranking based on subscores is presented in Table~\ref{tab:combranksubscoreGARCH-normalshort}, which has model \textbf{ranking by average subscores} across datasets rather than \textbf{average ranking} across datasets as in Table~\ref{tab:combrankGARCH-normalshort} . By looking at ranking based on subscores, we can find whether a model's overall rank is consistent across subscores, or whether good performance in one category is offset by bad performance in another category.  Since different models capture the relationships in the data differently, we expect some variation of ranking by subscores
\footnote{e.g. some models may capture correlation better, while others may capture distribution better.}
, otherwise columns in Table~\ref{tab:combrankGARCH-normalshort} would always show the same rankings.

\begin{table}
\centering
\caption{Model average sub-scores for dataset GARCH-normal}
\label{tab:combsubscoreGARCH-normalshort}
\begin{tabular}{lllcccc}
\toprule
 No. & Cat &      Model &  DIST &   ACF &    BT &  Composite \\
\midrule
   1 &  HS &        PHS & 0.360 & 0.341 & 0.416 &      1.116 \\
   2 &  PM & GARCHt-RET & 0.765 & 0.784 & 0.489 &      2.038 \\
   3 &  NN &    CGAN-FC & 0.714 & 0.878 & 0.579 &      2.171 \\
   4 &  NN &      CWGAN & 0.843 & 0.793 & 0.539 &      2.176 \\
   5 &  PM &     AR-RET & 0.999 & 0.622 & 0.564 &      2.185 \\
   6 &  NN &        VAE & 0.792 & 0.940 & 0.498 &      2.230 \\
   7 &  NN &        SIG & 0.894 & 1.000 & 0.563 &      2.457 \\
\bottomrule
\end{tabular}
\begin{tablenotes}\item[*] KPI/score: the smaller the value, the better the model.  These scores/subscores are the average scores across five datasets.  DIST: distribution score, ACF: autocorrelation score, BT: backtesting score.\end{tablenotes}\end{table}

\begin{table}
\centering
\caption{Model ranking by average sub-score for dataset GARCH-normal}
\label{tab:combranksubscoreGARCH-normalshort}
\begin{tabular}{llcccc}
\toprule
Cat &      Model &  Rank\_DIST &  Rank\_ACF &  Rank\_BT &  Rank \\
\midrule
 HS &        PHS &          1 &         1 &        1 &     1 \\
 PM & GARCHt-RET &          3 &         3 &        2 &     2 \\
 NN &    CGAN-FC &          2 &         5 &        7 &     3 \\
 NN &      CWGAN &          5 &         4 &        4 &     4 \\
 PM &     AR-RET &          7 &         2 &        6 &     5 \\
 NN &        VAE &          4 &         6 &        3 &     6 \\
 NN &        SIG &          6 &         7 &        5 &     7 \\
\bottomrule
\end{tabular}
\begin{tablenotes}\item[*] Model rank by subscores.  The lower the rank, the better the model.  Rank\_DIST: rank by distribution (DIST), Rank\_ACF: rank by ACF score, Rank\_BT: rank by backtesting score (BT), Rank: rank by composite score.\end{tablenotes}\end{table}

Table~\ref{tab:combsubscoreGARCH-normalshort} shows that (1) DIST and ACF scores have similar ranges (DIST: 0.360-0.999, ACF: 0.341-1.0), while BT has a smaller range (0.416-0.579).  This indicates that the models do not have much differentiation in backtest (BT) performance; (2) PHS performs especially well in Distribution and ACF, which guarantees its final ranking of being the best model (Table~\ref{tab:combranksubscoreGARCH-normalshort}).  (3) CGAN-FC is ranked quite well by Distribution (\#2) but ranked \#7 by backtesting (BT). (4) The performance of CWGAN model follows right after the CGAN-FC model. (5) AR-RET is ranked very badly by Distribution (\#7), and is ranked very well by ACF (\#2).  AR-RET tracks ACF well by the nature of AR model.
\subsubsection{Results using simulated GARCH-t(5) Dataset}
The models are run for five paths simulated from AR(1)+GARCH and t distribution with 5 degrees of freedom.  Compared with GARCH with normal distribution, GARCH with t(5) distribution has fat tails.  Similar to the analysis for GARCH-normal datasets, results for one path are presented in Tables~\ref{tab:combSIMGARCH2T5CR30Yshort}, with summary results across all five paths in Table~\ref{tab:combrankGARCH-t(5)short}, average subscores and corresponding model ranking in Tables~\ref{tab:combsubscoreGARCH-t(5)short} and \ref{tab:combranksubscoreGARCH-t(5)short}. 

The results are similar to those for GARCH-normal simulated data in the last subsection.
% GARCH t(3)
\begin{table}
\centering
\caption{Model comparison for dataset SIM\_GARCH2T5CR30Y}
\label{tab:combSIMGARCH2T5CR30Yshort}
\begin{tabular}{cclrrrr}
\toprule
 Rank & Cat &      Model &  DIST &   ACF &    BT &  Composite \\
\midrule
    1 &  HS &        PHS & 0.364 & 0.200 & 0.429 &      0.993 \\
    2 &  PM & GARCHt-RET & 0.830 & 0.795 & 0.445 &      2.071 \\
    3 &  NN &      CWGAN & 0.819 & 0.819 & 0.502 &      2.141 \\
    4 &  NN &    CGAN-FC & 0.853 & 0.776 & 0.593 &      2.222 \\
    5 &  PM &     AR-RET & 1.000 & 0.717 & 0.568 &      2.285 \\
    6 &  NN &        VAE & 0.886 & 0.997 & 0.500 &      2.384 \\
    7 &  NN &        SIG & 0.994 & 1.000 & 0.572 &      2.567 \\
\bottomrule
\end{tabular}
\begin{tablenotes}\item[*] KPI/score: the smaller the value, the better the model.  KPIs are calclulated for one dataset. DIST: distribution score, ACF: autocorrelation score, BT: backtesting score.\end{tablenotes}\end{table}

\begin{table}
\centering
\caption{Model ranking for dataset GARCH-t(5)}
\label{tab:combrankGARCH-t(5)short}
\begin{tabular}{cclcccccc}
\toprule
 Rank & Cat &      Model &  DS1 &  DS2 &  DS3 &  DS4 &  DS5 &  AVG \\
\midrule
    1 &  HS &        PHS &    1 &    1 &    1 &    1 &    1 &  1.0 \\
    2 &  PM & GARCHt-RET &    2 &    2 &    2 &    2 &    2 &  2.0 \\
    3 &  NN &    CGAN-FC &    4 &    4 &    4 &    3 &    3 &  3.6 \\
    4 &  NN &      CWGAN &    3 &    5 &    3 &    4 &    4 &  3.8 \\
    5 &  PM &     AR-RET &    5 &    3 &    6 &    7 &    5 &  5.2 \\
    6 &  NN &        VAE &    6 &    6 &    5 &    5 &    6 &  5.6 \\
    7 &  NN &        SIG &    7 &    7 &    7 &    6 &    7 &  6.8 \\
\bottomrule
\end{tabular}
\end{table}

\begin{table}
\centering
\caption{Model average sub-scores for dataset GARCH-t(5)}
\label{tab:combsubscoreGARCH-t(5)short}
\begin{tabular}{lllcccc}
\toprule
 No. & Cat &      Model &  DIST &   ACF &    BT &  Composite \\
\midrule
   1 &  HS &        PHS & 0.373 & 0.336 & 0.421 &      1.130 \\
   2 &  PM & GARCHt-RET & 0.743 & 0.797 & 0.415 &      1.955 \\
   3 &  NN &    CGAN-FC & 0.784 & 0.786 & 0.590 &      2.160 \\
   4 &  NN &      CWGAN & 0.909 & 0.796 & 0.508 &      2.213 \\
   5 &  PM &     AR-RET & 1.000 & 0.745 & 0.565 &      2.310 \\
   6 &  NN &        VAE & 0.907 & 0.992 & 0.510 &      2.410 \\
   7 &  NN &        SIG & 0.983 & 0.978 & 0.566 &      2.526 \\
\bottomrule
\end{tabular}
\begin{tablenotes}\item[*] KPI/score: the smaller the value, the better the model.  These scores/subscores are the average scores across five datasets.  DIST: distribution score, ACF: autocorrelation score, BT: backtesting score.\end{tablenotes}\end{table}

\begin{table}
\centering
\caption{Model ranking by average sub-score for dataset GARCH-t(5)}
\label{tab:combranksubscoreGARCH-t(5)short}
\begin{tabular}{llcccc}
\toprule
Cat &      Model &  Rank\_DIST &  Rank\_ACF &  Rank\_BT &  Rank \\
\midrule
 HS &        PHS &          1 &         1 &        2 &     1 \\
 PM & GARCHt-RET &          2 &         5 &        1 &     2 \\
 NN &    CGAN-FC &          3 &         3 &        7 &     3 \\
 NN &      CWGAN &          5 &         4 &        3 &     4 \\
 PM &     AR-RET &          7 &         2 &        5 &     5 \\
 NN &        VAE &          4 &         7 &        4 &     6 \\
 NN &        SIG &          6 &         6 &        6 &     7 \\
\bottomrule
\end{tabular}
\begin{tablenotes}\item[*] Model rank by subscores.  The lower the rank, the better the model.  Rank\_DIST: rank by distribution (DIST), Rank\_ACF: rank by ACF score, Rank\_BT: rank by backtesting score (BT), Rank: rank by composite score.\end{tablenotes}\end{table}

\subsubsection{Results using simulated GARCH-t(3) Dataset}
The models are run for five paths simulated from AR(1)+GARCH and t distribution with 3 degrees of freedom.  Compared with GARCH-normal and GARCH-t(5), GARCH with t(3) distribution has much fatter tails.  Similar to the analysis for GARCH-normal datasets, results for one path are presented in Tables~\ref{tab:combSIMGARCH2T3CR30Yshort}, with summary results across all five paths in Table~\ref{tab:combrankGARCH-t(3)short}, average subscores and model ranking in Tables~\ref{tab:combsubscoreGARCH-t(3)short} and \ref{tab:combranksubscoreGARCH-t(3)short}. 

The results are similar to those for GARCH-normal and GARCH-t(5) simulated data in the last subsections.

% GARCH t(3)
\begin{table}
\centering
\caption{Model comparison for dataset SIM\_GARCH2T3CR30Y}
\label{tab:combSIMGARCH2T3CR30Yshort}
\begin{tabular}{cclrrrr}
\toprule
 Rank & Cat &      Model &  DIST &   ACF &    BT &  Composite \\
\midrule
    1 &  HS &        PHS & 0.340 & 0.414 & 0.311 &      1.064 \\
    2 &  PM & GARCHt-RET & 0.845 & 0.741 & 0.455 &      2.042 \\
    3 &  NN &      CWGAN & 0.975 & 0.808 & 0.556 &      2.340 \\
    4 &  NN &    CGAN-FC & 0.946 & 0.900 & 0.568 &      2.414 \\
    5 &  PM &     AR-RET & 1.000 & 0.849 & 0.569 &      2.418 \\
    6 &  NN &        VAE & 0.999 & 0.999 & 0.482 &      2.479 \\
    7 &  NN &        SIG & 0.990 & 1.000 & 0.569 &      2.559 \\
\bottomrule
\end{tabular}
\begin{tablenotes}\item[*] KPI/score: the smaller the value, the better the model.  KPIs are calclulated for one dataset. DIST: distribution score, ACF: autocorrelation score, BT: backtesting score.\end{tablenotes}\end{table}

\begin{table}
\centering
\caption{Model ranking for dataset GARCH-t(3)}
\label{tab:combrankGARCH-t(3)short}
\begin{tabular}{cclcccccc}
\toprule
 Rank & Cat &      Model &  DS1 &  DS2 &  DS3 &  DS4 &  DS5 &  AVG \\
\midrule
    1 &  HS &        PHS &    1 &    1 &    1 &    1 &    1 &  1.0 \\
    2 &  PM & GARCHt-RET &    2 &    2 &    2 &    2 &    2 &  2.0 \\
    3 &  NN &    CGAN-FC &    4 &    3 &    3 &    4 &    3 &  3.4 \\
    4 &  NN &      CWGAN &    3 &    5 &    6 &    3 &    6 &  4.6 \\
    5 &  PM &     AR-RET &    5 &    4 &    5 &    5 &    4 &  4.6 \\
    6 &  NN &        VAE &    6 &    6 &    4 &    6 &    5 &  5.4 \\
    7 &  NN &        SIG &    7 &    7 &    7 &    7 &    7 &  7.0 \\
\bottomrule
\end{tabular}
\end{table}

We conjecture that the much fat tail in the t(3) distribution may have caused disruption to model training and performance.

\begin{table}
\centering
\caption{Model average sub-scores for dataset GARCH-t(3)}
\label{tab:combsubscoreGARCH-t(3)short}
\begin{tabular}{lllcccc}
\toprule
 No. & Cat &      Model &  DIST &   ACF &    BT &  Composite \\
\midrule
   1 &  HS &        PHS & 0.432 & 0.394 & 0.360 &      1.186 \\
   2 &  PM & GARCHt-RET & 0.754 & 0.781 & 0.456 &      1.990 \\
   3 &  NN &    CGAN-FC & 0.774 & 0.806 & 0.564 &      2.144 \\
   4 &  NN &      CWGAN & 0.870 & 0.866 & 0.545 &      2.282 \\
   5 &  PM &     AR-RET & 1.000 & 0.743 & 0.568 &      2.311 \\
   6 &  NN &        VAE & 0.873 & 0.943 & 0.496 &      2.312 \\
   7 &  NN &        SIG & 0.966 & 0.976 & 0.583 &      2.524 \\
\bottomrule
\end{tabular}
\begin{tablenotes}\item[*] KPI/score: the smaller the value, the better the model.  These scores/subscores are the average scores across five datasets.  DIST: distribution score, ACF: autocorrelation score, BT: backtesting score.\end{tablenotes}\end{table}

\begin{table}
\centering
\caption{Model ranking by average sub-score for dataset GARCH-t(3)}
\label{tab:combranksubscoreGARCH-t(3)short}
\begin{tabular}{llcccc}
\toprule
Cat &      Model &  Rank\_DIST &  Rank\_ACF &  Rank\_BT &  Rank \\
\midrule
 HS &        PHS &          1 &         1 &        1 &     1 \\
 PM & GARCHt-RET &          2 &         3 &        2 &     2 \\
 NN &    CGAN-FC &          3 &         4 &        5 &     3 \\
 NN &      CWGAN &          4 &         5 &        4 &     4 \\
 PM &     AR-RET &          7 &         2 &        6 &     5 \\
 NN &        VAE &          5 &         6 &        3 &     6 \\
 NN &        SIG &          6 &         7 &        7 &     7 \\
\bottomrule
\end{tabular}
\begin{tablenotes}\item[*] Model rank by subscores.  The lower the rank, the better the model.  Rank\_DIST: rank by distribution (DIST), Rank\_ACF: rank by ACF score, Rank\_BT: rank by backtesting score (BT), Rank: rank by composite score.\end{tablenotes}\end{table}

\subsubsection{Results using simulated CIR dataset}
The models are run for five simulated paths from CIR model.  Similar set of tables are presented: the results for one path in Tables~\ref{tab:combSIMCIRS1bishort}, summary of rankings across all five paths in Table~\ref{tab:combrankCIRshort}, the average subscores in Tables~\ref{tab:combsubscoreCIRshort} and model ranking based on average subscores in Table~\ref{tab:combranksubscoreCIRshort}.\\

There are some similarities and differences from the results for GARCH datasets. (1) HS model is the top performers consistently across the five paths (Table~\ref{tab:combrankCIRshort}) and across subscore categories (Table~\ref{tab:combranksubscoreCIRshort}). (2) GARCHt-NET is the 2nd best model.  (3) AR-NET is ranked \#3rd (botb Tables~\ref{tab:combrankCIRshort} and \ref{tab:combranksubscoreCIRshort}) and performs better than NN models (only slightly better than CWGAN with scores 2.115 and 2.181 in Tables~\ref{tab:combsubscoreGARCH-t(3)short}).
\begin{table}
\centering
\caption{Model comparison for dataset SIM\_CIRS1bi}
\label{tab:combSIMCIRS1bishort}
\begin{tabular}{cclrrrr}
\toprule
 Rank & Cat &      Model &  DIST &   ACF &    BT &  Composite \\
\midrule
    1 &  HS &        PHS & 0.071 & 0.411 & 0.456 &      0.938 \\
    2 &  NN &      CWGAN & 0.724 & 0.653 & 0.556 &      1.934 \\
    3 &  PM & GARCHt-RET & 0.709 & 0.761 & 0.469 &      1.939 \\
    4 &  PM &     AR-RET & 0.891 & 0.528 & 0.564 &      1.983 \\
    5 &  NN &    CGAN-FC & 0.996 & 0.407 & 0.606 &      2.009 \\
    6 &  NN &        SIG & 0.928 & 0.747 & 0.566 &      2.241 \\
    7 &  NN &        VAE & 0.881 & 0.791 & 0.595 &      2.268 \\
\bottomrule
\end{tabular}
\begin{tablenotes}\item[*] KPI/score: the smaller the value, the better the model.  KPIs are calclulated for one dataset. DIST: distribution score, ACF: autocorrelation score, BT: backtesting score.\end{tablenotes}\end{table}

\begin{table}
\centering
\caption{Model ranking for dataset CIR}
\label{tab:combrankCIRshort}
\begin{tabular}{cclcccccc}
\toprule
 Rank & Cat &      Model &  DS1 &  DS2 &  DS3 &  DS4 &  DS5 &  AVG \\
\midrule
    1 &  HS &        PHS &    1 &    1 &    1 &    1 &    2 &  1.2 \\
    2 &  PM & GARCHt-RET &    3 &    2 &    2 &    2 &    1 &  2.0 \\
    3 &  PM &     AR-RET &    4 &    3 &    3 &    5 &    3 &  3.6 \\
    4 &  NN &      CWGAN &    2 &    4 &    4 &    6 &    6 &  4.4 \\
    5 &  NN &    CGAN-FC &    5 &    6 &    7 &    3 &    5 &  5.2 \\
    6 &  NN &        VAE &    7 &    7 &    5 &    4 &    4 &  5.4 \\
    7 &  NN &        SIG &    6 &    5 &    6 &    7 &    7 &  6.2 \\
\bottomrule
\end{tabular}
\end{table}

\begin{table}
\centering
\caption{Model average sub-scores for dataset CIR}
\label{tab:combsubscoreCIRshort}
\begin{tabular}{lllcccc}
\toprule
 No. & Cat &      Model &  DIST &   ACF &    BT &  Composite \\
\midrule
   1 &  HS &        PHS & 0.382 & 0.378 & 0.489 &      1.249 \\
   2 &  PM & GARCHt-RET & 0.639 & 0.615 & 0.481 &      1.735 \\
   3 &  PM &     AR-RET & 0.969 & 0.571 & 0.575 &      2.115 \\
   4 &  NN &      CWGAN & 0.900 & 0.709 & 0.572 &      2.181 \\
   5 &  NN &        VAE & 0.908 & 0.758 & 0.579 &      2.245 \\
   6 &  NN &    CGAN-FC & 0.946 & 0.701 & 0.616 &      2.263 \\
   7 &  NN &        SIG & 0.981 & 0.796 & 0.585 &      2.362 \\
\bottomrule
\end{tabular}
\begin{tablenotes}\item[*] KPI/score: the smaller the value, the better the model.  These scores/subscores are the average scores across five datasets.  DIST: distribution score, ACF: autocorrelation score, BT: backtesting score.\end{tablenotes}\end{table}

\begin{table}
\centering
\caption{Model ranking by average sub-score for dataset CIR}
\label{tab:combranksubscoreCIRshort}
\begin{tabular}{llcccc}
\toprule
Cat &      Model &  Rank\_DIST &  Rank\_ACF &  Rank\_BT &  Rank \\
\midrule
 HS &        PHS &          1 &         1 &        2 &     1 \\
 PM & GARCHt-RET &          2 &         3 &        1 &     2 \\
 PM &     AR-RET &          6 &         2 &        4 &     3 \\
 NN &      CWGAN &          3 &         5 &        3 &     4 \\
 NN &        VAE &          4 &         6 &        5 &     5 \\
 NN &    CGAN-FC &          5 &         4 &        7 &     6 \\
 NN &        SIG &          7 &         7 &        6 &     7 \\
\bottomrule
\end{tabular}
\begin{tablenotes}\item[*] Model rank by subscores.  The lower the rank, the better the model.  Rank\_DIST: rank by distribution (DIST), Rank\_ACF: rank by ACF score, Rank\_BT: rank by backtesting score (BT), Rank: rank by composite score.\end{tablenotes}\end{table}

\subsubsection{Summary}
Results for simulated GARCH (with Normal and t distribution with 3 and 5 degrees of freedom) and CIR datasets can be summarized as follows.  

Plain historical simulation (PHS) model is the top performer for all 4 DGPs. 

GARCH with t distribution (GARCHt-NET) model is the 2nd best model.

CWGAN and CGAN-FC are the better performing NN models.

\subsection{Results using historical data\label{sec:hisresults}}
There are three sets of USD yield curve data, USDYC1 (Libor curve 2008-2022), USDYC2 (Par yield 2008-2023) and USDYC3 (Par yield 2000-2023).  See details in Sections~\ref{subsec:usd yield curve}. This section shows a summary of the results.  More detailed results are available in an extended appendix (available upon request).

KPIs measure the distance (difference) between the synthetic data and real data.  The smaller the KPI value, the better.  See Section~ref{subsec:evaluation metrics} for details.

\subsubsection{Results of model runs}
There are three sets of tables for results.  (1) Results for each dataset, Tables~\ref{tab:combUSDYC1prodshort} for USDYC1, Tables~\ref{tab:combUSDYC2prodshort} for USDYC2 and Table~\ref {tab:combUSDYC3prodshort} for USDYC3. (2) A summary of rankings across the three datasets is presented in Table~\ref{tab:combrankUSDYCshort}. (3)  The average subscores across the three datasets are presented in Table~\ref{tab:combsubscoreUSDYCshort} and ranking by subscores in Table~\ref{tab:combranksubscoreUSDYCshort}.

Observations are based mainly on Tables~\ref{tab:combrankUSDYCshort} and \ref{tab:combranksubscoreUSDYCshort}.

HS model performs the best, followed by GARCHt-RET model.  

CWGAN is the top performing NN model, followed by VAE and SIG models.

The simple AR-RET model has very decent performance and is ranked only after the best NN model (CWGAN). 

The difference between AR-RET and GARCHt-RET models is the GARCH dynamics of conditional variance for the errors of AR(1) model. The results show that modeling the GARCH effect (in returns) is important.

Recall that for simulated data, the simple CGAN-FC has comparable performance to CWGAN.  But for the USD yield curve, CGAN-FC has the worst performance.\\

\begin{table}
\centering
\caption{Model comparison for dataset USDYC1 (Libor curve)}
\label{tab:combUSDYC1prodshort}
\begin{tabular}{cclrrrr}
\toprule
 Rank & Cat &      Model &  DIST &   ACF &    BT &  Composite \\
\midrule
    1 &  HS &        PHS & 0.197 & 0.750 & 0.537 &      1.484 \\
    2 &  PM & GARCHt-RET & 0.734 & 0.789 & 0.545 &      2.068 \\
    3 &  NN &      CWGAN & 0.766 & 0.764 & 0.565 &      2.094 \\
    4 &  NN &        SIG & 0.807 & 0.876 & 0.551 &      2.234 \\
    5 &  PM &     AR-RET & 0.971 & 0.706 & 0.572 &      2.248 \\
    6 &  NN &        VAE & 0.997 & 0.931 & 0.556 &      2.484 \\
    7 &  NN &    CGAN-FC & 0.991 & 0.814 & 0.806 &      2.611 \\
\bottomrule
\end{tabular}
\begin{tablenotes}\item[*] KPI/score: the smaller the value, the better the model.  KPIs are calclulated for one dataset. DIST: distribution score, ACF: autocorrelation score, BT: backtesting score.\end{tablenotes}\end{table}

\begin{table}
\centering
\caption{Model comparison for dataset USDYC2 (Par curve 2008-2023)}
\label{tab:combUSDYC2prodshort}
\begin{tabular}{cclrrrr}
\toprule
 Rank & Cat &      Model &  DIST &   ACF &    BT &  Composite \\
\midrule
    1 &  HS &        PHS & 0.328 & 0.792 & 0.552 &      1.673 \\
    2 &  PM & GARCHt-RET & 0.837 & 0.824 & 0.538 &      2.199 \\
    3 &  NN &      CWGAN & 0.920 & 0.776 & 0.552 &      2.249 \\
    4 &  PM &     AR-RET & 0.958 & 0.829 & 0.581 &      2.368 \\
    5 &  NN &        VAE & 0.990 & 0.812 & 0.567 &      2.369 \\
    6 &  NN &        SIG & 0.956 & 0.876 & 0.573 &      2.406 \\
    7 &  NN &    CGAN-FC & 0.998 & 0.857 & 0.917 &      2.772 \\
\bottomrule
\end{tabular}
\begin{tablenotes}\item[*] KPI/score: the smaller the value, the better the model.  KPIs are calclulated for one dataset. DIST: distribution score, ACF: autocorrelation score, BT: backtesting score.\end{tablenotes}\end{table}

\begin{table}
\centering
\caption{Model comparison for dataset USDYC3 (Par curve 2000-2023)}
\label{tab:combUSDYC3prodshort}
\begin{tabular}{cclrrrr}
\toprule
 Rank & Cat &      Model &  DIST &   ACF &    BT &  Composite \\
\midrule
    1 &  HS &        PHS & 0.128 & 0.594 & 0.540 &      1.261 \\
    2 &  PM & GARCHt-RET & 0.831 & 0.844 & 0.542 &      2.217 \\
    3 &  NN &      CWGAN & 0.959 & 0.815 & 0.553 &      2.327 \\
    4 &  NN &        VAE & 0.998 & 0.817 & 0.586 &      2.401 \\
    5 &  PM &     AR-RET & 0.974 & 0.844 & 0.586 &      2.405 \\
    6 &  NN &        SIG & 0.962 & 0.945 & 0.562 &      2.469 \\
    7 &  NN &    CGAN-FC & 0.965 & 0.951 & 0.710 &      2.626 \\
\bottomrule
\end{tabular}
\begin{tablenotes}\item[*] KPI/score: the smaller the value, the better the model.  KPIs are calclulated for one dataset. DIST: distribution score, ACF: autocorrelation score, BT: backtesting score.\end{tablenotes}\end{table}

\begin{table}
\centering
\caption{Model ranking for dataset USDYC}
\label{tab:combrankUSDYCshort}
\begin{tabular}{cclcccc}
\toprule
 Rank & Cat &      Model &  USDYC1 &  USDYC2 &  USDYC3 &  AVG \\
\midrule
    1 &  HS &        PHS &       1 &       1 &       1 &  1.0 \\
    2 &  PM & GARCHt-RET &       2 &       2 &       2 &  2.0 \\
    3 &  NN &      CWGAN &       3 &       3 &       3 &  3.0 \\
    4 &  PM &     AR-RET &       5 &       4 &       5 &  4.7 \\
    5 &  NN &        VAE &       6 &       5 &       4 &  5.0 \\
    6 &  NN &        SIG &       4 &       6 &       6 &  5.3 \\
    7 &  NN &    CGAN-FC &       7 &       7 &       7 &  7.0 \\
\bottomrule
\end{tabular}
\end{table}

\begin{table}
\centering
\caption{Model average sub-scores for dataset USDYC}
\label{tab:combsubscoreUSDYCshort}
\begin{tabular}{lllcccc}
\toprule
 No. & Cat &      Model &  DIST &   ACF &    BT &  Composite \\
\midrule
   1 &  HS &        PHS & 0.218 & 0.712 & 0.543 &      1.473 \\
   2 &  PM & GARCHt-RET & 0.801 & 0.819 & 0.542 &      2.161 \\
   3 &  NN &      CWGAN & 0.882 & 0.785 & 0.557 &      2.223 \\
   4 &  PM &     AR-RET & 0.968 & 0.793 & 0.580 &      2.340 \\
   5 &  NN &        SIG & 0.908 & 0.899 & 0.562 &      2.370 \\
   6 &  NN &        VAE & 0.995 & 0.853 & 0.570 &      2.418 \\
   7 &  NN &    CGAN-FC & 0.984 & 0.874 & 0.811 &      2.669 \\
\bottomrule
\end{tabular}
\begin{tablenotes}\item[*] KPI/score: the smaller the value, the better the model.  These scores/subscores are the average scores across five datasets.  DIST: distribution score, ACF: autocorrelation score, BT: backtesting score.\end{tablenotes}\end{table}

\begin{table}
\centering
\caption{Model ranking by average sub-score for dataset USDYC}
\label{tab:combranksubscoreUSDYCshort}
\begin{tabular}{llcccc}
\toprule
Cat &      Model &  Rank\_DIST &  Rank\_ACF &  Rank\_BT &  Rank \\
\midrule
 HS &        PHS &          1 &         1 &        2 &     1 \\
 PM & GARCHt-RET &          2 &         4 &        1 &     2 \\
 NN &      CWGAN &          3 &         2 &        3 &     3 \\
 PM &     AR-RET &          5 &         3 &        6 &     4 \\
 NN &        SIG &          4 &         7 &        4 &     5 \\
 NN &        VAE &          7 &         5 &        5 &     6 \\
 NN &    CGAN-FC &          6 &         6 &        7 &     7 \\
\bottomrule
\end{tabular}
\begin{tablenotes}\item[*] Model rank by subscores.  The lower the rank, the better the model.  Rank\_DIST: rank by distribution (DIST), Rank\_ACF: rank by ACF score, Rank\_BT: rank by backtesting score (BT), Rank: rank by composite score.\end{tablenotes}\end{table}

Additonal model runs are made for USDYC1 dataset with 1-year condition in NN models.  The results in Appendix~\ref{sec:res1y} show that, with much more parameters, the NN models do not improve in performance.
\clearpage
\subsubsection{Additional results with random seeds}
Randomization is used in several steps in model training and testing.  For example, random selection is used in train-test split; it may also be used in model training for batch normalization and forming batches for some NN models; random numbers are used in generation and backtesting steps.

To assess the impact of random seeds on the models, additional model runs are made with different random seeds for USDYC1.  Summary of results across five random seeds are presented in Table~\ref{tab:combrankUSDYCrandshort}. Earlier results in Tables~\ref{tab:combUSDYC1prodshort} is just one set of results with a specific set of random seed.

There is some variation across the runs.  HS is the best model for all five random seeds.  GARCHt-RET is the 2nd best model for all but the 2nd seed.   The ranking of models for the top four models are consistent with the three USD yield curves in  Table~\ref{tab:combrankUSDYCshort} and \ref{tab:combranksubscoreUSDYCshort}.

Average subscores across random seeds are presented Tables~\ref{tab:combsubscoreUSDYCrandshort} with ranking results in Table~\ref{tab:combranksubscoreUSDYCrandshort}  The results are consistent with Table~\ref{tab:combrankUSDYCrandshort}.
\begin{table}
\centering
\caption{Model ranking for dataset USD\_YC with different random seeds}
\label{tab:combrankUSDYCrandshort}
\begin{tabular}{cllccccc}
\toprule
Cat &      Model &  rand1 &  rand2 &  rand3 &  rand4 &  rand5 &  AVG \\
\midrule
 HS &        PHS &      1 &      1 &      1 &      1 &      1 &  1.0 \\
 PM & GARCHt-RET &      2 &      4 &      2 &      2 &      2 &  2.4 \\
 NN &      CWGAN &      3 &      2 &      4 &      3 &      3 &  3.0 \\
 PM &     AR-RET &      5 &      3 &      5 &      4 &      4 &  4.2 \\
 NN &        SIG &      4 &      6 &      3 &      5 &      6 &  4.8 \\
 NN &        VAE &      6 &      5 &      6 &      6 &      5 &  5.6 \\
 NN &    CGAN-FC &      7 &      7 &      7 &      7 &      7 &  7.0 \\
\bottomrule
\end{tabular}
\end{table}

\begin{table}
\centering
\caption{Model average sub-scores for dataset USDYC (rand)}
\label{tab:combsubscoreUSDYCrandshort}
\begin{tabular}{lllcccc}
\toprule
 No. & Cat &      Model &  DIST &   ACF &    BT &  Composite \\
\midrule
   1 &  HS &        PHS & 0.304 & 0.733 & 0.537 &      1.574 \\
   2 &  PM & GARCHt-RET & 0.749 & 0.792 & 0.544 &      2.085 \\
   3 &  NN &      CWGAN & 0.859 & 0.780 & 0.551 &      2.189 \\
   4 &  PM &     AR-RET & 0.968 & 0.718 & 0.570 &      2.256 \\
   5 &  NN &        SIG & 0.916 & 0.858 & 0.561 &      2.335 \\
   6 &  NN &        VAE & 0.987 & 0.898 & 0.564 &      2.449 \\
   7 &  NN &    CGAN-FC & 0.949 & 0.785 & 0.830 &      2.564 \\
\bottomrule
\end{tabular}
\begin{tablenotes}\item[*] KPI/score: the smaller the value, the better the model.  These scores/subscores are the average scores across five model runs with different random seeds.  DIST: distribution score, ACF: autocorrelation score, BT: backtesting score.\end{tablenotes}\end{table}

\begin{table}
\centering
\caption{Model ranking by average sub-score for dataset USDYC (rand)}
\label{tab:combranksubscoreUSDYCrandshort}
\begin{tabular}{llcccc}
\toprule
Cat &      Model &  Rank\_DIST &  Rank\_ACF &  Rank\_BT &  Rank \\
\midrule
 HS &        PHS &          1 &         2 &        1 &     1 \\
 PM & GARCHt-RET &          2 &         5 &        2 &     2 \\
 NN &      CWGAN &          3 &         3 &        3 &     3 \\
 PM &     AR-RET &          6 &         1 &        6 &     4 \\
 NN &        SIG &          4 &         6 &        4 &     5 \\
 NN &        VAE &          7 &         7 &        5 &     6 \\
 NN &    CGAN-FC &          5 &         4 &        7 &     7 \\
\bottomrule
\end{tabular}
\begin{tablenotes}\item[*] Model rank by subscores.  The lower the rank, the better the model.  Rank\_DIST: rank by distribution (DIST), Rank\_ACF: rank by ACF score, Rank\_BT: rank by backtesting score (BT), Rank: rank by composite score.\end{tablenotes}\end{table}

\clearpage
\subsubsection{Results with a market risk dataset}
For further testing, we compile another dataset for market risk, which includes eight key time series spanning the period 2007-2022: SPX and VIX for equity market, 3M and 10Y forward Treasury rates for fixed income market, North American CDX IG and HY for credit market, USD/EUR exchange rate for foreign exchange market, 1M Oil futures for commodity.  For this dataset, log returns are calculated for four variables SPX, VIX, USD/EUR and Oil, and absolute returns are calculated for the rest four variable
\footnote{Logs of the level for the four variables SPX, VIX, USD/EUR and Oil are saved into the input dataset and the models are run as if absolute return is appropriate for all eight variables in the dataset.}
.  The results are presented in Table~\ref{tab:combMRDATAshort}.  The results are broadly consistent with those for the USDYC datasets in the sense that the top three models are the same.  The difference from the results of USDYC dataset is that the AR-RET model performs well for USDYC dataset but it does not perform well for this market risk dataset.
\begin{table}
\centering
\caption{Model comparison for dataset MRDATA}
\label{tab:combMRDATAshort}
\begin{tabular}{cclrrrr}
\toprule
 Rank & Cat &      Model &  DIST &   ACF &    BT &  Composite \\
\midrule
    1 &  HS &        PHS & 0.365 & 0.804 & 0.384 &      1.552 \\
    2 &  PM & GARCHt-RET & 0.813 & 0.829 & 0.502 &      2.144 \\
    3 &  NN &      CWGAN & 0.793 & 0.835 & 0.535 &      2.163 \\
    4 &  NN &        VAE & 0.883 & 0.849 & 0.612 &      2.345 \\
    5 &  NN &        SIG & 0.923 & 0.921 & 0.568 &      2.412 \\
    6 &  NN &    CGAN-FC & 0.934 & 0.837 & 0.644 &      2.415 \\
    7 &  PM &     AR-RET & 0.992 & 0.866 & 0.572 &      2.430 \\
\bottomrule
\end{tabular}
\begin{tablenotes}\item[*] KPI/score: the smaller the value, the better the model.  KPIs are calclulated for one dataset. DIST: distribution score, ACF: autocorrelation score, BT: backtesting score.\end{tablenotes}\end{table}

\clearpage
\subsubsection{Rationalization of results\label{sec:rationalization}}
\paragraph{Histogram of u-values}
The model ranking for both simulated and historical datasets are based on KPIs (or scores) that capture different aspects of model comparison.  The KPIs are calculated through multiple steps of combination (or aggregation) of individual scores.

\begin{figure}
  \includegraphics[width=\linewidth]{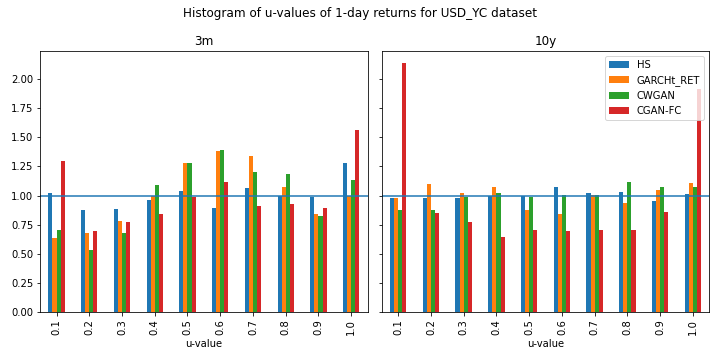}
  \caption{Histogram of u-values}
  \label{fig:histu}
\end{figure}

\begin{figure}
  \includegraphics[width=\linewidth]{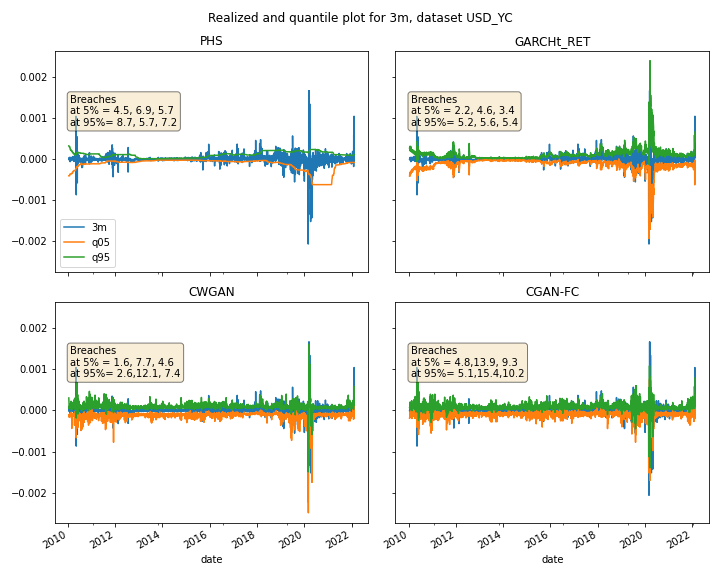}
  \caption{Envelope plot of 3M, USD\_YC}
  \label{fig:env3m}
\end{figure}

\begin{figure}
  \includegraphics[width=\linewidth]{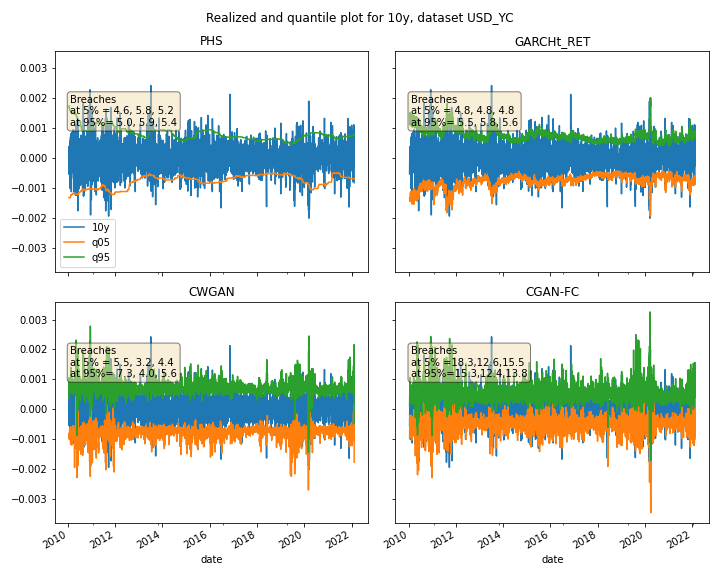}
  \caption{Envelope plot of 10Y, USD\_YC}
  \label{fig:env10y}
\end{figure}

To rationalize the results, we do additional analysis for four selected models, PHS, GARCH\_RET, CWGAN and NelsonSiegel model (NS) and shed more light on the model ranking and on the model features behind the KPIs.  We recall several key quantities for backtesting.\\[2ex]
For each business date $t$ in the backtest period, a condition (e.g. the previous 10 day's returns in period $t-10$ to $t-1$) is used along with a set of random numbers to generate a return distribution for every day in the next 10 days (including $t$).   The realized value $x_t$ is compared with the generated distribution for time $t, \tilde{x}_t,$ to calculate a u-value, $u_t=\hat{F_t}(x_t)$, where $\hat{F_t}$ is the generated distribution.  u-values, as probabilities, range from 0 to 1.  Histogram of u-values are presented in Figure~\ref{fig:histu}.  For a true model (or true distribution $\hat{F_t}$), the u-values should follow a uniform distribution, and the histogram should be flat at 1.0.  The following observations can be made based on Figure~\ref{fig:histu}:
\begin{itemize}
\item As the tenor gets longer, the histogram gets closer to uniform and there are less difference among the models (except for CGAN-FC).
\item Even for the top three models, the histogram seems rugged and different from uniform.  Statistical test for the null hypothesis of uniform distribution belong to formal model backtesting and is beyond the scope of this paper.
\end{itemize}
\paragraph{Envelope coverage}
The realized value $x_t$, and the $5\%$ and $95\%$ quantiles of the forecast distribution are plotted for every business date in the backtest period.  This is called an envelope plot, and it shows how well the generated distribution covers the realized returns.  Figure~\ref{fig:env3m} shows the envelope plot for 3M tenor and Figure~\ref{fig:env10y} for 10Y tenor.  The following observations can be made:
\begin{itemize}
\item The upper (95\%) and lower (5\%) quantiles for NN models (CWGAN) have very noisy day to day changes.
\item The upper (95\%) and lower (5\%) quantiles for HS and GARCHt-RET models and are less noisy and show clear trends over time corresponding to market episodes.
\item For 3M tenor, the upper quantiles for NN models become negative during the COVID-19 pandemic.  It is likely that, since the NN models depend on the condition (lagged values), when the condition has large values, the NN model prediction breaks down because there are not many such episodes in the training data.
\item The envelope plots for 10Y tenor is better and there is cleaner separation of upper and lower quantiles, since 10Y data are more stable and less challenging to model.
\item In the plot, we also show the breach rates for 5\% and 95\% quantiles.  For example, for GARCHt-RET model for 3m tenor, the meaning of ``at 5\%=2.2, 4.6, 3.4'' is, (1) in the first half of the date range (2008-2016, or first sub-period), 2.2\% of the realized return is lower than the 5\% quantile.  (2) In the second half of the date range (2016-2022, or second sub-period), the breach rate is 4.6\%. (3) In the whole date range, the breach rate is 3.4\%.  For the true model, the breach rate should be 5\%.  Thus the difference between the realized and expected breach rate (5\%) shows deviation of the model from the true model.  The breach rates for 95\% quantile are displayed similarly and show that the breach rates are close to 5\%.
\item For CWGAN model, the breach rates in the first sub-period (1.6\%) is much lower than 5\%, and those in the second sub-period (7.7\%) are much higher than 5\%, with the overall breach rates being much close to 5\%.  This serves as a reminder than sub-period analysis is important for studying model behavior.
\end{itemize}

\clearpage
\section{Conclusion and future work\label{sec:Model-performance-and-testing}}
This paper provides a comparative review of models for short horizon distribution forecasting, covering historical simulation models, parametric models as well as deep generative (NN) models. Model testing is conducted on both simulated and real-world historical data. The hyperparameters are set so that the performance ranking can be done consistently across all models. Comprehensive measures of model performance and aggregation of KPI schema are designed carefully to capture key features financial time series. 

The findings can be summarized as follows. (1) HS model perform the best across all simulated and historical datasets used in model testing.  HS VaR models are currently in use by many commercial banks, thus are included for comparison with parametric and NN models.  The limitation of HS models is that all future scenarios are drawn from and thus limited to realized historical scenarios.  The simple implementation of HS models do not generate a desired variety of future scenarios. (2) GARCH model (for returns, with conditional t distribution in particular) is consistently the best parametric model, and ranked only behind HS model. GARCH model captures volatility clustering, which is a salient feature of financial time series.  AR-RET also has good performance for some datasets, but not as good as GARCH-RET model. (3) CWGAN is consistently the best performing deep generative model across different datasets. (4) Other deep learning models such as SIG,  CGAN and VAE also have good performance for some datasets, however the performance is not consistent across all datasets. (5) Simulated data from GARCH and t distribution with three degrees of freedom have very different model ranking from other datasets.  This may be due to the very fat tails of the simulated data.

The contributions of our study include:
\begin{itemize}
\item A comprehensive review of three categories of models (historical simulation, parametric and deep generative models).
\item Innovations of deep learning methodologies, namely Encoder-Decoder CGAN (CGAN-LSTM) and continuous-conditional VAE.
\item A complete review of a collection of KPIs, and designed schema to combine KPIs.
\item Consistent model testing and performance ranking using simulated and real-world data.
%\item Model testing results show promising application of some deep learning models to financial time series modeling for short horizon market risk management.
\end{itemize}
Future work includes the following areas: 
(1) Expand the model list to include other NN architectures such as transformers; (2) Implement NN models such as Tail-GAN (\cite{cont2023tailgan}) that can improve modeling of the tails of the distribution; (3) Research on distribution forecasting for longer risk horizons, such as 5-, 10- and even 30-years; (4) Improve ways to integrate KPI's, such as KPIs for data variability and variety.
\\[4ex]
\textbf{Authors' affiliation and contacts}: all authors are with Wells Fargo.  Authors' email contacts are:\\
Lars Ericson: \href{mailto:Lars.Ericson@wellsfargo.com}{Lars.Ericson@wellsfargo.com}\\
Xuejun Zhu: \href{mailto:Xuejun.Zhu@wellsfargo.com}{Xuejun.Zhu@wellsfargo.com}\\
Xusi Han: \href{Xusi.Han@wellsfargo.com}{Xusi.Han@wellsfargo.com}\\
Rao Fu: \href{Rao.Fu@wellsfargo.com}{Rao.Fu@wellsfargo.com}\\
Shuang Li: \href{Shuang.Li2@wellsfargo.com}{Shuang.Li2@wellsfargo.com}\\
Steve Guo*: \href{Steve.Guo@wellsfargo.com}{Steve.Guo@wellsfargo.com}, corresponding author\\
Ping Hu: \href{Ping.Hu@wellsfargo.com}{Ping.Hu@wellsfargo.com}

\bibliographystyle{chicagoa}
\phantomsection\addcontentsline{toc}{section}{\refname}  %\nocite{*}
%\bibliography{main.bbl}
\bibliography{combined_bib_file_20240118}

\begin{thebibliography}{}

\bibitem[\protect\citeauthoryear{Arjovsky, Chintala, and Bottou}{Arjovsky
  et~al.}{2017}]{https://doi.org/10.48550/arxiv.1701.07875}
Arjovsky, M., S.~Chintala, and L.~Bottou (2017).
\newblock Wasserstein gan.
\newblock \url{https://arxiv.org/abs/1701.07875}.

\bibitem[\protect\citeauthoryear{BARONE-ADESI and GIANNOPOULOS}{BARONE-ADESI
  and GIANNOPOULOS}{2001}]{bg2001}
BARONE-ADESI, G. and K.~GIANNOPOULOS (2001).
\newblock Non-parametric var techniques. myths and realities.
\newblock {\em Economic Notes by Banca Monte dei Paschi di Siena SpA\/}~{\em
  30\/}(2), 167--181.


\bibitem[\protect\citeauthoryear{Barone-Adesi, Giannopoulos, and
  Vosper}{Barone-Adesi et~al.}{1999}]{bgv1999}
Barone-Adesi, G., K.~Giannopoulos, and L.~Vosper (1999).
\newblock Var without correlations for portfolios of derivatives securities.
\newblock {\em Journal of Futures Markets\/}~{\em 19}, 583--602.


\bibitem[\protect\citeauthoryear{{Basel Committee on Banking
  Supervision}}{{Basel Committee on Banking Supervision}}{2019}]{FRTB2019}
{Basel Committee on Banking Supervision} (2019).
\newblock Minimum capital requirements for market risk.
\newblock {\em Bank for international settlements\/}.


\bibitem[\protect\citeauthoryear{Berkowitz}{Berkowitz}{2001}]{berkowitz2001}
Berkowitz, J. (2001).
\newblock Testing density forecasts, with applications to risk management.
\newblock {\em Journal of Business \& Economic Statistics\/}~{\em 19\/}(4),
  465--474.


\bibitem[\protect\citeauthoryear{Bollerslev}{Bollerslev}{1986}]{garch}
Bollerslev, T. (1986).
\newblock {Generalized Autoregressive Conditional Heteroskedasticity}.
\newblock {\em Journal of Economics\/}~{\em 31\/}(3), 307--327.


\bibitem[\protect\citeauthoryear{Brophy, Wang, She, and Ward}{Brophy
  et~al.}{2021}]{https://doi.org/10.48550/arxiv.2107.11098}
Brophy, E., Z.~Wang, Q.~She, and T.~Ward (2021).
\newblock Generative adversarial networks in time series: A survey and
  taxonomy.
\newblock \url{https://arxiv.org/abs/2107.11098}.

\bibitem[\protect\citeauthoryear{Brophy, Wang, She, and Ward}{Brophy
  et~al.}{2023}]{10.1145/3559540}
Brophy, E., Z.~Wang, Q.~She, and T.~Ward (2023, feb).
\newblock Generative adversarial networks in time series: A systematic
  literature review.
\newblock {\em ACM Comput. Surv.\/}~{\em 55\/}(10).
\newblock \url{https://doi.org/10.1145/3559540}.


\bibitem[\protect\citeauthoryear{Chevyrev and Kormilitzin}{Chevyrev and
  Kormilitzin}{2016}]{chevyrev2016primer}
Chevyrev, I. and A.~Kormilitzin (2016).
\newblock A primer on the signature method in machine learning.
\newblock {\em arXiv preprint arXiv:1603.03788\/}.


\bibitem[\protect\citeauthoryear{Cont, Cucuringu, Xu, and Zhang}{Cont
  et~al.}{2023}]{cont2023tailgan}
Cont, R., M.~Cucuringu, R.~Xu, and C.~Zhang (2023).
\newblock Tail-gan: Learning to simulate tail risk scenarios.

\bibitem[\protect\citeauthoryear{Cox, Ingersoll, and Rossi}{Cox
  et~al.}{1985}]{cir1985}
Cox, J.~C., J.~E. Ingersoll, and S.~A. Rossi (1985).
\newblock {A Theory of the Term Structure of Interest Rates}.
\newblock {\em Econometrica\/}~{\em 53\/}(2), 385--407.


\bibitem[\protect\citeauthoryear{Crnkovic and Drachman}{Crnkovic and
  Drachman}{1996}]{crnkovic1996}
Crnkovic, C. and J.~Drachman (1996).
\newblock Quality control.
\newblock {\em Risk\/}~{\em 9\/}(9), 138--143.


\bibitem[\protect\citeauthoryear{Desai, Freeman, Wang, and Beaver}{Desai
  et~al.}{2021}]{Desai2021}
Desai, A., C.~Freeman, Z.~Wang, and I.~Beaver (2021).
\newblock Timevae: A variational auto-encoder for multivariate time series
  generation.
\newblock \url{https://arxiv.org/abs/2111.08095}.

\bibitem[\protect\citeauthoryear{Dhariwal and Nichol}{Dhariwal and
  Nichol}{2021}]{NEURIPS2021_49ad23d1}
Dhariwal, P. and A.~Nichol (2021).
\newblock Diffusion models beat gans on image synthesis.
\newblock In M.~Ranzato, A.~Beygelzimer, Y.~Dauphin, P.~Liang, and J.~W.
  Vaughan (Eds.), {\em Advances in Neural Information Processing Systems},
  Volume~34, pp.\  8780--8794. Curran Associates, Inc.
\newblock
  \url{https://proceedings.neurips.cc/paper/2021/file/49ad23d1ec9fa4bd8d77d02681df5cfa-Paper.pdf}.

\bibitem[\protect\citeauthoryear{Dickey and Fuller}{Dickey and
  Fuller}{1979}]{df1979}
Dickey, D. and W.~Fuller (1979).
\newblock Distribution of the estimators for autoregressive time series with a
  unit root.
\newblock {\em Journal of the American Statistical Association\/}~{\em
  74\/}(366), 427--431.


\bibitem[\protect\citeauthoryear{Diebold and Li}{Diebold and
  Li}{2006}]{dieboldli2006}
Diebold, F.~X. and C.~Li (2006).
\newblock {Forecasting the Term Structure of Government Bond Yields}.
\newblock {\em Journal of Economics\/}~{\em 130\/}(2), 337--364.


\bibitem[\protect\citeauthoryear{Dragulescu and Yakovenko}{Dragulescu and
  Yakovenko}{2002}]{DragulescuEtAl02}
Dragulescu, A.~A. and V.~M. Yakovenko (2002).
\newblock Probability distribution of returns in the heston model with
  stochastic volatility.
\newblock \url{https://arxiv.org/abs/cond-mat/0203046}.


\bibitem[\protect\citeauthoryear{Engle}{Engle}{1982}]{arch}
Engle, R.~F. (1982).
\newblock "autoregressive conditional heteroskedasticity with estimates of the
  variance of {United Kingdom} inflation".
\newblock {\em Econometrica\/}~{\em 50\/}(4), 987--1007.


\bibitem[\protect\citeauthoryear{Fu}{Fu}{2022}]{Fu22}
Fu, R. (2022).
\newblock An introduction of encoder-and-decoder cgan framework.

\bibitem[\protect\citeauthoryear{Fu, Chen, Zeng, Zhuang, and Sudjianto}{Fu
  et~al.}{2019}]{https://doi.org/10.48550/arxiv.1904.11419}
Fu, R., J.~Chen, S.~Zeng, Y.~Zhuang, and A.~Sudjianto (2019).
\newblock Time series simulation by conditional generative adversarial net.
\newblock \url{https://arxiv.org/abs/1904.11419}.

\bibitem[\protect\citeauthoryear{Graves}{Graves}{2013}]{https://doi.org/10.48550/arxiv.1308.0850}
Graves, A. (2013).
\newblock Generating sequences with recurrent neural networks.
\newblock \url{https://arxiv.org/abs/1308.0850}.

\bibitem[\protect\citeauthoryear{Ho, Jain, and Abbeel}{Ho
  et~al.}{2020}]{NEURIPS2020_4c5bcfec}
Ho, J., A.~Jain, and P.~Abbeel (2020).
\newblock Denoising diffusion probabilistic models.
\newblock In H.~Larochelle, M.~Ranzato, R.~Hadsell, M.~Balcan, and H.~Lin
  (Eds.), {\em Advances in Neural Information Processing Systems}, Volume~33,
  pp.\  6840--6851. Curran Associates, Inc.
\newblock
  \url{https://proceedings.neurips.cc/paper/2020/file/4c5bcfec8584af0d967f1ab10179ca4b-Paper.pdf}.

\bibitem[\protect\citeauthoryear{Johnson}{Johnson}{2020}]{Johnson20}
Johnson, S. (2020).
\newblock Synthetic data for finance - from theory to practice.
\newblock
  \url{https://www.ubs.com/global/en/investment-bank/in-focus/research-focus/quant-answers/quant-insight-series/_jcr_content/mainpar/toplevelgrid_7262680_322968126/col3/actionbutton_3358030.0949818704.file/PS9jb250ZW50L2RhbS9pbnRlcm5ldGhvc3RpbmcvaW52ZXN0bWVudGJhbmsvZW4vZXF1aXRpZXMvcWlzLXZpcnR1YWwtZXZlbnQtZGVja3MtMjAyMC9zeW50aGV0aWMtZGF0YS1mb3ItZmluYW5jZS1zdGVmYW4tamFuc2VuLXVicy0yMDIxLnBkZg==/synthetic-data-for-finance-stefan-jansen-ubs-2021.pdf}.

\bibitem[\protect\citeauthoryear{Kingma and Welling}{Kingma and
  Welling}{2019}]{Kingma_2019}
Kingma, D.~P. and M.~Welling (2019).
\newblock An introduction to variational autoencoders.
\newblock {\em Foundations and Trends{\textregistered} in Machine
  Learning\/}~{\em 12\/}(4), 307--392.
\newblock \url{https://doi.org/10.1561%2F2200000056}.


\bibitem[\protect\citeauthoryear{Kuznetsov and Mariet}{Kuznetsov and
  Mariet}{2018}]{https://doi.org/10.48550/arxiv.1805.03714}
Kuznetsov, V. and Z.~Mariet (2018).
\newblock Foundations of sequence-to-sequence modeling for time series.
\newblock \url{https://arxiv.org/abs/1805.03714}.

\bibitem[\protect\citeauthoryear{McInnes, Healy, and Melville}{McInnes
  et~al.}{2020}]{mcinnes2020umap}
McInnes, L., J.~Healy, and J.~Melville (2020).
\newblock Umap: Uniform manifold approximation and projection for dimension
  reduction.

\bibitem[\protect\citeauthoryear{Morrill, Fermanian, Kidger, and Lyons}{Morrill
  et~al.}{2020}]{morrill2020generalised}
Morrill, J., A.~Fermanian, P.~Kidger, and T.~Lyons (2020).
\newblock A generalised signature method for multivariate time series feature
  extraction.
\newblock {\em arXiv preprint arXiv:2006.00873\/}.


\bibitem[\protect\citeauthoryear{Nelson and Siegel}{Nelson and
  Siegel}{1987}]{ns1987}
Nelson, C.~R. and A.~F. Siegel (1987).
\newblock Parsimonious modeling of yield curve.
\newblock {\em Journal of Business\/}~{\em 60}, 473--489.


\bibitem[\protect\citeauthoryear{Perignon and Smith}{Perignon and
  Smith}{2010}]{perignon2010}
Perignon, C. and D.~R. Smith (2010).
\newblock The level and quality of value-at-risk disclosure by commercial
  banks.
\newblock {\em Journal of Banking \& Finance\/}~{\em 34\/}(2), 362--377.


\bibitem[\protect\citeauthoryear{Ramdas, Garcia, and Cuturi}{Ramdas
  et~al.}{2015}]{ramdas2015wasserstein}
Ramdas, A., N.~Garcia, and M.~Cuturi (2015).
\newblock On wasserstein two sample testing and related families of
  nonparametric tests.
\newblock \url{https://arxiv.org/abs/1509.02237}.

\bibitem[\protect\citeauthoryear{Ramya Malur~Srinivasan}{Ramya
  Malur~Srinivasan}{2020}]{Fujitsu20}
Ramya Malur~Srinivasan, A.~C. (2020).
\newblock Generating user-friendly explanations for loan denials using
  generative adversarial networks.
\newblock
  \url{https://www.risk.net/investing/quant-investing/7833326/in-fake-data-quants-see-a-fix-for-backtesting}.

\bibitem[\protect\citeauthoryear{Rasul, Seward, Schuster, and Vollgraf}{Rasul
  et~al.}{2021}]{https://doi.org/10.48550/arxiv.2101.12072}
Rasul, K., C.~Seward, I.~Schuster, and R.~Vollgraf (2021).
\newblock Autoregressive denoising diffusion models for multivariate
  probabilistic time series forecasting.
\newblock \url{https://arxiv.org/abs/2101.12072}.


\bibitem[\protect\citeauthoryear{Sabat\'e}{Sabat\'e}{2021}]{Sabate21}
Sabat\'e, M. (2021).
\newblock The authors' official pytorch sigcwgan implementation.

\bibitem[\protect\citeauthoryear{Song, Meng, and Ermon}{Song
  et~al.}{2020}]{https://doi.org/10.48550/arxiv.2010.02502}
Song, J., C.~Meng, and S.~Ermon (2020).
\newblock Denoising diffusion implicit models.
\newblock \url{https://arxiv.org/abs/2010.02502}.

\bibitem[\protect\citeauthoryear{Srinivasan and Knottenbelt}{Srinivasan and
  Knottenbelt}{2022}]{Srinivasan2022}
Srinivasan, P. and W.~J. Knottenbelt (2022).
\newblock Time-series transformer generative adversarial networks.
\newblock \url{https://arxiv.org/abs/2205.11164}.

\bibitem[\protect\citeauthoryear{Sutskever, Vinyals, and Le}{Sutskever
  et~al.}{2014}]{NIPS2014_a14ac55a}
Sutskever, I., O.~Vinyals, and Q.~V. Le (2014).
\newblock Sequence to sequence learning with neural networks.
\newblock In Z.~Ghahramani, M.~Welling, C.~Cortes, N.~Lawrence, and
  K.~Weinberger (Eds.), {\em Advances in Neural Information Processing
  Systems}, Volume~27. Curran Associates, Inc.
\newblock
  \url{https://proceedings.neurips.cc/paper/2014/file/a14ac55a4f27472c5d894ec1c3c743d2-Paper.pdf}.

\bibitem[\protect\citeauthoryear{van~den Berg}{van~den
  Berg}{2011}]{VasicekCalibration}
van~den Berg, T. (2011).
\newblock Calibrating the ornstein-uhlenbeck ({Vasicek}) model.
\newblock
  \url{https://www.statisticshowto.com/wp-content/uploads/2016/01/Calibrating-the-Ornstein.pdf}.


\bibitem[\protect\citeauthoryear{van~der Maaten and Hinton}{van~der Maaten and
  Hinton}{2008}]{JMLR:v9:vandermaaten08a}
van~der Maaten, L. and G.~Hinton (2008).
\newblock Visualizing data using t-sne.
\newblock {\em Journal of Machine Learning Research\/}~{\em 9\/}(86),
  2579--2605.


\bibitem[\protect\citeauthoryear{Vasicek}{Vasicek}{1977}]{vasicek1977}
Vasicek, O. (1977).
\newblock An equilibrium characterization of the term structure.
\newblock {\em Journal of Financial Economics\/}~{\em 5\/}(2), 177--188.


\bibitem[\protect\citeauthoryear{Villani}{Villani}{2008}]{villani2008optimal}
Villani, C. (2008).
\newblock {\em Optimal Transport: Old and New}.
\newblock Grundlehren der mathematischen Wissenschaften. Springer Berlin
  Heidelberg.


\bibitem[\protect\citeauthoryear{Yu, Srivastava, and Canales}{Yu
  et~al.}{2021}]{Yu_2021}
Yu, Y., A.~Srivastava, and S.~Canales (2021, feb).
\newblock Conditional {LSTM}-{GAN} for melody generation from lyrics.
\newblock {\em {ACM} Transactions on Multimedia Computing, Communications, and
  Applications\/}~{\em 17\/}(1), 1--20.
\newblock \url{https://doi.org/10.1145%2F3424116}.


\end{thebibliography}

\clearpage{}
\appendix
\thispagestyle{empty}\null

\pagenumbering{arabic}

\renewcommand{\thefootnote}{A\arabic{footnote}}
\renewcommand{\thepage}{A\arabic{page}}
\renewcommand{\thetable}{A\arabic{table}}
\renewcommand{\thefigure}{A\arabic{figure}}

\setcounter{footnote}{0} 
\setcounter{section}{0}
\setcounter{table}{0}
\setcounter{figure}{0}
\section{Appendix\label{sec:Appendix-1}}
\subsection{List of githubs for source code\label{sec:githubs}}
The following list of githubs is used as the source code for several NN models.  We made some changes for our specific purpose.
\begin{itemize}
\item \url{https://github.com/FernandoDeMeer/Hierarchical-SigCWGAN}
\item \url{https://github.com/SigCGANs/Conditional-Sig-Wasserstein-GANs}
\item \url{https://github.com/abudesai/timeVAE}
\item \url{https://github.com/luphord/nelson_siegel_svensson}
\end{itemize}
\clearpage
\subsection{Results with 1Y condition for NN models\label{sec:res1y}}
The results in the Tables~\ref{tab:combUSDYC1nn1ycond} and \ref{tab:sizeUSDYC1nn1ycond} show that with 1-year condition, the NN models have much more parameters but do not have improved model performance.

\begin{table}
\centering
\caption{Model comparison for dataset USDYC1 (Libor curve) (cond=1y)}
\label{tab:combUSDYC1nn1ycond}
\begin{tabular}{cclrrrr}
\toprule
 Rank & Cat &     Model &  DIST &   ACF &    BT &  Composite \\
\midrule
    1 &  NN & CGAN-LSTM & 0.995 & 0.969 & 0.606 &      2.570 \\
    2 &  NN &       VAE & 0.981 & 0.933 & 0.674 &      2.589 \\
    3 &  NN &     CWGAN & 0.993 & 0.647 & 1.573 &      3.213 \\
    4 &  NN & DIFFUSION & 0.899 & 0.740 & 1.802 &      3.441 \\
    5 &  NN &   CGAN-FC & 0.999 & 0.724 & 1.752 &      3.475 \\
\bottomrule
\end{tabular}
\begin{tablenotes}\item[*] KPI/score: the smaller the value, the better the model.  KPIs are calclulated for one dataset. DIST: distribution score, ACF: autocorrelation score, BT: backtesting score.\end{tablenotes}\end{table}

\begin{table}
\centering
\caption{Number of model parameters for USD Libor dataset (nn1ycond)}
\label{tab:sizeUSDYC1nn1ycond}
\begin{tabular}{lrrrr}
\toprule
    Model &   GEN/DEC &    DIS/ENC &      TOTAL & Code Library \\
\midrule
  CGAN-FC &   413,914 &    665,345 &  1,079,259 &   Tensorflow \\
    CWGAN & 1,074,138 &  1,264,385 &  2,338,523 &        Torch \\
DIFFUSION &    39,279 &          0 &     39,279 &      GluonTS \\
CGAN-LSTM &     9,929 &      8,502 &     18,431 &   Tensorflow \\
      SIG &   182,316 & 15,220,260 & 15,402,576 &        Torch \\
\bottomrule
\end{tabular}
\end{table}

\end{document}